\documentclass[a4paper,11pt]{article}
\pdfoutput=1 

\usepackage{jheppub} 

\usepackage[T1]{fontenc} 

\usepackage[acronym]{glossaries}
\glsdisablehyper

\usepackage{color}
\usepackage[dvipsnames]{xcolor}
\usepackage{graphicx}
\usepackage{wrapfig}
\usepackage{verbatim}
\usepackage{amsmath}
\usepackage{amssymb}
\usepackage{url}
\usepackage{bbold}
\usepackage{xspace}
\usepackage{slashed}
\usepackage{multirow}
\usepackage{threeparttable}
\usepackage{paralist}
\usepackage{subcaption}
\usepackage{floatrow}
\usepackage{placeins}
\usepackage{soul}
\usepackage[export]{adjustbox}
\usepackage[normalem]{ulem}
\usepackage{braket}
\usepackage{feynmp-auto}
\usepackage{ytableau}

\usepackage{MnSymbol}

\usepackage{relsize}

\usepackage{scalerel,stackengine}

\usepackage{tikz}
\usetikzlibrary{patterns,patterns.meta}
\usetikzlibrary{arrows,shapes}
\usetikzlibrary{arrows.meta,bending}
\usetikzlibrary{trees}
\usetikzlibrary{matrix,arrows} 				
\usetikzlibrary{calc} 
\usetikzlibrary{positioning}				
\usetikzlibrary{calc,through}				
\usetikzlibrary{decorations.pathreplacing}  
\usepackage[tikz]{bclogo} 					
\usepackage{pgffor}							

\usetikzlibrary{decorations.pathmorphing}	
\usetikzlibrary{decorations.markings}
\usetikzlibrary{intersections}				
\tikzset{
    vector/.style={decorate, decoration={snake}, draw},
    graviton/.style={decorate, decoration={snake,amplitude=1.5pt}, draw},
    fermion/.style={postaction={decorate},
        decoration={markings,mark=at position .55 with {\arrow{>}}}},
    fermionbar/.style={draw, postaction={decorate},
        decoration={markings,mark=at position .55 with {\arrow{<}}}},
    fermionnoarrow/.style={},
    gluon/.style={decorate,
        decoration={coil,amplitude=4pt, segment length=5pt}},
    scalar/.style={dashed, postaction={decorate},
        decoration={markings,mark=at position .55 with {\arrow{>}}}},
    scalarbar/.style={dashed, postaction={decorate},
        decoration={markings,mark=at position .55 with {\arrow{<}}}},
    scalarnoarrow/.style={dashed,draw},
	provector/.style={decorate, decoration={snake,amplitude=2.5pt}, draw},
	antivector/.style={decorate, decoration={snake,amplitude=-2.5pt}, draw},
	    electron/.style={draw=black, postaction={decorate},
        decoration={markings,mark=at position .55 with {\arrow[draw=black]{>}}}},
	bigvector/.style={decorate, decoration={snake,amplitude=4pt}, draw},
	vectorscalar/.style={loosely dotted,draw=black, postaction={decorate}},
}

\tikzset{
    partial ellipse/.style args={#1:#2:#3}{
        insert path={+ (#1:#3) arc (#1:#2:#3)}
    }
}

\newcommand{\SSYTshape}[3]{%
    \begingroup
    \pgfmathtruncatemacro{\ssytwhitecols}{#2/2}%
    \pgfmathtruncatemacro{\ssytbluecols}{#3/2}%
    \pgfmathtruncatemacro{\ssytblueheight}{2}%
    \pgfmathtruncatemacro{\ssyttotalcols}{\ssytwhitecols+\ssytbluecols}%
    \ifnum\ssyttotalcols=0
        \ensuremath{\mathbf 1}%
    \else
        \begin{tikzpicture}[baseline=-0.5ex,x=6.5pt,y=6.5pt]
            \ifnum\ssytbluecols>0
                \foreach \ssytx in {1,...,\ssytbluecols}{
                    \foreach \ssyty in {1,...,\ssytblueheight}{
                        \path[fill=blue!40,draw=black,line width=0.3pt]
                            ({\ssytx-1},{\ssyty-1}) rectangle (\ssytx,\ssyty);
                    }
                }
                \draw[line width=0.35pt]
                    (-0.08,{\ssytblueheight+0.25}) --
                    ({\ssytbluecols+0.08},{\ssytblueheight+0.25});
            \fi
            \ifnum\ssytwhitecols>0
                \foreach \ssytx in {1,...,\ssytwhitecols}{
                    \foreach \ssyty in {1,2}{
                        \path[fill=white,draw=black,line width=0.3pt]
                            ({\ssytbluecols+\ssytx-1},{\ssyty-1}) rectangle
                            ({\ssytbluecols+\ssytx},\ssyty);
                    }
                }
            \fi
        \end{tikzpicture}%
    \fi
    \endgroup
}

\usepackage{accents}

\newcommand{\U}{\mathrm{U}}
\newcommand{\SU}{\mathrm{SU}}

\newcommand{\OO}{\mathrm{O}}

\definecolor{pastelpurp}{RGB}{140,84,208}

\newcommand{\ie}{\emph{i.e.}\xspace}

\newcommand{\be}{\begin{eqnarray}}
\newcommand{\ee}{\end{eqnarray}}

\newcommand{\CC}{\mathbb{C}}

\newcommand\reallywidehat[1]{%
\savestack{\tmpbox}{\stretchto{%
  \scaleto{%
    \scalerel*[\widthof{\ensuremath{#1}}]{\kern-.6pt\bigwedge\kern-.6pt}%
    {\rule[-\textheight/2]{1ex}{\textheight}}
  }{\textheight}%
}{0.5ex}}%
\stackon[1pt]{#1}{\tmpbox}%
}

\newacronym{sm}{SM}{Standard Model}
\newacronym[shortplural=EFTs, longplural=effective field theories]{eft}{EFT}{effective field theory}
\newacronym{smeft}{SMEFT}{Standard Model effective field theory}
\newacronym[shortplural=ADMs, longplural=anomalous-dimension matrices]{adm}{ADM}{anomalous-dimension matrix}
\newacronym[shortplural=RGEs, longplural=renormalization group equations]{rge}{RGE}{renormalization group equation}
\newacronym[shortplural=YTs, longplural=Young tableaux]{yt}{YT}{Young tableau}
\newacronym[shortplural=SSYTs, longplural=semi-standard Young tableaux]{ssyt}{SSYT}{semi-standard Young tableau}

\DeclareRobustCommand{\Sec}[1]{Sec.~\ref{#1}}
\DeclareRobustCommand{\Secs}[2]{Secs.~\ref{#1} and \ref{#2}}
\DeclareRobustCommand{\App}[1]{App.~\ref{#1}}
\DeclareRobustCommand{\Apps}[2]{Apps.~\ref{#1} and \ref{#2}}

\DeclareRobustCommand{\Fig}[1]{Fig.~\ref{#1}}
\DeclareRobustCommand{\Figs}[2]{Figs.~\ref{#1} and \ref{#2}}
\DeclareRobustCommand{\Eq}[1]{Eq.~(\ref{#1})}
\DeclareRobustCommand{\Eqs}[2]{Eqs.~(\ref{#1}) and (\ref{#2})}

\DeclareRobustCommand{\okina}{%
  \raisebox{\dimexpr\fontcharht\font`A-\height}{%
    \scalebox{0.8}{`}%
  }%
}

\title{\resizebox{\textwidth}{!}{\shortstack[l]{The Harmonic Structure of Multiparticle Scattering:\\Amplitude Projection and Non-Renormalization}}}

\author[a,b]{Nathaniel Craig}
\author[c]{Brian Henning}
\author[d,e,f]{Michael Stadlbauer}

\affiliation[a]{Department of Physics, University of California, Santa Barbara, CA 93106, USA}
\affiliation[b]{Kavli Institute for Theoretical Physics, University of California, Santa Barbara, CA 93106, USA}
\affiliation[c]{Department of Physics and Astronomy, University of Hawai\okina i, Honolulu, HI 96822, USA}
\affiliation[d]{Technical University of Munich, TUM School of Natural Sciences, Physics Department, James-Franck-Str. 1, 85748 Garching, Germany}
\affiliation[e]{Max Planck Institute for Physics, Boltzmannstr. 8, 85748 Garching, Germany}
\affiliation[f]{Theoretical Particle Physics Laboratory (LPTP), Institute of Physics, EPFL, Lausanne, Switzerland}

\emailAdd{ncraig@ucsb.edu}
\emailAdd{bhenning@hawaii.edu}
\emailAdd{michael.stadlbauer@epfl.ch}

\abstract{
We develop a projection formalism for four-dimensional massless scattering amplitudes based on the construction of harmonics on multi-particle phase space.
This provides a generalization of the partial-wave decomposition valid for any number of particles.
Applying this formalism to calculations of on-shell cuts, we show that any cut calculation can be rewritten as a sum of products of projected coefficients and derive a condition for this sum to vanish.
Using the on-shell renormalization group equation in combination with our projection formalism, we derive new non-renormalization theorems for effective field theories valid at general loop orders.
}

\begin{document}
\DeclareRobustCommand{\Ref}[1]{Ref.~\cite{#1}}
\DeclareRobustCommand{\Refs}[1]{Refs.~\cite{#1}}
\maketitle
\flushbottom

\newpage

\section{Introduction}

\Glspl{eft} have become a standard framework for studying physics beyond the \gls{sm}.
They describe the low-energy effects of heavy particles through local interactions among the light fields, without requiring a specific model of the underlying physics.
These interactions are organized in an expansion in inverse powers of the heavy scale, with Wilson coefficients encoding the underlying dynamics.
The \gls{smeft} implements this expansion using the \gls{sm} fields and gauge symmetries, with the Warsaw basis providing a complete set of independent operators at dimension six~\cite{Grzadkowski:2010es}.
Renormalization group running couples the Wilson coefficients through operator mixing, allowing interactions absent at the matching scale to be generated radiatively.
The \gls{adm} governs this evolution and determines which operators can renormalize one another at a given loop order.
For baryon-number-conserving dimension-six \gls{smeft} operators, the complete one-loop \gls{adm} was first calculated in \Refs{Jenkins:2013zja,Jenkins:2013wua,Alonso:2013hga}.
More recently, the two-loop \glspl{rge} of the baryon-number-conserving \gls{smeft} through dimension six were presented in \Ref{Born:2026xkr}.

Nontrivial zeros in dimension-six operator mixing were already observed in partial one-loop calculations, before the complete \gls{adm} was available~\cite{Elias-Miro:2013gya}.
In particular, the dimension-six \gls{smeft} \gls{adm} was found to exhibit an approximate holomorphic structure~\cite{Alonso:2014rga}.
Several one-loop zeros were also understood by embedding the effective theory into a supersymmetric framework~\cite{Elias-Miro:2014eia}.
This structure was understood better by Cheung and Shen, who derived a general one-loop helicity theorem, explaining a broad class of one-loop zeros at fixed mass dimension using unitarity and the helicity structure of tree amplitudes~\cite{Cheung:2015aba}.
Their non-renormalization theorem constrains mixing through the holomorphic and anti-holomorphic weights of the source and target operators, with exceptional non-holomorphic Yukawa contributions excluded.
Beyond one loop, \Ref{Bern:2019wie} established restrictions based on operator length, showing that mixing from operators with more fields into operators with fewer fields can vanish even at the first order at which Feynman diagrams exist.
Further two-loop restrictions arise from color selection rules and the structure of one-loop rational terms~\cite{Craig:2019wmo}, with some additional zeros depending on the renormalization scheme~\cite{Bern:2020ikv}.
Additional three-loop zeros beyond the general operator-length bound have been found in complex scalar \glspl{eft}~\cite{Cao:2021cdt}.
Conservation of total angular momentum in a specified scattering channel also constrains operator mixing~\cite{Jiang:2020rwz}.
These results identify helicity, field content, and channel quantum numbers as useful organizing principles for the \gls{adm}.

On-shell methods make these principles accessible by relating renormalization to physical intermediate states.
The coefficients of ultraviolet logarithms also determine discontinuities, which unitarity expresses as phase-space integrals of amplitudes and form factors.
Caron-Huot and Wilhelm formulated this relation as an on-shell \gls{rge}~\cite{Caron-Huot:2016cwu}.
Its applications to effective theories have provided both direct calculations of anomalous dimensions and explanations of their zeros~\cite{Bern:2019wie,Baratella:2020lzz,EliasMiro:2020tdv,Jiang:2020mhe,Bern:2020ikv,AccettulliHuber:2021uoa,Machado:2022ozb}.
In this description, operator mixing is determined by the phase-space overlap of amplitudes and form factors, together with the infrared and lower-order terms required by the \gls{rge}~\cite{Baratella:2022nog}.
A suitable choice of basis for the amplitudes and form factors can make restrictions on mixing more transparent, which are otherwise difficult to see in individual Feynman diagrams~\cite{Jiang:2020rwz,Baratella:2020dvw}.

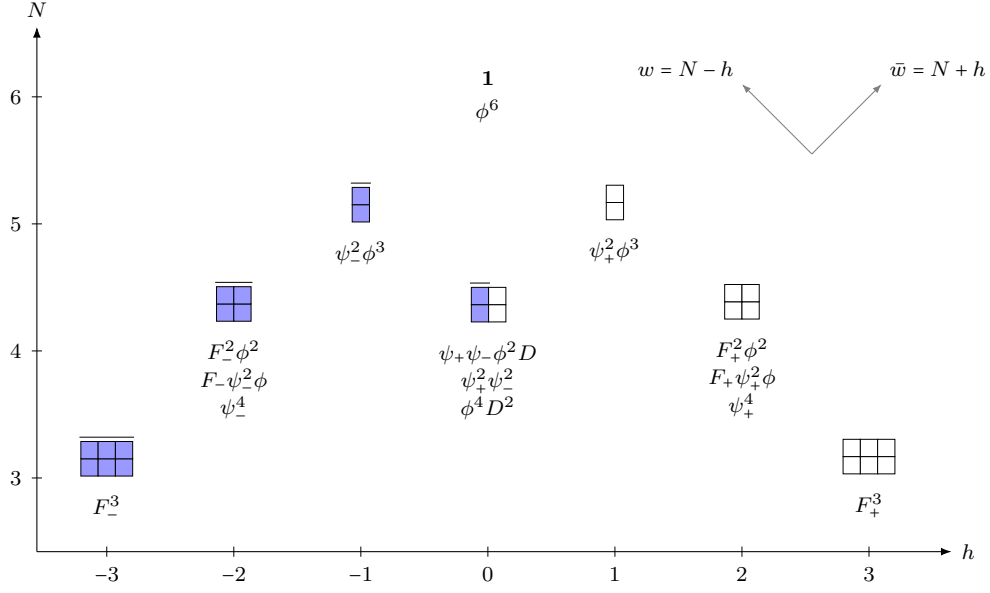
\begin{figure}[t]
    \centering
    \begin{tikzpicture}[
        x=1.68cm,
        y=1.68cm,
        operator entry/.style={
            align=center,
            text width=2.65cm,
            inner sep=0pt,
            font=\scriptsize
        }
    ]
        \draw[-{Latex[length=4pt]},line width=0.4pt]
            (-3.55,-0.58) -- (3.65,-0.58)
            node[right,font=\scriptsize] {$h$};
        \draw[-{Latex[length=4pt]},line width=0.4pt]
            (-3.55,-0.58) -- (-3.55,3.55)
            node[above,font=\scriptsize] {$N$};

        \foreach \x in {-3,-2,-1,0,1,2,3}{
            \draw[line width=0.3pt] (\x,-0.62) -- (\x,-0.54);
            \node[below=2pt,font=\scriptsize] at (\x,-0.58) {$\x$};
        }
        \foreach \y/\N in {0/3,1/4,2/5,3/6}{
            \draw[line width=0.3pt] (-3.59,\y) -- (-3.51,\y);
            \node[left=2pt,font=\scriptsize] at (-3.55,\y) {$\N$};
        }

        \node[operator entry] at (-3,0)
            {\SSYTshape{3}{0}{6}\\[2pt]$F_-^3$};
        \node[operator entry] at (-2,1)
            {\SSYTshape{4}{0}{4}\\[2pt]
             $F_-^2\phi^2$\\$F_-\psi_-^2\phi$\\$\psi_-^4$};
        \node[operator entry] at (-1,2)
            {\SSYTshape{5}{0}{2}\\[2pt]$\psi_-^2\phi^3$};

        \node[operator entry] at (0,1)
            {\SSYTshape{4}{2}{2}\\[2pt]
             $\psi_+\psi_-\phi^2D$\\$\psi_+^2\psi_-^2$\\$\phi^4D^2$};
        \node[operator entry] at (0,3)
            {\SSYTshape{6}{0}{0}\\[2pt]$\phi^6$};

        \node[operator entry] at (1,2)
            {\SSYTshape{5}{2}{0}\\[2pt]$\psi_+^2\phi^3$};
        \node[operator entry] at (2,1)
            {\SSYTshape{4}{4}{0}\\[2pt]
             $F_+^2\phi^2$\\$F_+\psi_+^2\phi$\\$\psi_+^4$};
        \node[operator entry] at (3,0)
            {\SSYTshape{3}{6}{0}\\[2pt]$F_+^3$};

        \coordinate (weightorigin) at (2.55,2.55);
        \draw[-{Latex[length=3.5pt]},gray,line width=0.4pt]
            (weightorigin) -- ++(-0.55,0.55)
            node[above left=-1pt,font=\scriptsize,text=black]
            {$w=N-h$};
        \draw[-{Latex[length=3.5pt]},gray,line width=0.4pt]
            (weightorigin) -- ++(0.55,0.55)
            node[above right=-1pt,font=\scriptsize,text=black]
            {$\bar w=N+h$};
    \end{tikzpicture}
    \caption{\Gls{yt} shapes associated with dimension-six \gls{smeft} contact amplitudes, organized by the number of external particles $N$ and total helicity $h$. The field-content classification follows \Ref{Cheung:2015aba}. Each point shows a tableau shape and its compatible field contents, with gauge and flavor contractions and tableau fillings suppressed.
    White and barred blue columns represent pairs of holomorphic and anti-holomorphic spinors, respectively, using the notation explained below \Eq{eq:li-ssyt-pictorial}.
    Here $F_\pm$, $\psi_\pm$, $\phi$, and $D$ denote field-strength components of helicity $\pm1$, fermions of helicity $\pm\frac12$, scalars, and derivatives, while $1$ denotes the trivial Lorentz polynomial. The diagonal arrows indicate the weights $w=N-h$ and $\bar w=N+h$. The helicity convention is specified in \Sec{sec:on-shell-rge-form-factors}.}
    \label{fig:dimension-six-ssyt-weight-lattice}
\end{figure}

There is a deep relation between the kinematic structures governing local on-shell amplitudes and the phase space over which the amplitude is defined. In particular, phase-space harmonics~\cite{Henning:2019mcv} provide a complete operator basis in the \gls{eft}~\cite{Henning:2019enq}. For \(N\)-particle massless Lorentz-invariant phase space (LIPS), where these results are best understood, the harmonics have a representation-theoretic origin as harmonics on the Stiefel manifold \(V_2(\mathbb{C}^N) \simeq U(N)/U(N-2)\), with Lorentz-scalar operators living on the Grassmannian submanifold \(G_2(\mathbb{C}^N) \simeq V_2(\mathbb{C}^N)/U(2)\). A major aspect of this work is to port this representation theory into a framework for understanding and evaluating the cuts that govern operator mixing.

The phase-space harmonics lead to a basis in terms of \glspl{ssyt}, organized by operator mass dimension, particle number, and helicity~\cite{Henning:2019mcv,Henning:2019enq}. 
The operator mass dimension $\Delta$, particle number $N$, and total helicity $h$ determine the \gls{yt} shape, while the allowed fillings distinguish independent Lorentz structures for specified individual helicities.
Once gauge and flavor structures and the exchange symmetries of identical particles are included, the resulting contact amplitudes correspond to independent \gls{eft} operators up to equations of motion and integration by parts~\cite{Henning:2017fpj,Henning:2019enq}.
For the field content and gauge symmetries of the \gls{sm}, this construction yields the \gls{smeft} operator bases.

At fixed operator mass dimension $\Delta$, the \gls{ssyt} basis organizes definite-helicity leading contact amplitudes by their \gls{yt} shapes on a lattice labeled by particle number $N$ and total helicity $h$.
Equivalently, this lattice can be labeled by the weights $w=N-h$ and $\bar w=N+h$, reproducing the helicity-weight classification of Cheung and Shen~\cite{Cheung:2015aba}.
Figure~\ref{fig:dimension-six-ssyt-weight-lattice} illustrates this correspondence for dimension-six \gls{smeft} contact amplitudes with definite external helicities, displaying the \gls{yt} shape associated with each lattice point.
For each operator field content listed below a tableau, the allowed \gls{ssyt} fillings distinguish independent Lorentz structures compatible with the specified external helicities.
The helicity-based restrictions on operator mixing of \Ref{Cheung:2015aba} can thus be interpreted as restrictions on mixing between tableau shapes.
With exceptional Yukawa contributions excluded, one-loop mixing from a source $j$ into a target $i$ requires $w_i\geq w_j$ and $\bar w_i\geq\bar w_j$~\cite{Cheung:2015aba}.
In \Fig{fig:dimension-six-ssyt-weight-lattice}, these inequalities restrict the possible targets to the source's upward-pointing ``forward light cone'', including its boundary.

At this level, the shapes in \Fig{fig:dimension-six-ssyt-weight-lattice} encode the same information as the helicity weights.
Further information is contained in the independent polynomials within these sectors and in their angular-momentum content in a chosen scattering channel.
Partial-wave methods organize amplitudes by angular momentum and express cut integrals algebraically, with general constructions available for multiparticle states~\cite{Jiang:2020rwz,Baratella:2020dvw,Shu:2021qlr}.
The general construction in \Ref{Henning:2019mcv} provides a basis of Stiefel harmonics, labeled by \glspl{ssyt}, for multiparticle states at fixed total momentum.
We develop a channel projection onto these Stiefel harmonics, together with a joint \gls{ssyt} and partial-wave projection that resolves their angular-momentum content.
This projection formalism constitutes one of the main results of this work.
Both projections act separately on the incoming and outgoing states of a chosen scattering channel.
In the joint formulation, the channel harmonics have definite total angular momentum $J$, and their tableau shape and filling labels explicitly resolve the kinematic degeneracy at fixed $J$ in the general multiparticle partial-wave expansion~\cite{Jiang:2020rwz,Shu:2021qlr}.

We apply this formalism to project both sides of the unitarity cuts: the \gls{sm} amplitude on one side and the amplitude or form factor containing an insertion of a higher-dimensional \gls{eft} operator on the other side.
In the joint \gls{ssyt} and partial-wave basis, the phase-space integrals entering the cut \glspl{rge} become overlaps of channel harmonics.
Orthonormality with respect to the physical Lorentz-invariant phase-space measure reduces the cut contractions to matrix multiplication over the intermediate channel labels.
A cut can therefore be nonzero only if its two sides have common angular-momentum and tableau components for at least one allowed intermediate particle state.
Matching these labels is necessary, but the corresponding contributions may still cancel.
When one side is a local \gls{eft} contact amplitude, its finite harmonic expansion selects only a finite set of projections of the \gls{sm} amplitude.
These projections suffice to evaluate the cut without reconstructing the generally infinite harmonic expansion of the \gls{sm} amplitude.
Singular projections require a common regulator and an infrared treatment consistent with the cut \gls{rge}, as discussed in \Sec{sec:on-shell-rge-infrared}.
The projection thus provides both a method for evaluating cuts and a criterion for their vanishing beyond helicity counting.

We study operator mixing in the unbroken, massless \gls{sm}, at fixed operator dimension and to first order in the higher-dimensional Wilson coefficients.
The channel projection itself applies more generally to massless four-dimensional scattering.
We use this framework to study non-renormalization at one and higher loops, recovering the one-loop helicity theorem~\cite{Cheung:2015aba} and deriving additional higher-loop restrictions beyond those implied by operator length~\cite{Bern:2019wie}.
With exceptional Yukawa contributions excluded, maximal tree cuts at loop order $L$ can contribute only to targets inside or on the boundary of the forward light cone of source operators in \Fig{fig:dimension-six-ssyt-weight-lattice}, with its vertex shifted downward from the source component by $2(L-1)$ units in $N$.
For mixing from longer into shorter operators at the first order allowed by operator length, we establish the conditions under which targets outside this shifted cone have vanishing mixing in the full \gls{adm}.
We discuss the weaker bounds in the presence of exceptional Yukawa contributions separately.

The paper is organized as follows.
\Sec{sec:rge} reviews the on-shell \gls{rge} and establishes the conditions needed to extract ultraviolet mixing from cuts.
\Sec{sec:sm-helicity-structure} discusses the \gls{sm} helicity restrictions, introduces the channel decomposition, and applies it to cut integrals.
\Sec{sec:non-renormalization} presents the resulting selection rules at one and higher loops, and \Sec{sec:conclusions} summarizes our conclusions.
\App{app:basis-projection} develops the detailed projection and reconstruction formulas, and \App{app:phase-space} provides the phase-space measures and integrals.
\App{app:examples} presents worked examples of channel projections and cut integrals, while \App{app:two-particle-casimirs} discusses the relations between conformal, auxiliary, and angular-momentum Casimirs.

\section{On-shell renormalization} \label{sec:rge}

We use the on-shell renormalization group equation to relate operator mixing to cuts of amplitudes and form factors~\cite{Caron-Huot:2016cwu,Bern:2019wie,Bern:2020ikv}.
Throughout, the \gls{sm} is taken in its unbroken, massless limit, with the Higgs mass parameter neglected.
We work to first order in the higher-dimensional Wilson coefficients.
We use dimensional regularization with minimal subtraction and consider operators of fixed mass dimension $\Delta$.
For Wilson coefficients $C_i$ multiplying operators $\mathcal O_i$, we define the anomalous-dimension matrix by
\begin{equation}
    \mu\frac{d C_i}{d\mu}
    =\sum_j\gamma_{ij}C_j.
    \label{eq:adm-convention}
\end{equation}
We refer to $\mathcal O_j$ as the source operator and to $\mathcal O_i$ as the target operator. An insertion of $\mathcal O_j$ can generate an ultraviolet counterterm proportional to $\mathcal O_i$, corresponding to mixing of $\mathcal O_j$ into $\mathcal O_i$.

\subsection{The on-shell renormalization group equation}
\label{sec:on-shell-rge-form-factors}

The on-shell \gls{rge} combines analyticity, which relates the scale dependence of form factors to their discontinuities, with unitarity, which expresses these discontinuities in terms of on-shell intermediate states~\cite{Caron-Huot:2016cwu}.
Using the Feynman prescription $s\to s+i0$, with $s>0$, and choosing the branch cut of the logarithm along the negative real axis, we have
\begin{equation}
    \log\frac{-s-i0}{\mu^2}
    =\log\frac{s}{\mu^2}-i\pi.
    \label{eq:rge-log-convention}
\end{equation}
Thus the same coefficient that determines its scale dependence determines also its imaginary part.
To generalize this observation, we introduce the dilatation operator
\begin{equation}
    D=\sum_i p_i^\mu\frac{\partial}{\partial p_i^\mu},
    \label{eq:rge-dilatation-definition}
\end{equation}
which generates a common rescaling of the external momenta.

For an operator $\mathcal O_j$, define the form factor
\begin{equation}
    \langle p_1,h_1;\ldots;p_n,h_n|\mathcal O_j(q)|0\rangle
    =(2\pi)^4\delta^{(4)}\!\left(q-\sum_{i=1}^{n}p_i\right)
    F_j(1,\ldots,n;q).
    \label{eq:rge-form-factor-definition}
\end{equation}
Throughout this paper, the labels $h_i$ follow the operator-helicity convention, where each undotted spinor $\lambda_i$ contributes $+1/2$, while each dotted spinor $\widetilde\lambda_i$ contributes $-1/2$.\footnote{For the same spinor expression, these helicity signs are opposite to those used in \Ref{Baratella:2020lzz}.}
The operator injects a generally off-shell momentum $q$, while all external particles are on shell.
We use $F_j$ with this overall delta function removed throughout.
We suppress the momentum argument $q=\sum_i p_i$ below unless needed explicitly.
The amplitudes with one operator insertion considered below correspond to $q=0$, after crossing the appropriate external legs~\cite{Bern:2020ikv}.
The complex rescaling generated by $D$ acts as
\begin{equation}
    e^{i\alpha D}F_j(p_1,\ldots,p_n)
    =F_j(p_1e^{i\alpha},\ldots,p_ne^{i\alpha}).
    \label{eq:rge-complex-rescaling}
\end{equation}
Starting in the region with all external energies outgoing and positive, this rotates every invariant $s_{\mathcal I}=(\sum_{i\in\mathcal I}p_i)^2$ by $e^{2i\alpha}$, where $\mathcal I$ is a subset of the external legs.
At $\alpha=\pi$, the invariants return to their original values on the opposite side of the cuts, giving the conjugate boundary value $F_j^*$.
Equivalently~\cite{Caron-Huot:2016cwu},
\begin{equation}
    F_j=e^{-i\pi D}F_j^*.
    \label{eq:rge-analytic-continuation}
\end{equation}
Here $F_j^*$ is the form factor with anti-time-ordered propagators.

Unitarity provides a second identity for $F_j$.
Throughout this paper, we write the scattering operator as
\begin{equation}
    S=\mathbf1+i\mathcal M,
    \label{eq:scattering-operator-convention}
\end{equation}
where $\mathcal M$ is the transition operator.
In the form-factor \gls{rge}, $S$ contains the \gls{sm} interactions, while the operator insertion is carried by $F_j$.
Treating this insertion as an infinitesimal perturbation and expanding $SS^\dagger=\mathbf1$ to first order gives, between the vacuum and the outgoing state,
\begin{equation}
    F_j=SF_j^*.
    \label{eq:rge-form-factor-unitarity}
\end{equation}
The product includes the sum and phase-space integral over intermediate on-shell states, as we will see below.
Combining this unitarity relation with \Eq{eq:rge-analytic-continuation} yields the on-shell master relation
\begin{equation}
    e^{-i\pi D}F_j^*
    =SF_j^*.
    \label{eq:rge-s-matrix-master}
\end{equation}

After separating the canonical momentum factors, the dependence on the overall scale enters through ratios $s_{\mathcal I}/\mu^2$, so that $D=-\mu\partial_\mu$.
The form-factor \gls{rge} therefore relates $D$ to the anomalous dimensions~\cite{Caron-Huot:2016cwu,Bern:2020ikv},
\begin{equation}
    DF_j
    =-\mu\partial_\mu F_j
    =\sum_i\gamma_{ij}F_i
    -\Gamma_{\mathrm{IR}}F_j
    +\sum_a\beta_a\frac{\partial F_j}{\partial g_a},
    \qquad
    \beta_a=\mu\frac{d g_a}{d\mu}.
    \label{eq:rge-dilatation-form-factor}
\end{equation}
The three terms describe UV operator mixing, soft and collinear contributions, and the running of the renormalizable \gls{sm} couplings $g_a$, respectively.
Unlike $\gamma_{ij}$, the universal IR operator $\Gamma_{\mathrm{IR}}$ acts on the external states, including their color indices, rather than mixing the inserted operators~\cite{Baratella:2022nog}.

For later use, the length $N_j$ denotes the number of elementary fields in $\mathcal O_j$. Its minimal tree-level form factor has $N_j$ external particles and is a local polynomial.
We normalize operators without explicit \gls{sm} couplings, so this minimal form factor is coupling independent.
Nonminimal form factors can contain additional \gls{sm} vertices and exchange poles.

\subsection{The one-loop cut \texorpdfstring{\glsentryshort{rge}}{RGE}}
\label{sec:on-shell-rge-one-loop}

Expanding \Eq{eq:rge-s-matrix-master} using \Eq{eq:rge-dilatation-form-factor} to one-loop order gives in general
\begin{equation}
\begin{aligned}
    \sum_i\gamma_{ij}^{(1)}F_i^{(0)}
    ={}&-\frac{1}{\pi}\bigl(\mathcal M\otimes F_j\bigr)^{(1)}
    +\left[\Gamma_{\mathrm{IR}}^{(1)}
    +\frac{i\pi}{2}D\Gamma_{\mathrm{IR}}^{(1)}\right]F_j^{(0)}
    -\sum_a\beta_a^{(1)}\frac{\partial F_j^{(0)}}{\partial g_a},
    \label{eq:rge-one-loop-uv}
\end{aligned}
\end{equation}
where the superscripts denote loop order.
The first term on the right-hand side denotes the sum of two-particle cuts joining a tree-level \gls{sm} amplitude to a tree-level source form factor~\cite{Caron-Huot:2016cwu,Bern:2020ikv}.

We denote an on-shell multiparticle state by $\ket{\alpha}=\ket{p_1,h_1;\ldots;p_N,h_N}$, with species, gauge, and flavor labels suppressed.
The connected scattering amplitude is defined by
\begin{equation}
    \langle\alpha_f|\mathcal M|\alpha_i\rangle_{\mathrm{conn}}
    =(2\pi)^4\delta^{(4)}(P_f-P_i)\,
    A(\alpha_f;\alpha_i),
    \label{eq:scattering-amplitude-convention}
\end{equation}
where $P_i$ and $P_f$ are the total incoming and outgoing momenta.
For matrix elements at a common fixed total momentum $P$, we suppress this overall factor and keep the bra-ket notation.
The same convention applies to the basis overlaps and projected matrix elements in \App{app:basis-projection}.
Whenever a scattering amplitude or its projected coefficients are written as matrix elements of $\mathcal M$, the connected contribution is understood.

For each cut, $\alpha_L$ and $\alpha_R$ denote the uncut external configurations on the \gls{sm} and form-factor sides, respectively, while $\alpha_c$ denotes the two-particle intermediate configuration.
The tree-level \gls{sm} amplitude is
\begin{equation}
    A_{\mathrm{SM}}^{(0)}(\alpha_L;\alpha_c)
    =
    \bra{\alpha_L}\mathcal M^{(0)}\ket{\alpha_c},
\end{equation}
where $\mathcal M^{(0)}$ contains only tree-level \gls{sm} interactions, and the matrix element uses the fixed-momentum convention above.
In an amplitude argument, a semicolon separates the outgoing configuration from the incoming configuration.
An amplitude written as a single list, $A(1,\ldots,N)$, instead has all legs outgoing.
These are two representations of the same amplitude, related by crossing.
All arguments of $F_j$ are outgoing, as in \Eq{eq:rge-form-factor-definition}.
The factor $F_j^{(0)}(\alpha_c,\alpha_R)$ is the tree-level form factor defined in \Eq{eq:rge-form-factor-definition}, evaluated on the outgoing intermediate states $\alpha_c$ together with the uncut external states $\alpha_R$.
The sum runs over all allowed two-particle cuts, including the assignments of external legs to the two factors and the sums over intermediate species, helicities, and internal quantum numbers.
The two cut momenta are $\ell_1,\ell_2$, with total momentum $P=\ell_1+\ell_2$.
The intermediate states are outgoing in the form factor and incoming in the \gls{sm} amplitude.

Write the ordered intermediate state, including particle labels and positive-energy momenta, as $\alpha_c=(\ell_1,\ldots,\ell_{N_c})$, where $N_c$ is the number of intermediate particles.
A crossed leg $-\ell_r$ has reversed momentum and helicity and conjugate particle quantum numbers.
We use
\begin{equation}
    |{-\ell}\rangle=i|\ell\rangle,
    \qquad
    |{-\ell}]=i|\ell].
    \label{eq:rge-crossing-convention}
\end{equation}
With the reversed cut-leg ordering shown below, we choose the external-state phases so that
\[
    A(\alpha_L;\ell_1,\ldots,\ell_{N_c})
    =i^{n_F(c)}A(-\ell_{N_c},\ldots,-\ell_1,\alpha_L).
\]
Here $n_F(c)$ counts the cut fermions and antifermions.
Thus $\sigma_c=i^{n_F(c)}$ occurs when sewing these all-outgoing expressions, but not as an additional factor when sewing physical matrix elements.
The ordering and spinor continuation follow the sewing prescription of \Ref{Baratella:2020lzz}.
Throughout, symmetry factors for identical intermediate particles are included in the sums over cuts or cut states, not in $\sigma_c$ or the phase-space measure.
The measure $d\Phi_2(P)$ is physical LIPS, defined in \App{app:phase-space}.

At this order, $\mathcal M^{(0)\dagger}=\mathcal M^{(0)}$, and the cut contribution can be written as
\begin{equation}
\begin{aligned}
    \bigl(\mathcal M\otimes F_j\bigr)^{(1)}
    ={}&\sum_{\text{cuts }c}
    \int d\Phi_2(P)\,
    A_{\mathrm{SM}}^{(0)}(\alpha_L;\alpha_c)\,
    F_j^{(0)}(\alpha_c,\alpha_R)\\[1ex]
    ={}&\sum_{\text{cuts }c}\sigma_c
    \int d\Phi_2(P)\,
    A_{\mathrm{SM}}^{(0)}(-\ell_2,-\ell_1,\alpha_L)\,
    F_j^{(0)}(\ell_1,\ell_2,\alpha_R).
    \label{eq:rge-one-loop-convolution}
\end{aligned}
\end{equation}
The first line sews physical states.
The second is the same cut written with all-outgoing spinor amplitudes.
Both use the same weighted cut sum and physical LIPS.

The remaining terms in \Eq{eq:rge-one-loop-uv} account for infrared contributions and the running of the \gls{sm} couplings.
For a minimal source tree-level form factor, our operator normalization implies independence of the \gls{sm} couplings, so the coupling-derivative term vanishes.
In the absence of infrared contributions to the matrix element under consideration, \Eq{eq:rge-one-loop-uv} reduces to
\begin{equation}
    \sum_i\gamma_{ij}^{(1)}F_i^{(0)}
    =
    -\frac{1}{\pi}
    \bigl(\mathcal M\otimes F_j\bigr)^{(1)}.
    \label{eq:rge-one-loop-ir-free}
\end{equation}
The UV mixing coefficients are then obtained by matching the complete cut sum onto the target tree-level form factors.
The treatment of infrared contributions when they are present is discussed in \Sec{sec:on-shell-rge-infrared}.

\subsection{The cut \texorpdfstring{\glsentryshort{rge}}{RGE} beyond one loop}
\label{sec:on-shell-rge-higher-loops}

At higher loops, the cut \gls{rge} follows by expanding \Eq{eq:rge-s-matrix-master} and using \Eq{eq:rge-dilatation-form-factor} at the desired perturbative order.
Equivalently, the master relation can be written as
\begin{equation}
    \left(e^{i\pi D}-1\right)F_j
    =
    -i\,\mathcal M^\dagger\otimes F_j.
    \label{eq:rge-higher-loop-master}
\end{equation}
The right-hand side now includes cuts involving loop amplitudes and form factors, while higher powers of $D$ generate iterations of lower-order anomalous dimensions, together with coupling-running and infrared terms~\cite{Bern:2020ikv}.
For a cut into an amplitude with $L_A$ loops and a form factor with $L_F$ loops, the total loop order is $L_A+L_F+N_c-1$, where $N_c$ is the number of intermediate particles.
Consequently, a generic higher-loop anomalous dimension cannot be obtained from tree-level cuts alone.

A substantial simplification occurs for the renormalization of a longer operator $\mathcal O_l$ into a shorter operator $\mathcal O_s$~\cite{Bern:2019wie}.
We therefore specialize the source and target indices to $j=l$ and $i=s$, with lengths $N_l=\ell(\mathcal O_l)$ as well as $N_s=\ell(\mathcal O_s)$, and define
\begin{equation}
    \delta N=N_l-N_s>0.
    \label{eq:rge-length-difference}
\end{equation}
All form factors are evaluated on $N_s$ external particles, so that $F_s^{(0)}$ is minimal.
When several independent target operators contribute to these external states, we obtain a sum over $s$, with their common length $N_s$ held fixed.

We first consider loop order $\delta N$ in a massless theory without relevant couplings.
The lower-loop source form factors vanish on the chosen external states.
Their coupling derivatives therefore vanish as well, and the infrared terms are absent for the reason discussed in \Sec{sec:on-shell-rge-infrared}.
Following the argument of \Ref{Bern:2019wie}, the \gls{rge} at this order reduces to
\begin{equation}
    \sum_s\gamma_{sl}^{(\delta N)}F_s^{(0)}
    =
    -\frac{1}{\pi}
    \bigl(\mathcal M^\dagger\otimes F_l\bigr)^{(\delta N)}.
    \label{eq:rge-length-bound-cut}
\end{equation}
At this order, however, the required cuts are absent or contain scaleless subintegrals.
In the absence of any infrared contribution canceling a UV anomalous dimension, this gives the length selection rule $\gamma_{sl}^{(\delta N)}=0$~\cite{Bern:2019wie}, where $\delta N$ is defined in \Eq{eq:rge-length-difference}.
Thus, although candidate diagrams can exist at $\delta N$ loops, the first loop order not excluded by length alone is $L_\star=\delta N+1$.

At order $L_\star$, the counting instead permits cuts through $N_c=L_\star+1=\delta N+2$ intermediate particles, with a tree-level \gls{sm} amplitude and a minimal tree-level source form factor~\cite{Bern:2019wie,Bern:2020ikv}.
In fact, only cuts between two tree-level factors contribute to the anomalous dimension at this order.
These are the maximal tree cuts discussed in \Sec{sec:higher-loop-cut-topologies}.

This also simplifies the cut \gls{rge} at order $L_\star$.
The source form factors on the chosen $N_s$-particle states vanish through order $\delta N$, removing all terms associated with the running of other couplings and avoiding infrared terms.
Higher powers of $D$ contain successive mixings through intermediate operators.
By the length bound, two nonzero mixing steps that together reduce the length by $\delta N$ require at least $\delta N+2$ loops, so these terms cannot contribute at order $L_\star$.
The length counting similarly excludes lower-order mixing followed by a loop form factor.
The cut \gls{rge} therefore becomes
\begin{equation}
\begin{aligned}
    \sum_s\gamma_{sl}^{(L_\star)}F_s^{(0)}
    &=
    -\frac{1}{\pi}
    \bigl(\mathcal M\otimes F_l\bigr)_{\mathrm{max.\ cut}}^{(L_\star)}
    \\
    &=
    -\frac{1}{\pi}
    \sum_{\text{cuts }c}
    \int d\Phi_{L_\star+1}(P)\,
    A_{\mathrm{SM}}^{(0)}(\alpha_L;\alpha_c)\,
    F_l^{(0)}(\alpha_c,\alpha_R).
    \label{eq:rge-maximal-tree-cut}
\end{aligned}
\end{equation}
Here $\alpha_c$ contains $L_\star+1$ intermediate particles, and the remaining notation is the same as in \Eq{eq:rge-one-loop-convolution}.
The all-outgoing form follows from the same conversion as in \Eq{eq:rge-one-loop-convolution}, now with $N_c=L_\star+1$.
Matching the complete cut sum onto the independent target form factors determines $\gamma_{sl}^{(L_\star)}$.
The reduction to maximal tree cuts is specific to this length-changing order and does not hold for a generic higher-loop mixing.

After crossing and setting the injected momentum to zero, $q=0$, the source form factor becomes an amplitude with an operator insertion.
The phase-space contractions in \Eqs{eq:rge-one-loop-convolution}{eq:rge-maximal-tree-cut} can then be organized using the channel projection developed in \Sec{sec:cut-integrals-channel-basis}.
Setting $q=0$ is not required for projecting the cut variables alone, as explained in \App{appsec:one-sided-projection}.
Before turning to that construction, we clarify when infrared contributions must be included and why they are absent in the length-changing case considered here.

\subsection{Infrared contributions}
\label{sec:on-shell-rge-infrared}

The infrared terms in the on-shell \gls{rge} account for the soft and collinear contributions contained in the cut integrals.
Extracting $\gamma_{ij}$ from \Eq{eq:rge-one-loop-uv} and its higher-loop analogs therefore requires either including these terms explicitly or establishing that they vanish for the matrix element under consideration.
Their universal structure is determined by infrared factorization~\cite{Catani:1998bh,Sterman:2002qn,Becher:2009cu,Gardi:2009qi,Becher:2009qa}.
In dimensional regularization with $d=4-2\varepsilon$, the UV-renormalized on-shell form factor $F_j^{\mathrm{UV}}$ still contains infrared poles.
The form factor entering the differential \gls{rge} in \Eq{eq:rge-dilatation-form-factor} is obtained by infrared renormalization,
\begin{equation}
    F_j=\lim_{\varepsilon\to0}Z_{\mathrm{IR}}^{-1}F_j^{\mathrm{UV}},
    \qquad
    \Gamma_{\mathrm{IR}}=-Z_{\mathrm{IR}}^{-1}\mu\frac{dZ_{\mathrm{IR}}}{d\mu}.
    \label{eq:rge-ir-renormalization}
\end{equation}
The same convention must be implemented in the cut relation, including the corresponding counterterm and subtraction contributions.

The operator $\Gamma_{\mathrm{IR}}$ in \Eq{eq:rge-dilatation-form-factor} acts on the external-state indices, in particular color, rather than mixing the inserted operators.
It preserves the external particle species and helicities, although it can mix color structures.
It contains the cusp contributions associated with soft exchange and logarithms of kinematic invariants, together with collinear contributions associated with the external particles.
At higher orders, additional correlations between external color charges can occur~\cite{Almelid:2015jia}.
We use $\Gamma_{\mathrm{IR}}$ for this full operator, not only its collinear part.

Since $\Gamma_{\mathrm{IR}}$ starts at one loop, its explicit product with $F_j$ in \Eq{eq:rge-dilatation-form-factor} at loop order $\ell$ reads
\begin{equation}
    \bigl(\Gamma_{\mathrm{IR}}F_j\bigr)^{(\ell)}
    =
    \sum_{r=1}^{\ell}
    \Gamma_{\mathrm{IR}}^{(r)}F_j^{(\ell-r)}.
    \label{eq:rge-ir-lower-loop}
\end{equation}
If all relevant lower-loop source form factors vanish on the chosen external states, this product vanishes even though $\Gamma_{\mathrm{IR}}$ itself is generally nonzero.
A nonzero but infrared-finite lower-loop form factor can still generate infrared contributions at the next order.
In the cut \gls{rge}, expanding $e^{i\pi D}$ generates scale derivatives of $\Gamma_{\mathrm{IR}}$ as well as iterations of lower-order \gls{rge} terms~\cite{Stadlbauer:2025psi}.
The explicit scale dependence of the cusp contribution produces the $D\Gamma_{\mathrm{IR}}^{(1)}$ term already displayed in \Eq{eq:rge-one-loop-uv}.

When infrared contributions are present, the required anomalous dimensions can be taken from the literature or determined independently from form factors of UV-protected operators, such as the conserved energy-momentum tensor, see, e.g., \Refs{Moch:2005id,Ahmed:2015qia,Caron-Huot:2016cwu,Baratella:2022nog,Stadlbauer:2024jij}.

To extract the UV anomalous dimensions, the infrared contributions to the cut integrals must be subtracted consistently with the form-factor renormalization in \Eq{eq:rge-ir-renormalization}.
For two-particle cuts at one loop, the universal soft contribution can be subtracted from the integrand before phase-space integration, while the collinear anomalous dimensions of the external particles enter as a separate correction in the \gls{rge}~\cite{Caron-Huot:2016cwu,Baratella:2020dvw}.
In dimensional regularization, pole subtractions and finite terms must be treated consistently with the form-factor renormalization~\cite{Baratella:2022nog}.
At higher loops, finite lower-loop matrix elements also enter, so their normalization and subtraction scheme must be specified~\cite{Bern:2020ikv}.

Scaleless massless bubbles require particular care.
Even though they vanish in dimensional regularization, they can contain cancellations between UV and IR poles, so that this does not by itself establish the absence of operator mixing.
For the one-loop non-renormalization theorems reproduced in \Sec{sec:one-loop-ssyt-selection}, \Ref{Cheung:2015aba} closes this loophole using soft and collinear factorization together with the cancellation between virtual and inclusive real-emission infrared contributions~\cite{Kinoshita:1962ur,Lee:1964is}.
Leading soft emission preserves the hard external helicities and therefore cannot generate the interference required to violate the helicity exclusion.
In the collinear limit, the interference between the relevant splitting amplitudes carries an azimuthal phase that averages to zero upon integration.
The omitted massless bubbles therefore cannot generate the operator mixing excluded by the theorem.

For the longer-to-shorter mixing of \Sec{sec:on-shell-rge-higher-loops}, the absence of infrared contributions instead follows from length counting~\cite{Bern:2019wie}.
With $\delta N=N_l-N_s>0$, the source form factors evaluated on the $N_s$ target external particles vanish at all loop orders up to and including $\delta N$.
Consequently, \Eq{eq:rge-ir-lower-loop} vanishes both at order $\delta N$ and at the first length-allowed order $L_\star=\delta N+1$.
Together with the exclusion of iteration terms by length counting in \Sec{sec:on-shell-rge-higher-loops}, this allows us to determine the mixing at loop order $L_\star$ using \Eq{eq:rge-maximal-tree-cut}.

These infrared considerations also apply to the channel projections in \Sec{sec:cut-integrals-channel-basis} and \App{app:basis-projection}.
The projection integrals probe the same soft and collinear regions as the cut integrals and can diverge for singular \gls{sm} amplitudes.
For example, in massless $2\to2$ scattering, a $t$-channel pole proportional to $1/t$, with $t=-s(1-\cos\theta)/2$, can produce a divergent angular overlap even at fixed $s>0$.
Completeness of the harmonic basis does not remove these singularities, and the divergent overlaps require regularization before they can be evaluated separately.
Regulating these overlaps does not by itself perform the infrared subtraction required by the \gls{rge}.

At one loop, the universal soft contribution can be subtracted before phase-space integration.
For massless $2\to2$ amplitudes, \Ref{Baratella:2020dvw} implements this subtraction in the partial-wave coefficients and shows that their contraction reproduces the soft-subtracted cut, with the collinear contribution included separately in the \gls{rge}.
In this case, our channel projection reduces to the same helicity partial-wave projection, and the prescription applies directly.
For two-particle cuts involving higher-point amplitudes, a BCFW prescription for extracting bubble contributions through partial-wave projections is discussed in \Ref{Shu:2021qlr}.

For channels with more than two particles, a consistent treatment of singular projections is equally necessary.
An explicit example is the cancellation of a three-particle-cut divergence against a two-particle cut containing the counterterm for a one-loop UV subdivergence~\cite{Baratella:2022nog}.
We do not derive a general regularization and infrared-subtraction prescription for multiparticle harmonic projection coefficients.
The four-dimensional projection and contraction formulas below therefore apply directly to convergent overlaps.
For the one-loop helicity exclusions and the longer-to-shorter mixing at the loop orders discussed above, the relevant infrared contributions to the \gls{rge} vanish by the arguments given there.

Even when a cut is infrared finite, individual projection coefficients can diverge, with their singular parts canceling in the complete contraction.
Such coefficients must be combined using a common regulator, or treated with a subtraction prescription that preserves the value of the cut.
The absence of infrared contributions to the relevant mixing therefore does not by itself establish that each projection integral is finite.

\section{Helicity bounds and channel decomposition} \label{sec:sm-helicity-structure}

We now study the helicity structure of the \gls{sm} tree amplitudes entering the cut relations of \Sec{sec:rge}. As in the rest of this paper, we use the helicity convention introduced in \Sec{sec:on-shell-rge-form-factors}.
The total helicity $h(A^{\mathrm{SM}}_N)$ of a massless tree-level \gls{sm} amplitude with $N$ legs is known to be constrained~\cite{Cheung:2015aba,Azatov:2016sqh}.\footnote{The higher-point bound of \Ref{Azatov:2016sqh} is used here with exceptional Yukawa contributions excluded, following the discussion in \Ref{Cheung:2015aba}.}

For nonvanishing \gls{sm} three-point amplitudes, the total helicity satisfies $\lvert h(A_{3}^{\text{SM}}) \rvert = 1$.
For nonvanishing four-point amplitudes, the total helicity is $h=0$, except for all-plus and all-minus four-fermion amplitudes, which have $h=\pm2$ and always involve both up- and down-type Yukawa interactions. 
Thus, for $N=4$, the allowed total helicities are
\begin{equation}
    h(A_{4}^{\text{SM}}) = 0, \quad \text{except for} \quad h(A_{4}^{\text{SM}}(\psi^\pm, \psi^\pm, \psi^\pm, \psi^\pm)) = \pm 2.
\end{equation}
Unless stated otherwise, our helicity-based selection rules assume that the Yukawa combinations generating exceptional four-fermion amplitudes are absent.
Under this assumption, higher-point \gls{sm} tree amplitudes with $N\geq5$ satisfy
\begin{equation} \label{eq:constraint_higher_point}
    \lvert h(A_{N}^{\mathrm{SM}}) \rvert \le N - 4, \quad \text{with $h$ even (odd) for $N$ even (odd).}
\end{equation}
In particular, the 5- and 6-point amplitudes that will be important in the following can only have total helicities $h(A_{5}^{\mathrm{SM}})=\{-1,1\}$ and $h(A_{6}^{\mathrm{SM}})=\{-2,0,2\}$, respectively. 

If the exceptional Yukawa contributions are present, their higher-point extensions need not satisfy \Eq{eq:constraint_higher_point}.
By soft-gluon factorization~\cite{Dixon:1996wi}, a nonzero exceptional four-quark amplitude with $h=2$ implies a nonzero five-point amplitude with an additional positive-helicity gluon and $h=3$.
The general bound including exceptional contributions is instead $|h(A_N^{\mathrm{SM}})|\leq N-2$~\cite{Cheung:2015aba}, with $h$ still even (odd) for $N$ even (odd).

\subsection{Channel decomposition and Lorentz-invariant \texorpdfstring{\glsentryshortpl{ssyt}}{SSYTs}}
\label{sec:ssyt-decomposition}

Having established the possible total helicities of \gls{sm} amplitudes, we now organize their remaining kinematic dependence in a basis of Stiefel harmonics constructed using \glspl{ssyt}~\cite{Henning:2019mcv,Henning:2019enq}, refined by their total angular momentum.
The detailed construction and projection technique are given in \App{app:basis-projection}. Here we only summarize the ingredients needed below. Consider a timelike channel $i\to f$ with fixed external helicities and
\begin{equation} \label{eq:fixed-P-channel}
    P=\sum_{r\in i}p_r=\sum_{r\in f}p_r,
    \qquad s=P^2>0.
\end{equation}
Let $N_i$ and $N_f$ be the numbers of particles on the two sides of the
channel. After extracting the overall scale and expanding the spinors in a
common massive spin frame, \ie $\lambda_{i \alpha} = Z_i^{\;I} x_{I \alpha}$, we collect the resulting dimensionless
coefficients into an $N_r\times2$ matrix $Z_r$, defined in \Eq{eq:bp-Lambda-Zx}, that satisfies $Z_r^\dagger Z_r=\mathbf1_2$.
On either side $r\in\{i,f\}$, a $\U(N_r)$ \gls{yt} shape $\Lambda_r$ and an \gls{ssyt} filling $T_r$ uniquely specify a covariant Stiefel harmonic $Y^{(r)}_{\Lambda_r,T_r}(Z_r)$.
Starting from the Lorentz-covariant polynomials of \Ref{Henning:2019mcv}, we contract all open Lorentz indices with the massive spinors $x^{I\alpha}$ and $\widetilde x_J{}^{\dot\alpha}$. The resulting harmonics carry only open $\SU(2)_P$ indices, where $\SU(2)_P$ denotes the massive little group leaving $P$ invariant.
For timelike $P$, this is an invertible change of basis, not a projection, so no information is lost, see \App{app:basis-projection}.

For example, consider the following tableau with $N=3$ particles
\begin{equation}
\begin{aligned}
    \begin{array}{@{}c@{}}
        \begin{ytableau}
            *(blue!40) 1 & 1 & 1 \\
            *(blue!40) 2 & 2
        \end{ytableau}
    \end{array}
    \quad & \sim \;
    \langle12\rangle\,
    \lambda_{1\alpha}\widetilde\lambda_{3\dot\beta}\,
    x^{I\alpha}\widetilde x_J{}^{\dot\beta}
    \\
    &= \;
    -\sqrt{s}\,
    \epsilon_{K_1K_2}Z_1{}^{K_1}Z_2{}^{K_2}
    Z_1{}^K\overline Z_{3L}\,
    \bigl(x_{K\alpha}x^{I\alpha}\bigr)
    \bigl(\widetilde x^L{}_{\dot\beta}
          \widetilde x_J{}^{\dot\beta}\bigr)
    \\
    &= \;
    s^{3/2}\,
    \epsilon_{K_1K_2}Z_1{}^{K_1}Z_2{}^{K_2}
    Z_1{}^I\overline Z_{3J} \, .
\end{aligned}
\label{eq:covariant-ssyt-example}
\end{equation}
As in \Refs{Henning:2019mcv,Henning:2019enq}, white and blue distinguish the holomorphic and anti-holomorphic parts of the diagrams, respectively. With our helicity convention, these contribute positively and negatively to the total helicity.
Here $\overline Z_{iK}=(Z_i{}^K)^*$ and we use the same massive-spinor conventions as in \Ref{Durieux:2019eor}, which give $x_{K\alpha}x^{I\alpha}=\sqrt{s}\,\delta_K{}^I$ and $\widetilde x^L{}_{\dot\beta}\widetilde x_J{}^{\dot\beta}=-\sqrt{s}\,\delta_J{}^L$.
For each \gls{ssyt} polynomial, these identities reduce all explicit massive-spinor dependence to an overall power of $\sqrt{s}$ fixed by the tableau.
After extracting the factor $s^{3/2}$ from the last line, we obtain a dimensionless polynomial representative before imposing the physical-LIPS normalization of the harmonic.
Its open indices contain spin $J=0$ and $J=1$ components.
Resolving its tensor components into irreducible spin-$J$ representations gives the channel harmonics defined in \App{appsec:joint_basis},
\begin{equation}
    Y^{(r)}_{\Lambda_r,T_r;J,M}(Z_r)
    =\langle Z_r;\mathbf h_r\vert P;\Lambda_r,T_r;J,M\rangle .
    \label{eq:def-ssyt-harmonic}
\end{equation}
Here, $J$ is the physical total angular momentum in the chosen channel, and the harmonics satisfy the eigenvalue equation
\begin{equation}
    W_r^2 Y^{(r)}_{\Lambda_r,T_r;J,M} =-sJ(J+1)Y^{(r)}_{\Lambda_r,T_r;J,M},
\end{equation}
where $W_r^2$ is the Pauli--Lubanski Casimir for the channel side $r$, given explicitly in \Eq{eq:app-pauli-lubanski-spinor-form}.
Note that $(J,M)$ do not label additional \glspl{ssyt}. They resolve the open massive-spinor tensor of the single harmonic fixed by $(\Lambda_r,T_r)$, with $M$ denoting a component in the spin-$J$ carrier. We orthonormalize the finite fixed-helicity blocks using physical LIPS, as specified in \App{app:basis-projection}. The resulting normalization depends on the fixed channel momentum $P$, which is suppressed in the arguments of $Y$ and $\mathcal B^J$. Independent species, gauge, flavor, and permutation labels are suppressed throughout.

In this notation, the amplitude is given by
\begin{equation}
    A_{i\to f}(Z_f;Z_i)
    =\langle Z_f;\mathbf h_f\vert\mathcal M\vert Z_i;\mathbf h_i\rangle .
    \label{eq:amplitude-kernel}
\end{equation}
As explained in \App{app:basis-projection}, the channel states form a complete orthonormal basis at fixed $P$, with the resolution of the identity given in \Eq{eq:bp-joint-resolution-identity}. Poincaré invariance implies that $\mathcal M$ commutes with $\SU(2)_P$ rotations, hence
\begin{equation}
\langle P;\Lambda_f,T_f;J_f,M_f\vert\mathcal M\vert P;\Lambda_i,T_i;J_i,M_i\rangle=\delta_{J_fJ_i}\delta_{M_fM_i}\,c^J_{\Lambda_fT_f;\Lambda_iT_i}(s),
\end{equation}
where $J=J_f=J_i$ in the nonzero blocks.
The identity $\delta_{M_fM_i}$ contracts the open spin-$J$ indices of the initial and final harmonics, so the reduced coefficient $c^J$ carries no open massive-spinor indices.
Inserting the identity on each side of $\mathcal M$ in \Eq{eq:amplitude-kernel} and using \Eq{eq:def-ssyt-harmonic}, we obtain
\begin{equation}
\begin{aligned}
A_{i\to f}(Z_f;Z_i)
&=\langle Z_f;\mathbf h_f\vert
\mathbf1_{P,\mathbf h_f}\mathcal M\mathbf1_{P,\mathbf h_i}
\vert Z_i;\mathbf h_i\rangle\\
&=\sum_{J,M}\sum_{\Lambda_f,T_f}\sum_{\Lambda_i,T_i}
\langle Z_f;\mathbf h_f\vert P;\Lambda_f,T_f;J,M\rangle\,
c^J_{\Lambda_fT_f;\Lambda_iT_i}(s)\,
\langle P;\Lambda_i,T_i;J,M\vert Z_i;\mathbf h_i\rangle\\
&=\sum_{J,M}\sum_{\Lambda_f,T_f}\sum_{\Lambda_i,T_i}
Y^{(f)}_{\Lambda_f,T_f;J,M}(Z_f)\,
c^J_{\Lambda_fT_f;\Lambda_iT_i}(s)\,
\bigl(Y^{(i)}_{\Lambda_i,T_i;J,M}(Z_i)\bigr)^*\\
&=\sum_J\sum_{\Lambda_f,T_f}\sum_{\Lambda_i,T_i}
c^J_{\Lambda_fT_f;\Lambda_iT_i}(s)\,
\mathcal B^J_{\Lambda_fT_f;\Lambda_iT_i}(Z_f,Z_i),
\end{aligned}
    \label{eq:ssyt-channel-expansion}
\end{equation}
where the channel function is defined by
\begin{equation}
    \mathcal B^J_{\Lambda_fT_f;\Lambda_iT_i}(Z_f,Z_i)
    =\sum_{M=-J}^{J}
    Y^{(f)}_{\Lambda_f,T_f;J,M}(Z_f)
    \bigl(Y^{(i)}_{\Lambda_i,T_i;J,M}(Z_i)\bigr)^*.
    \label{eq:def-ssyt-basis-amplitude}
\end{equation}
The coefficient $c^J$ in \Eq{eq:ssyt-channel-expansion} is independent of $M$ but need not be diagonal in the \gls{yt} labels, because the auxiliary $\U(N)$ is not a symmetry of the $S$-matrix. Once both sides have been projected completely, it depends only on $s$.
Note that after a one-sided projection, the expansion coefficients can still depend on the unprojected same-side invariants. Equation~\eqref{eq:ssyt-channel-expansion} is the partial-wave expansion of \Refs{Jiang:2020rwz,Shu:2021qlr,Li:2022abx}, with its degeneracy labels resolved by the $\U(N_r)$ \gls{yt} shapes and fillings and with both sides projected.
Related multiparticle partial-wave constructions have been applied to \gls{eft} unitarity bounds in \Refs{Bresciani:2025toe,Degrande:2025uil}.

To relate this channel expansion to the polynomial basis of \Ref{Henning:2019enq}, let $\mathcal Y^{(N)}_{\Lambda,T}$ denote a Lorentz-invariant spinor polynomial constructed from all $N=N_i+N_f$ external legs and labeled by the \gls{ssyt} $(\Lambda,T)$. We refer to these polynomials as Lorentz-invariant \glspl{ssyt}. After crossing the appropriate legs into the chosen initial and final channels, we write the polynomial as $\mathcal Y^{(N)}_{\Lambda,T}(Z_f,Z_i)$.
If the polynomial contains $n$ holomorphic and $\widetilde n$ anti-holomorphic spinors, we denote its shape by $\Lambda_N(n,\widetilde n)$. Pictorially, this shape can be represented as
\begin{equation}
        \mathcal Y^{(N)}_{\Lambda,T}(Z_f,Z_i)
    \quad \sim \quad
    {\scriptstyle
    \mathord{\underbrace{
    \overline{
    \begin{array}{@{}c@{}}
    \begin{ytableau}
        *(blue!40) \\ *(blue!40)
    \end{ytableau}
    \end{array}
    \mkern3mu\cdots\mkern3mu
    \begin{array}{@{}c@{}}
    \begin{ytableau}
        *(blue!40) \\ *(blue!40)
    \end{ytableau}
    \end{array}}}_{\widetilde n/2}}
    \mathord{\underbrace{
    \begin{array}{@{}c@{}}
    \begin{ytableau}
        \, \\ \,
    \end{ytableau}
    \end{array}
    \mkern3mu\cdots\mkern3mu
    \begin{array}{@{}c@{}}
    \begin{ytableau}
        \, \\ \,
    \end{ytableau}
    \end{array}}_{n/2}}},
    \label{eq:li-ssyt-pictorial}
\end{equation}
where the label $\Lambda$ fixes the numbers of blue and white columns, while the filling $T$ is not shown.
The blue part is shown in the barred notation, where the overline denotes the conjugate $\U(N)$ representation~\cite{Henning:2019mcv,Henning:2019enq}.
Contracting the particle indices with epsilon tensors converts a barred column of height $r$ into an unbarred column of height $N-r$.
For Lorentz-invariant polynomials, each barred blue column represents two anti-holomorphic spinors and therefore has height two, corresponding to an unbarred column of height $N-2$.
Explicitly filled tableaux, such as the covariant example in \Eq{eq:covariant-ssyt-example}, are displayed in the unbarred notation.
Since $\mathcal Y^{(N)}_{\Lambda,T}(Z_f,Z_i)$ is invariant under simultaneous $\SU(2)_P$ rotations of $Z_f$ and $Z_i$, the argument leading to \Eq{eq:ssyt-channel-expansion} also applies here. 
The polynomial $\mathcal Y^{(N)}_{\Lambda,T}$ has finite degree, so we obtain the finite decomposition
\begin{equation}
    \mathcal Y^{(N)}_{\Lambda,T}(Z_f,Z_i)
    =\sum_J\sum_{\Lambda_f,T_f}\sum_{\Lambda_i,T_i}
    y^{J;\Lambda,T}_{\Lambda_fT_f;\Lambda_iT_i}(s)\,
    \mathcal B^J_{\Lambda_fT_f;\Lambda_iT_i}(Z_f,Z_i).
    \label{eq:li-ssyt-in-channel-basis}
\end{equation}
This is not a linear dependence among \glspl{ssyt} of the same $\U(N)$ basis. 
The left-hand side is described by a single $\U(N)$ tableau for the full external configuration, whereas each term on the right-hand side couples channel harmonics labeled by separate $\U(N_f)$ and $\U(N_i)$ tableaux. It is the analog of the change between coupled and uncoupled angular-momentum bases.

Conversely, after restoring the appropriate powers of $s$, each channel function $\mathcal B^J$ can be written as a finite sum of Lorentz-invariant \glspl{ssyt} for the full amplitude at the corresponding mass dimension.
This gives the schematic expansion
\begin{equation}
    A_{i\to f}(Z_f;Z_i)
    =\sum_{\Lambda,T}d_{\Lambda,T}(s)\,
    \mathcal Y^{(N)}_{\Lambda,T}(Z_f,Z_i),
    \label{eq:sm-amplitude-in-li-ssyt}
\end{equation}
of a general \gls{sm} amplitude in the Lorentz-invariant \gls{ssyt} basis. The coefficients $d_{\Lambda,T}(s)$ are obtained by combining the channel coefficients $c^J$ with this reverse expansion, see also \App{app:basis-projection}. The sum is finite for a local contact amplitude and generally infinite for an exchange or loop amplitude.
The polynomials $\mathcal Y^{(N)}_{\Lambda,T}(Z_f,Z_i)$ are evaluated with the initial legs crossed to the all-outgoing configuration.
Their arguments label the two sets of coordinates and do not define a physical matrix element.
The constant external-state crossing phases are included in the amplitude coefficients $d_{\Lambda,T}(s)$.
The same applies to the coefficients of a cut expansion.
Soft or collinear singularities can make the projection integrals divergent. Their regularization and relation to infrared subtraction are discussed in \Sec{sec:on-shell-rge-infrared}.

For \gls{sm} tree amplitudes, the helicity bounds above determine which shapes of \glspl{yt} can occur in \Eq{eq:sm-amplitude-in-li-ssyt}.
The helicity convention of \Sec{sec:on-shell-rge-form-factors} gives
\begin{equation}
    n-\widetilde n=2h.
    \label{eq:helicity-ssyt-degree}
\end{equation}
Thus, the total helicity fixes the difference between the two parts of the Young diagram, but not their common size, the filling, or the mass dimension.
Nonexceptional four-point \gls{sm} tree amplitudes have $n=\widetilde n$, while the exceptional four-fermion tree amplitudes have $n-\widetilde n=\pm4$.
In the absence of exceptional Yukawa contributions, five-point \gls{sm} tree amplitudes can have only total helicities $h=\pm1$, while six-point amplitudes can have $h\in\{-2,0,2\}$.
This is the \gls{ssyt} form of the helicity information that will enter the phase-space integrals below.

We can display the helicity pattern directly at the level of \gls{yt} shapes.
When only the shape is shown, we leave the sum over its allowed fillings implicit, using the shorthand notation
\begin{equation}
    d_{\Lambda}(s)\,
    \mathcal Y^{(N)}_{\Lambda}
    \equiv
    \sum_T
    d_{\Lambda,T}(s)\,
    \mathcal Y^{(N)}_{\Lambda,T}.
    \label{eq:implicit-ssyt-filling-sum}
\end{equation}
Thus, each term below represents the sum over all fillings compatible with the individual external helicities, with fully projected coefficients that depend only on the squared momentum of the respective channel, $s$.
When the particle number is explicit in the shape label $\Lambda_N(n,\widetilde n)$, we omit the superscript $(N)$ on $\mathcal Y$.

To make this notation concrete, we first show two representative towers
pictorially. 
Pictorially, the first terms in the expansion of the $h=0$ four-point tree amplitude are
\begin{equation}
    A^{\mathrm{SM}}_{4,h=0} = d^{(4,0)}_0(s)
    +d^{(4,0)}_1(s)\,
    {\scriptstyle
    \overline{\begin{array}{@{}c@{}}
    \begin{ytableau}
        *(blue!40) \\ *(blue!40)
    \end{ytableau}
    \end{array}}
    \begin{array}{@{}c@{}}
    \begin{ytableau}
        \, \\ \,
    \end{ytableau}
    \end{array}}
    +d^{(4,0)}_2(s)\,
    {\scriptstyle
    \overline{\begin{array}{@{}c@{}}
    \begin{ytableau}
        *(blue!40) & *(blue!40) \\
        *(blue!40) & *(blue!40)
    \end{ytableau}
    \end{array}}
    \begin{array}{@{}c@{}}
    \begin{ytableau}
        \, & \, \\ \, & \,
    \end{ytableau}
    \end{array}}
    +\ldots\, .
\label{eq:pictorial-ssyt-tower-four-zero}
\end{equation}
If exceptional Yukawa contributions are present, the all-plus four-fermion tree amplitude with $h=+2$ can be expanded as
\begin{equation}
    A^{\mathrm{SM}}_{4,h=+2}
    (\psi^+,\psi^+,\psi^+,\psi^+) = d^{(4,+2)}_0(s)\,
    {\scriptstyle
    \begin{array}{@{}c@{}}
    \begin{ytableau}
        \, & \, \\ \, & \,
    \end{ytableau}
    \end{array}}
    +d^{(4,+2)}_1(s)\,
    {\scriptstyle
    \overline{\begin{array}{@{}c@{}}
    \begin{ytableau}
        *(blue!40) \\ *(blue!40)
    \end{ytableau}
    \end{array}}
    \begin{array}{@{}c@{}}
    \begin{ytableau}
        \, & \, & \, \\
        \, & \, & \,
    \end{ytableau}
    \end{array}}
    +\ldots\, .
\label{eq:pictorial-ssyt-tower-four-plus}
\end{equation}
In the all-minus-helicity case, the white and blue boxes have to be exchanged.
Again, each diagram in these equations denotes the filling sum in \Eq{eq:implicit-ssyt-filling-sum}. If no allowed filling exists for the specified external states, that term is absent. 
Using the shape notation $\Lambda_N(n,\widetilde n)$, these expansions, including the $h=-2$ amplitude, read
\begin{align}
    A^{\mathrm{SM}}_{4,h=0}
    &=
    \sum_k
    d^{(4,0)}_k(s)\,
    \mathcal Y_{\Lambda_4(2k,2k)},
    \label{eq:ssyt-shape-tower-four-zero}\\
    A^{\mathrm{SM}}_{4,h=2}
    &=
    \sum_k
    d^{(4,2)}_k(s)\,
    \mathcal Y_{\Lambda_4(2k+4,2k)},
    &
    A^{\mathrm{SM}}_{4,h=-2}
    &=
    \sum_k
    d^{(4,-2)}_k(s)\,
    \mathcal Y_{\Lambda_4(2k,2k+4)}.
    \label{eq:ssyt-shape-tower-four-exceptional}
\end{align}
Here and in all subsequent tableau expansions, the sums over $k$ run over $k=0,1,2,\ldots$.
Returning to the nonexceptional case, the two five-point helicities allowed by \Eq{eq:constraint_higher_point} give
\begin{equation}
\begin{aligned}
    A^{\mathrm{SM}}_{5,h=1}
    &=
    \sum_k
    d^{(5,1)}_k(s)\,
    \mathcal Y_{\Lambda_5(2k+2,2k)},
    &
    A^{\mathrm{SM}}_{5,h=-1}
    &=
    \sum_k
    d^{(5,-1)}_k(s)\,
    \mathcal Y_{\Lambda_5(2k,2k+2)}.
\end{aligned}
\label{eq:ssyt-shape-tower-five}
\end{equation}
More generally, for $H\geq0$ and $N\geq5$,
\begin{equation}
\begin{aligned}
    A^{\mathrm{SM}}_{N,h=H}
    &=
    \sum_k
    d^{(N,H)}_k(s)\,
    \mathcal Y_{\Lambda_N(2k+2H,2k)},
    &
    A^{\mathrm{SM}}_{N,h=-H}
    &=
    \sum_k
    d^{(N,-H)}_k(s)\,
    \mathcal Y_{\Lambda_N(2k,2k+2H)}.
\end{aligned}
\label{eq:ssyt-shape-tower-general}
\end{equation}
Some terms in these sums may be absent if no tableau filling is compatible with the specified external helicities and particle species.
Increasing $k$ adds equal numbers of holomorphic and anti-holomorphic spinors, increasing the mass dimension while leaving the total helicity unchanged.

\subsection{Cut integrals in the channel basis} \label{sec:cut-integrals-channel-basis}

We now apply the decomposition of the previous subsection to the phase-space integrals in \Eqs{eq:rge-one-loop-convolution}{eq:rge-maximal-tree-cut}.
Consider a cut through $N_c$ massless states with total momentum $P$ and let $Z_c$ denote the spinor coefficients of their momentum configuration.
For a fixed assignment of the external legs to the two amplitudes, the corresponding cut integrals can be written as
\begin{equation}
    \mathcal C_{i\to f}(Z_f;Z_i)
    =
    \sum_{\text{cut states}}
    \int d\Phi_{N_c}(P)\,
    A_L(Z_f;Z_c)\,
    A_R(Z_c;Z_i).
    \label{eq:physical-cut-kernel}
\end{equation}
As in \Sec{sec:on-shell-rge-one-loop}, identical-particle symmetry factors are included in the state sum, while the external-leg assignment is fixed.
The same physical intermediate state is outgoing on the right and incoming on the left.
Since both factors are physical matrix elements, no additional all-outgoing sewing phase $\sigma_c$ appears, as in the first line of \Eq{eq:rge-one-loop-convolution}.
The measure $d\Phi_{N_c}(P)$ is physical LIPS, with respect to which the channel harmonics are orthonormal.

The two physical factors have opposite little-group phases on each intermediate state, so their product is independent of those phases.
Using \Eq{eq:appPS_LIPS_normalization}, we can express the phase-space integral as an integral over the Stiefel manifold,
\begin{equation}
    \mathcal C_{i\to f}(Z_f;Z_i)
    =
    \mathcal N_{N_c}(s)
    \sum_{\text{cut states}}
    \int d\mu_{N_c}(Z_c)\,
    A_L(Z_f;Z_c)\,
    A_R(Z_c;Z_i),
    \label{eq:cut-kernel-stiefel-measure}
\end{equation}
where $\mathcal N_{N_c}(s)\equiv\int d\Phi_{N_c}(P)$.
To evaluate this integral, it is useful to expand both amplitudes completely in the channel basis before performing the cut. For compactness, we write $a_r=(\Lambda_r,T_r)$, with $r\in\{i,f,c\}$, while keeping the spin labels $J,M$ explicit. Equation~\eqref{eq:ssyt-channel-expansion} then gives
\begin{align}
    A_L(Z_f;Z_c)
    ={}&
    \sum_{J,a_f,a_c}c^J_{L;a_fa_c}(s)
    \sum_M
    Y^{(f)}_{a_f;J,M}(Z_f)
    Y^{(c)}_{a_c;J,M}(Z_c)^*,
    \label{eq:left-amplitude-cut-expansion}\\
    A_R(Z_c;Z_i)
    ={}&
    \sum_{J',a_c',a_i}c^{J'}_{R;a_c'a_i}(s)
    \sum_{M'}
    Y^{(c)}_{a_c';J',M'}(Z_c)
    Y^{(i)}_{a_i;J',M'}(Z_i)^*.
    \label{eq:right-amplitude-cut-expansion}
\end{align}
Even though the on-shell cut integrates only over $Z_c$, we still project both sides of each amplitude. The harmonics, depending on $Z_f$ and $Z_i$, are, however, not affected by the cut integral.
Projecting only the cut legs of each amplitude, as in \Ref{Shu:2021qlr}, gives the same result, as shown explicitly in \App{appsec:one-sided-projection}.
Orthonormality with respect to physical LIPS\footnote{An orthonormal basis can be obtained by Gram--Schmidt at fixed $\Lambda$, $J$ and external helicities. The label $T$ then refers to the resulting orthonormal combinations of tableau fillings, see \App{app:basis-projection}.} implies
\begin{equation}
    \int d\Phi_{N_c}(P)\,
    Y^{(c)}_{a_c;J,M}(Z_c)^*
    Y^{(c)}_{a_c';J',M'}(Z_c)
    =
    \delta_{a_ca_c'}\delta_{JJ'}\delta_{MM'}.
    \label{eq:cut-harmonic-orthogonality}
\end{equation}
Inserting \Eqs{eq:left-amplitude-cut-expansion}{eq:right-amplitude-cut-expansion} into \Eq{eq:physical-cut-kernel} therefore yields
\begin{equation}
\begin{aligned}
    \mathcal C_{i\to f}(Z_f;Z_i)
    ={}&
    \sum_{\text{cut states}}
    \sum_{J,a_f,a_i}
    \left[
        \sum_{a_c}
        c^J_{L;a_fa_c}(s)c^J_{R;a_ca_i}(s)
    \right]
    \mathcal B^J_{a_f;a_i}(Z_f,Z_i).
\end{aligned}
\label{eq:cut-channel-composition}
\end{equation}
The phase-space integral contracts the two cut-channel harmonics.
Their orthogonality identifies $a_c'=a_c$, $J'=J$, and $M'=M$, requiring the same tableau shape $\Lambda_c$ and filling $T_c$ on both sides of the cut.
The remaining sum over the common spin component $M$ then contracts the external spin-$J$ carriers and is precisely the channel function $\mathcal B^J_{a_f;a_i}$.
We can collect the coefficient in front of $\mathcal{B}^J$ by defining
\begin{equation}
    \boxed{
    c^J_{a_f a_i}(s)
    =
    \sum_{\text{cut states}} \sum_{a_c}
    c^J_{L;a_fa_c}(s)c^J_{R;a_ca_i}(s)
    },
    \label{eq:cut-coefficient-matrix-product}
\end{equation}
such that we can write
\begin{equation} \label{eq:cut-channel-composition_2}
    \mathcal C_{i\to f}(Z_f;Z_i) =
    \sum_{J,a_f,a_i}
        c^J_{a_f a_i}(s)
    \mathcal B^J_{a_f;a_i}(Z_f,Z_i) .
\end{equation}

Only cut-channel components $(J,a_c)$ with nonzero coefficients in both amplitudes contribute.
If no such component exists for any allowed intermediate state, the cut vanishes.
The contributions from common components can still cancel in the sum.

For a cut with a tree-level \gls{sm} amplitude $A_L$ and a local \gls{smeft} contact amplitude $A_R$, only finitely many projections of $A_L$ onto the cut-channel harmonics enter the cut.
At fixed operator dimension, external states, and helicities, $A_R$ is a homogeneous polynomial and therefore a finite linear combination of Lorentz-invariant \glspl{ssyt}.
Its expansion in the cut-adapted channel consequently has support on only finitely many $(J,\Lambda_c,T_c)$ components.
Equation~\eqref{eq:cut-coefficient-matrix-product} selects the matching components from the generally infinite channel expansion of $A_L$.
The selected \gls{sm} projection coefficients need not be finite individually, as discussed in \Sec{sec:on-shell-rge-infrared}.

As in \Eq{eq:sm-amplitude-in-li-ssyt}, the result in \Eq{eq:cut-channel-composition_2} can be expanded in Lorentz-invariant \glspl{ssyt} for its $N=N_L+N_R-2N_c$ uncut external legs,
\begin{equation}
    \mathcal C_{i\to f}(Z_f;Z_i)
    =
    \sum_{\Lambda,T}
    d^{\mathrm{cut}}_{\Lambda,T}(s)\,
    \mathcal Y^{(N)}_{\Lambda,T}(Z_f,Z_i).
    \label{eq:cut-in-li-ssyt-basis}
\end{equation}
Here $N_L$ and $N_R$ count the legs of the two input amplitudes, including the $N_c$ cut legs on each side.
The helicities $h_L$, $h_R$, and $h_{\mathcal C}$ refer to the all-outgoing representations.
Crossing the incoming intermediate particles reverses their helicities, so the cut-leg contributions cancel and
\begin{equation}
    h_{\mathcal C}=h_L+h_R,
    \qquad
    n_{\mathcal C}-\widetilde n_{\mathcal C}
    =2(h_L+h_R).
    \label{eq:cut-helicity-ssyt-degree}
\end{equation}
The helicity bounds of the \gls{sm} amplitude derived above, therefore, restrict which Lorentz-invariant \glspl{ssyt} can occur after performing the cut, and together with \Eq{eq:cut-coefficient-matrix-product}, this provides the basis for the non-renormalization theorems of the next section.

For $H=h_L+h_R\geq0$, the full cut expansion in \Eq{eq:cut-in-li-ssyt-basis} can be displayed pictorially as
\begin{equation}
\begin{aligned}
    \mathcal C_{i\to f}(Z_f;Z_i) = \, & d^{\mathrm{cut}}_0(s)\,
    {\scriptstyle
    \mathord{\underbrace{
    \begin{array}{@{}c@{}}
    \begin{ytableau}
        \, \\ \,
    \end{ytableau}
    \end{array}
    \mkern3mu\cdots\mkern3mu
    \begin{array}{@{}c@{}}
    \begin{ytableau}
        \, \\ \,
    \end{ytableau}
    \end{array}}_{H}}}
    +d^{\mathrm{cut}}_1(s)\,
    {\scriptstyle
    \overline{
    \begin{array}{@{}c@{}}
    \begin{ytableau}
        *(blue!40) \\ *(blue!40)
    \end{ytableau}
    \end{array}}
    \mathord{\underbrace{
    \begin{array}{@{}c@{}}
    \begin{ytableau}
        \, \\ \,
    \end{ytableau}
    \end{array}
    \mkern3mu\cdots\mkern3mu
    \begin{array}{@{}c@{}}
    \begin{ytableau}
        \, \\ \,
    \end{ytableau}
    \end{array}}_{H+1}}}
    +d^{\mathrm{cut}}_2(s)\,
    {\scriptstyle
    \overline{
    \begin{array}{@{}c@{}}
    \begin{ytableau}
        *(blue!40) & *(blue!40) \\
        *(blue!40) & *(blue!40)
    \end{ytableau}
    \end{array}}
    \mathord{\underbrace{
    \begin{array}{@{}c@{}}
    \begin{ytableau}
        \, \\ \,
    \end{ytableau}
    \end{array}
    \mkern3mu\cdots\mkern3mu
    \begin{array}{@{}c@{}}
    \begin{ytableau}
        \, \\ \,
    \end{ytableau}
    \end{array}}_{H+2}}}
    +\ldots
    \\
    = &
    \sum_k d^{\mathrm{cut}}_k(s)\,
    \mathcal Y_{\Lambda_N(2k+2H,2k)}.
\end{aligned}
\label{eq:pictorial-cut-ssyt-tower}
\end{equation}
Each diagram again denotes the sum over allowed fillings.
For $H<0$, the blue and white parts are exchanged.
The coefficients $d^{\mathrm{cut}}_k(s)$ collect the contributions from all pairs of terms in the two amplitude expansions, including possible cancellations.

After summing the relevant cut contributions and performing the required infrared subtractions, the local ultraviolet contribution has the same operator mass dimension as the \gls{smeft} insertion and helicity $h_{\mathcal C}=h_{\mathrm{SM}}+h_{\mathrm{SMEFT}}$.
At fixed external states, it is a polynomial and can therefore be written as a finite sum of Lorentz-invariant \glspl{ssyt}.
For mixing into a given target operator, the local ultraviolet contribution must match its contact amplitude in external species, gauge quantum numbers, helicity, and mass dimension.

\section{Non-renormalization theorems}
\label{sec:non-renormalization}

Throughout this section, we work in the unbroken, massless \gls{sm} and neglect the Higgs mass parameter.
The renormalizable gauge, Yukawa, and scalar couplings are dimensionless, so the counterterms can be organized at fixed canonical mass dimension.
Although we focus on \gls{smeft}, the projection construction does not depend on the \gls{sm} field content.
The helicity-based selection rules extend to other four-dimensional massless theories with marginal renormalizable interactions, provided the same tree-level helicity bounds hold.
We now combine the cut \glspl{rge} of \Sec{sec:rge} with the \gls{ssyt} decomposition of \Sec{sec:sm-helicity-structure} to derive selection rules for operator mixing.\footnote{For three-particle targets, whose $q=0$ kinematics has no timelike channel, we use form factors with $q\neq0$ and project only the cut variables as in \App{appsec:one-sided-projection}.}
The source and target labels follow the convention in \Eq{eq:adm-convention}.

\subsection{\texorpdfstring{\glsentryshort{ssyt}}{SSYT} shapes and helicity weights of \texorpdfstring{\glsentryshort{smeft}}{SMEFT} operators}
\label{sec:ssyt-weight-lattice}

We first relate the helicity-weight classification of local operators introduced in \Ref{Cheung:2015aba} to the Lorentz-invariant \gls{ssyt} basis introduced in \Sec{sec:ssyt-decomposition}. The objects classified below are fixed-helicity leading contact amplitudes, obtained from the corresponding minimal tree-level form factors by crossing and setting $q=0$.
A single operator can generate contact amplitudes with different total helicities. These correspond to different tableau shapes and are classified separately below.

At fixed operator mass dimension $\Delta$, let $N_r=\ell(\mathcal O_r)$ denote the length of a contact component, \ie the number of fields in its minimal form factor, and let $h_r$ be the total helicity of its external states. We use the helicity convention of \Sec{sec:on-shell-rge-form-factors}.
For a Lorentz-invariant \gls{ssyt} of shape $\Lambda_{N_r}(n_r,\widetilde n_r)$, the labels $n_r$ and $\widetilde n_r$ count the holomorphic and anti-holomorphic spinors in its polynomial. The corresponding operator mass dimension and total helicity satisfy~\cite{Henning:2019enq}
\begin{equation}
    \Delta=N_r+\frac{n_r+\widetilde n_r}{2},
    \qquad
    h_r=\frac{n_r-\widetilde n_r}{2}.
    \label{eq:dimension-helicity-ssyt-counts}
\end{equation}
In four spacetime dimensions, the contact amplitude including its Wilson-coefficient prefactor has mass dimension $4-N_r$. Removing this prefactor, whose mass dimension is $4-\Delta$, leaves a polynomial of mass dimension $\Delta-N_r$. Since each spinor has mass dimension $1/2$, this gives the first relation.
For a Lorentz scalar, $n_r$ and $\widetilde n_r$ are non-negative even integers. As illustrated in \Eq{eq:li-ssyt-pictorial}, the shape $\Lambda_{N_r}(n_r,\widetilde n_r)$ consists of $n_r/2$ white holomorphic columns and $\widetilde n_r/2$ barred blue conjugate columns.

\begin{figure}[t]
    \centering
    \begin{tikzpicture}[
        x=1.50cm,
        y=1.50cm,
        operator entry/.style={
            align=center,
            anchor=north,
            text width=2.55cm,
            inner sep=0pt,
            font=\scriptsize
        }
    ]
        \newcommand{\dimEightEntry}[5]{%
            \node[inner sep=0pt] at (#1,{#2-4}) {\SSYTshape{#2}{#3}{#4}};
            \node[operator entry] at (#1,{#2-4-0.35}) {#5};
        }
        \draw[-{Latex[length=4pt]},line width=0.4pt]
            (-4.65,-2.55) -- (4.65,-2.55)
            node[right,font=\scriptsize] {$h$};
        \draw[-{Latex[length=4pt]},line width=0.4pt]
            (-4.65,-2.55) -- (-4.65,4.35)
            node[above,font=\scriptsize] {$N$};

        \foreach \x in {-4,-3,-2,-1,0,1,2,3,4}{
            \draw[line width=0.3pt] (\x,-2.595) -- (\x,-2.505);
            \node[below=2pt,font=\scriptsize] at (\x,-2.55) {$\x$};
        }
        \foreach \y/\N in {0/4,1/5,2/6,3/7,4/8}{
            \draw[line width=0.3pt] (-4.703,\y) -- (-4.597,\y);
            \node[left=2pt,font=\scriptsize] at (-4.65,\y) {$\N$};
        }

        \dimEightEntry{-4}{4}{0}{8}{$F_-^4$}
        \dimEightEntry{-2}{4}{2}{6}{
             $F_-^2\phi^2D^2$\\$F_-^2\psi_+\psi_-D$\\
             $F_-\psi_-^2\phi D^2$\\$\psi_-^4D^2$}
        \dimEightEntry{0}{4}{4}{4}{
             $F_+^2F_-^2$\\$F_+F_-\phi^2D^2$\\
             $F_+F_-\psi_+\psi_-D$\\$F_+\psi_-^2\phi D^2$\\
             $F_-\psi_+^2\phi D^2$\\$\psi_+^2\psi_-^2D^2$\\
             $\psi_+\psi_-\phi^2D^3$\\$\phi^4D^4$}
        \dimEightEntry{2}{4}{6}{2}{
             $F_+^2\phi^2D^2$\\$F_+^2\psi_+\psi_-D$\\
             $F_+\psi_+^2\phi D^2$\\$\psi_+^4D^2$}
        \dimEightEntry{4}{4}{8}{0}{$F_+^4$}

        \dimEightEntry{-3}{5}{0}{6}{
             $F_-^3\phi^2$\\$F_-^2\psi_-^2\phi$\\$F_-\psi_-^4$}
        \dimEightEntry{-1}{5}{2}{4}{
             $F_-^2\psi_+^2\phi$\\$F_-\psi_+^2\psi_-^2$\\
             $F_-\psi_+\psi_-\phi^2D$\\$F_-\phi^4D^2$\\
             $\psi_+\psi_-^3\phi D$\\$\psi_-^2\phi^3D^2$}
        \dimEightEntry{1}{5}{4}{2}{
             $F_+^2\psi_-^2\phi$\\$F_+\psi_+^2\psi_-^2$\\
             $F_+\psi_+\psi_-\phi^2D$\\$F_+\phi^4D^2$\\
             $\psi_+^3\psi_-\phi D$\\$\psi_+^2\phi^3D^2$}
        \dimEightEntry{3}{5}{6}{0}{
             $F_+^3\phi^2$\\$F_+^2\psi_+^2\phi$\\$F_+\psi_+^4$}

        \dimEightEntry{-2}{6}{0}{4}{
             $F_-^2\phi^4$\\$F_-\psi_-^2\phi^3$\\$\psi_-^4\phi^2$}
        \dimEightEntry{0}{6}{2}{2}{
             $\psi_+^2\psi_-^2\phi^2$\\$\psi_+\psi_-\phi^4D$\\$\phi^6D^2$}
        \dimEightEntry{2}{6}{4}{0}{
             $F_+^2\phi^4$\\$F_+\psi_+^2\phi^3$\\$\psi_+^4\phi^2$}
        \dimEightEntry{-1}{7}{0}{2}{$\psi_-^2\phi^5$}
        \dimEightEntry{1}{7}{2}{0}{$\psi_+^2\phi^5$}
        \dimEightEntry{0}{8}{0}{0}{$\phi^8$}

        \coordinate (weightorigin) at (2.90,3.15);
        \draw[-{Latex[length=3.5pt]},gray,line width=0.4pt]
            (weightorigin) -- ++(-0.55,0.55)
            node[above left=-1pt,font=\scriptsize,text=black]
            {$w=N-h$};
        \draw[-{Latex[length=3.5pt]},gray,line width=0.4pt]
            (weightorigin) -- ++(0.55,0.55)
            node[above right=-1pt,font=\scriptsize,text=black]
            {$\bar w=N+h$};
    \end{tikzpicture}
    \caption{\Gls{yt} shapes associated with dimension-eight \gls{smeft} contact amplitudes in the $(N,h)$ plane, using the conventions of \Fig{fig:dimension-six-ssyt-weight-lattice}. The field contents follow the dimension-eight classifications of \Refs{Murphy:2020rsh,Li:2020gnx}. Each point shows a tableau shape and its compatible field contents, with gauge and flavor contractions and individual tableau fillings suppressed. The diagonal arrows indicate the weights $w=N-h$ and $\bar w=N+h$.}
    \label{fig:dimension-eight-ssyt-weight-lattice}
\end{figure}
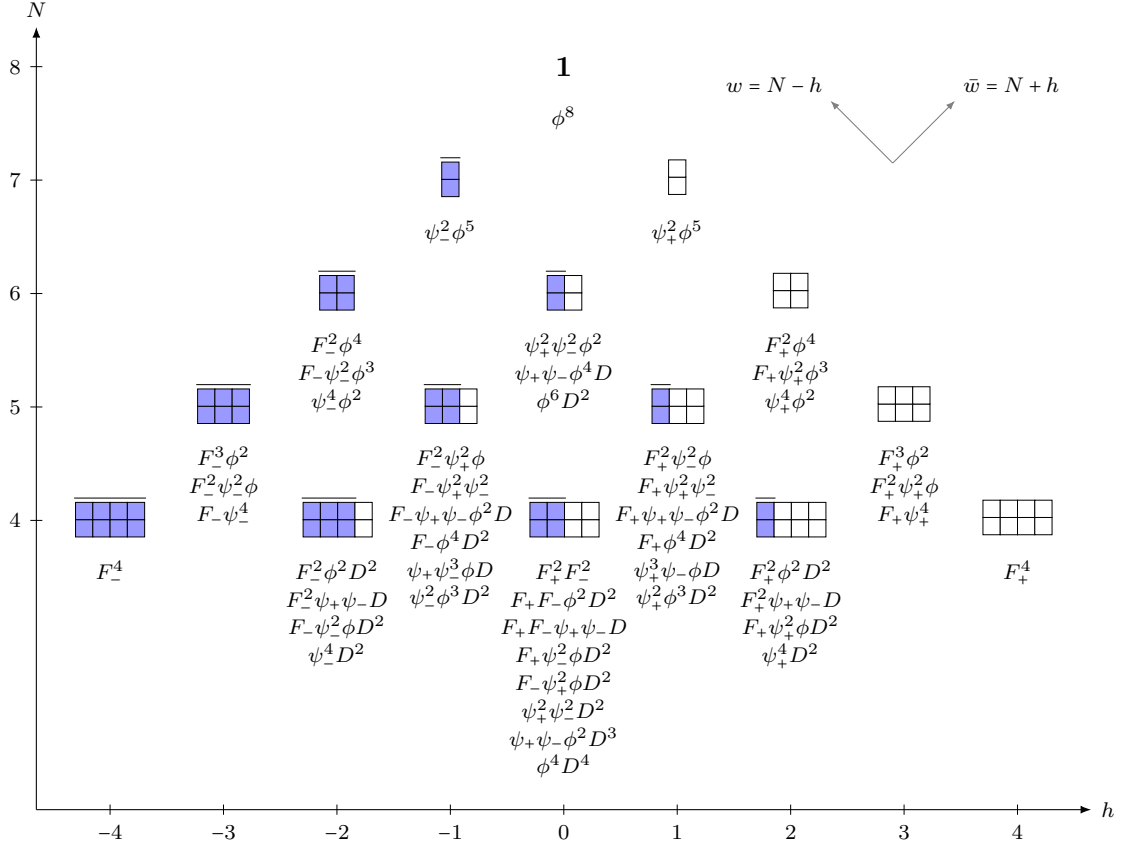

The holomorphic and anti-holomorphic weights of an amplitude with $N_r$ external states and total helicity $h_r$ are given by
\begin{equation}
    w_r=N_r-h_r,
    \qquad
    \bar w_r=N_r+h_r.
    \label{eq:def-holomorphic-weights}
\end{equation}
The weights of an operator are defined more generally by the two separate
minima
\begin{equation}
    w(\mathcal O_r)=\min_{A\,\supset\,\mathcal O_r}w(A),
    \qquad
    \bar w(\mathcal O_r)=\min_{A\,\supset\,\mathcal O_r}
    \bar w(A),
    \label{eq:def-operator-weights}
\end{equation}
where the minimization is over tree-level amplitudes with one insertion of $\mathcal O_r$~\cite{Cheung:2015aba}. These minima can be determined from the helicity components of its minimal tree-level form factor. Figures~\ref{fig:dimension-six-ssyt-weight-lattice} and \ref{fig:dimension-eight-ssyt-weight-lattice} display the \gls{ssyt} shapes separately, including those associated with different helicity components of the same Hermitian operator.

Combining \Eq{eq:dimension-helicity-ssyt-counts} with \Eq{eq:def-holomorphic-weights} gives
\begin{equation}
    n_r=\Delta-w_r=\Delta-N_r+h_r,
    \qquad
    \widetilde n_r=\Delta-\bar w_r=\Delta-N_r-h_r.
    \label{eq:ssyt-cheung-weight-dictionary}
\end{equation}
Thus, at fixed $\Delta$, the $(N,h)$ and $(w,\bar w)$ planes contain the same information about the \gls{ssyt} shapes.
At fixed $N_r$, the allowed helicities range from $-(\Delta-N_r)$ to $+(\Delta-N_r)$ in steps of two. 

Figure~\ref{fig:dimension-six-ssyt-weight-lattice} illustrates this organization at dimension six, while \Fig{fig:dimension-eight-ssyt-weight-lattice} shows the corresponding shapes and field contents at dimension eight.
In both figures, each point specifies a Lorentz/helicity sector rather than an individual gauge-invariant \gls{smeft} operator, since its \gls{ssyt} fillings, gauge and flavor contractions, and channel labels $(J,\Lambda_c,T_c)$ remain unresolved. Conjugation reflects the lattice across $h=0$, exchanging $n\leftrightarrow\widetilde n$ and $w\leftrightarrow\bar w$, so a Hermitian operator may combine components in two reflected cells.

The following subsections first derive shape-level constraints and then use the channel decomposition to refine them.

\subsection{One-loop selection rules from the \texorpdfstring{\glsentryshort{ssyt}}{SSYT} decomposition}
\label{sec:one-loop-ssyt-selection}

\begin{figure}[t]
    \centering
    \begin{tikzpicture}[
        every node/.style={font=\footnotesize},
        amplitude/.style={circle,draw,minimum size=1.25cm,align=center,inner sep=0pt,outer sep=0pt},
        cut/.style={densely dashed,black!80,line width=0.8pt}
    ]
        \node[amplitude,fill=gray!10] (sm) at (-0.90,0) {$A^{\mathrm{SM}}$};
        \node[amplitude,fill=blue!8] (source) at (0.90,0) {$A_j^{\mathrm{tree}}$};

        \draw (sm.25) -- (source.155);
        \draw (sm.335) -- (source.205);
        \draw[cut] (0,-0.75) -- (0,0.75);
        \node at (0,-1.25) {$N_c=2$};

        \draw (sm.155) -- (-2.05,0.70);
        \draw (sm.205) -- (-2.05,-0.70);
        \draw (source.25) -- (2.05,0.70);
        \draw (source.335) -- (2.05,-0.70);
        \foreach \dotangle in {158,169,180,191,202}{
            \fill ($(sm.center)+(\dotangle:1cm)$) circle[radius=0.6pt];
        }
        \foreach \dotangle in {-22,-11,0,11,22}{
            \fill ($(source.center)+(\dotangle:1cm)$) circle[radius=0.6pt];
        }
    \end{tikzpicture}
    \caption{Two-particle cut entering the one-loop selection-rule analysis. The gray and blue blobs denote an \gls{sm} tree amplitude and a source tree amplitude with one insertion of $\mathcal O_j$, respectively. The dashed line cuts two on-shell intermediate states. Dots indicate variable external multiplicities. Sums over allowed intermediate states and external-leg assignments are implicit.}
    \label{fig:one-loop-cut-topology}
\end{figure}
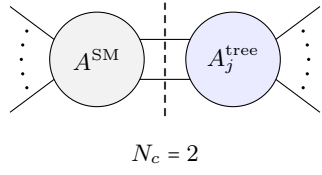

We now combine the one-loop cut \gls{rge} in \Eq{eq:rge-one-loop-uv} with the decomposition of \Sec{sec:cut-integrals-channel-basis} to identify which operator mixings can occur at one loop and thereby reproduce the results of \Ref{Cheung:2015aba}.
The two tree-level amplitudes in \Fig{fig:one-loop-cut-topology} are first expanded in the cut-adapted channel basis, and the phase-space integral contracts their common components according to \Eq{eq:cut-coefficient-matrix-product}.
The result can then be expressed in the Lorentz-invariant \gls{ssyt} basis as in \Eq{eq:cut-in-li-ssyt-basis}.
After summing over all possible cuts and subtracting IR divergences, its local ultraviolet part can contribute to a target operator only if the corresponding contact amplitude occurs in this expansion.

We first illustrate this statement with an example using the dimension-six shapes in \Fig{fig:dimension-six-ssyt-weight-lattice}.
Consider a source contact component with field content $\psi_-^2\phi^3$, total helicity $h_j=-1$, and shape $\Lambda_5(0,2)=\vcenter{\hbox{\SSYTshape{5}{0}{2}}}$.
As the target, we consider the $\psi_+^2\phi^3$ component of either the same or a different operator which has $h_i=+1$ and shape $\Lambda_5(2,0)=\vcenter{\hbox{\SSYTshape{5}{2}{0}}}$.
Both have mass dimension $\Delta=6$ and five external legs.

At one loop, particle counting fixes the non-scaleless two-particle cuts to combine the minimal five-point source amplitude $A_j$ with a four-point \gls{sm} tree amplitude.
First, we consider a nonexceptional \gls{sm} amplitude with $h_{\mathrm{SM}}=0$.
We insert the \gls{ssyt} expansion of the \gls{sm} amplitude into the cut and evaluate the integral through the channel contraction of \Eq{eq:cut-coefficient-matrix-product}.
By \Eq{eq:cut-helicity-ssyt-degree}, the cut has $h_{\mathcal C}=h_j+h_{\mathrm{SM}}=-1$, so every output tableau contains one more barred blue column than white columns.

In both examples below, $\mathcal C$ denotes the all-outgoing cut, with kinematic arguments suppressed throughout.
The \gls{sm} amplitude and its tableau expansion are evaluated on $(-\ell_2,-\ell_1,\alpha_L)$, while the source amplitude and its tableau are evaluated on $(\ell_1,\ell_2,\alpha_R)$, as in \Eq{eq:rge-one-loop-convolution}.
The output tableaux contain only the uncut external legs.

The input amplitudes and their possible output shapes can be displayed together as
\begin{equation}
\begin{aligned}
    \mathcal C%
    ={}&
    \sum_{\text{cut states}}\sigma_c
    \int d\Phi_2(P)\,
    A^{\mathrm{SM}}_{4,h=0}\,
    A_j
    \\[1ex]
    ={}&
    \sum_{\text{cut states}}\sigma_c
    \int d\Phi_2(P)\,
    \biggl[
    d^{(4,0)}_0(s)
    +d^{(4,0)}_1(s)\,
    {\scriptstyle
    \overline{\begin{array}{@{}c@{}}
    \begin{ytableau}
        *(blue!40) \\
        *(blue!40)
    \end{ytableau}
    \end{array}}
    \begin{array}{@{}c@{}}
    \begin{ytableau}
        \, \\
        \,
    \end{ytableau}
    \end{array}}
    +\ldots
    \biggr]
    \times
    {\scriptstyle
    \overline{\begin{array}{@{}c@{}}
    \begin{ytableau}
        *(blue!40) \\
        *(blue!40)
    \end{ytableau}
    \end{array}}}
    \\[1ex]
    ={}&
    d_0(s)\,
    {\scriptstyle
    \overline{\begin{array}{@{}c@{}}
    \begin{ytableau}
        *(blue!40) \\
        *(blue!40)
    \end{ytableau}
    \end{array}}}
    +d_1(s)\,
    {\scriptstyle
    \overline{\begin{array}{@{}c@{}}
    \begin{ytableau}
        *(blue!40) & *(blue!40) \\
        *(blue!40) & *(blue!40)
    \end{ytableau}
    \end{array}}
    \begin{array}{@{}c@{}}
    \begin{ytableau}
        \, \\
        \,
    \end{ytableau}
    \end{array}}
    +d_2(s)\,
    {\scriptstyle
    \overline{\begin{array}{@{}c@{}}
    \begin{ytableau}
        *(blue!40) & *(blue!40) & *(blue!40) \\
        *(blue!40) & *(blue!40) & *(blue!40)
    \end{ytableau}
    \end{array}}
    \begin{array}{@{}c@{}}
    \begin{ytableau}
        \, & \, \\
        \, & \,
    \end{ytableau}
    \end{array}}
    +\ldots
    \\
    ={}&
    \sum_k d_k(s)\,
    \mathcal Y_{\Lambda_5(2k,2k+2)}.
\end{aligned}
\label{eq:one-loop-example-ordinary}
\end{equation}
The source diagram represents the chosen contact amplitude $A_j$, including its normalization.
As in \Eq{eq:implicit-ssyt-filling-sum}, each coefficient multiplying a shape implies a sum over allowed fillings.
Terms without fillings compatible with the specified external states are absent.
None of the output shapes in \Eq{eq:one-loop-example-ordinary} matches the target shape $\Lambda_5(2,0)=\vcenter{\hbox{\SSYTshape{5}{2}{0}}}$, so this nonexceptional cut does not contribute to $\gamma_{ij}^{(1)}$.

Equation~\eqref{eq:one-loop-example-ordinary} serves as an example where the \glspl{ssyt} are displayed explicitly. Below, we use the compact notation shown in the last line of \Eq{eq:one-loop-example-ordinary}.

If the exceptional Yukawa contributions are present, the four-fermion \gls{sm} amplitude with $h_{\mathrm{SM}}=+2$ changes this conclusion.
Its \gls{ssyt} expansion contains two more white columns than barred blue columns.
Together with the same source amplitude, the cut now has $h_{\mathcal C}=-1+2=+1$.
The corresponding cut is
\begin{equation}
\begin{aligned}
    \mathcal C
    ={}&
    \sum_{\text{cut states}}\sigma_c
    \int d\Phi_2(P)\,
    A^{\mathrm{SM}}_{4,h=+2}\,
    A_j
    \\
    ={}&
    \sum_{\text{cut states}}\sigma_c
    \int d\Phi_2(P)\,
    \left[
    \sum_k d_k^{(4,+2)}(s)\,
    \mathcal Y_{\Lambda_4(2k+4,2k)}
    \right]
    A_j
    \\
    ={}&
    \sum_k \widehat d_k(s)\,
    \mathcal Y_{\Lambda_5(2k+2,2k)}.
\end{aligned}
\label{eq:one-loop-example-exceptional}
\end{equation}
The target shape $\Lambda_5(2,0)$ is now present among the possible outputs.
The mixing is therefore no longer excluded by helicity, although a nonzero result still requires compatible species, fillings, gauge contractions, and nonvanishing channel coefficients.

Note that in both examples, the column counting reflects helicity conservation and does not prescribe how to multiply the input tableaux or delete their boxes. The contraction is performed in the channel basis of \Sec{sec:cut-integrals-channel-basis}.

Unlike the examples above, general mixing can involve source and target operators with different numbers of external legs.
Denote the \gls{sm} tree amplitude by $A_L$ and the source tree amplitude by $A_R=A_j^{\mathrm{tree}}$, with $N_L$ and $N_R$ legs, respectively.
The source amplitude need not be minimal.
For a pair of tableaux in their Lorentz-invariant \gls{ssyt} expansions, denote the spinor degrees by $(n_L,\widetilde n_L)$ and $(n_R,\widetilde n_R)$.

We now derive the general selection rule in the absence of exceptional Yukawa contributions.
For an \gls{sm} tree amplitude with $N_L\geq4$, the four-point result $h_L=0$ and the higher-point helicity bound in \Eq{eq:constraint_higher_point} restrict its tableaux to $\lvert n_L-\widetilde n_L\rvert\leq2(N_L-4)$.
The channel contraction preserves the sum of the external helicities, as expressed in \Eq{eq:cut-helicity-ssyt-degree}.
A two-particle cut leaves $N_{\mathcal C}=N_L+N_R-4$ uncut external legs, so these relations restrict the output tableaux to
\begin{equation}
    \left|
    (n_{\mathcal C}-\widetilde n_{\mathcal C})
    -(n_R-\widetilde n_R)
    \right|
    \leq 2(N_L-4)
    =2(N_{\mathcal C}-N_R).
\end{equation}
A target tableau that violates this condition cannot appear in the projected cut, regardless of the projection coefficients.

The corresponding weight bounds now follow from \Eq{eq:def-holomorphic-weights}.
For a general cut through $N_c$ intermediate states, particle and helicity counting give
\begin{equation}
    w_{\mathcal C}=w_L+w_R-2N_c,
    \qquad
    \bar w_{\mathcal C}=\bar w_L+\bar w_R-2N_c.
    \label{eq:cut-weight-composition}
\end{equation}
The allowed \gls{sm} tableaux have $w_L,\bar w_L\geq4$, while the source amplitude satisfies $w_R\geq w_j$ and $\bar w_R\geq\bar w_j$ by \Eq{eq:def-operator-weights}.
At one loop, $N_c=2$ therefore implies $w_{\mathcal C}\geq w_j$ and $\bar w_{\mathcal C}\geq\bar w_j$.
In an operator basis with definite-helicity leading contact amplitudes and at fixed operator dimension $\Delta$, the exclusion above gives
\begin{equation}
    \boxed{
    \gamma_{ij}^{(1)}=0
    \quad\text{if}\quad
    w_i<w_j
    \quad\text{or}\quad
    \bar w_i<\bar w_j
    }.
    \label{eq:cheung-shen-selection}
\end{equation}
This reproduces the one-loop selection rule of \Ref{Cheung:2015aba} and holds when exceptional Yukawa contributions are excluded.

Using \Eq{eq:ssyt-cheung-weight-dictionary}, the non-renormalization theorem can be written in the equivalent forms
\begin{equation}
    \boxed{
    \begin{gathered}
        \gamma_{ij}^{(1)}=0
        \quad\text{if}\quad
        n_i>n_j
        \quad\text{or}\quad
        \widetilde n_i>\widetilde n_j,
        \quad\text{or, equivalently,}\quad
        N_i-N_j<\lvert h_i-h_j\rvert
    \end{gathered}
    }.
    \label{eq:one-loop-ssyt-degree-selection}
\end{equation}
For nonzero mixing, the target can therefore have no more white columns and no more barred blue columns than the source.
In \Figs{fig:dimension-six-ssyt-weight-lattice}{fig:dimension-eight-ssyt-weight-lattice}, this restricts the possible targets of a source component at $(N_j,h_j)$ to its upward-pointing ``forward light cone'', including its boundary.
The two boundaries satisfy $N_i-N_j=\pm(h_i-h_j)$ and follow the diagonal weight directions.
Moving along the upper-left boundary decreases the number of white columns while leaving the barred blue count unchanged, whereas moving along the upper-right boundary does the reverse.
The cone therefore gives a graphical representation of the tableau restriction in \Eq{eq:one-loop-ssyt-degree-selection}.

All targets outside this forward cone, including those below the source, are excluded.
At equal particle numbers, nonzero mixing requires equal total helicities, reproducing the exclusion in the first example.
Targets inside the cone are not excluded by this theorem, but their mixing can still vanish because of additional constraints.
For operators containing several helicity components, the relevant source and target components must be examined separately.

The exceptional four-fermion \gls{sm} amplitudes have $n_L-\widetilde n_L=\pm4$ at $N_L=4$ and therefore evade the \gls{sm} tableau bound used above.
Their weights $(w_L,\bar w_L)=(2,6)$ or $(6,2)$ allow the cut to lower one weight by two units relative to the source tree amplitude, as illustrated by \Eq{eq:one-loop-example-exceptional}.
Including these amplitudes and their higher-point extensions, the general \gls{sm} tree-level bound is $w_L,\bar w_L\geq2$~\cite{Cheung:2015aba}.
The same cut argument therefore still implies $\gamma_{ij}^{(1)}=0$ if $w_i+2<w_j$ or $\bar w_i+2<\bar w_j$.

A three-point factor instead gives a massless bubble.
Such scaleless integrals require care because their vanishing in dimensional regularization can hide cancellations between UV and IR poles.
The infrared analysis of \Ref{Cheung:2015aba}, reviewed in \Sec{sec:on-shell-rge-infrared}, shows that these contributions do not restore the entries excluded by the nonexceptional one-loop theorem.

Even inside the cone, mixing can be excluded by the channel-matching condition of \Sec{sec:cut-integrals-channel-basis}.
For a two-particle cut, the auxiliary $\SU(2)$ Casimir fixes the physical $\SU(2)_P$ spin $J$, despite the distinct group actions, as derived in \App{app:two-particle-casimirs}.
\Ref{Jiang:2020rwz} derives non-renormalization selection rules from this physical spin using the Pauli--Lubanski Casimir.
For $N_c>2$, an auxiliary $\U(N_c)$ representation can contain several spins $J$, so its Casimir does not in general determine the physical spin.
The absence of channel overlap in an individual cut does not by itself establish a vanishing anomalous-dimension entry.
We next discuss when higher-loop mixing is determined entirely by cuts involving tree-level amplitudes.

\subsection{Higher-loop cuts and length selection rules}
\label{sec:higher-loop-cut-topologies}

\begin{figure}[t]
    \centering
    \begin{tikzpicture}[
        every node/.style={font=\footnotesize},
        amplitude/.style={circle,draw,minimum size=1.05cm,align=center,inner sep=0pt,outer sep=0pt},
        cut/.style={densely dashed,black!80,line width=0.8pt}
    ]
        \foreach \panel/\smloop/\ffloop/\cutlegs/\offset in {a/1/0/2/0,b/0/1/2/5,c/0/0/3/10} {
            \begin{scope}[shift={(\offset,0)}]
                \node[amplitude,fill=gray!10] (sm) at (-0.90,0) {$A_{\mathrm{SM}}^{(\smloop)}$};
                \node[amplitude,fill=blue!8] (source) at (0.90,0) {$F_j^{(\ffloop)}$};
                \ifnum\cutlegs=3\relax
                    \draw (sm.35) .. controls (-0.20,0.62) and (0.20,0.62) .. (source.145);
                    \draw (sm.0) -- (source.180);
                    \draw (sm.325) .. controls (-0.20,-0.62) and (0.20,-0.62) .. (source.215);
                \else
                    \draw (sm.25) -- (source.155);
                    \draw (sm.335) -- (source.205);
                \fi
                \draw (sm.155) -- (-2.05,0.70);
                \draw (sm.205) -- (-2.05,-0.70);
                \draw (source.25) -- (2.05,0.70);
                \draw (source.335) -- (2.05,-0.70);
                \foreach \dotangle in {158,169,180,191,202} {
                    \fill ($(sm.center)+(\dotangle:1cm)$) circle[radius=0.6pt];
                }
                \foreach \dotangle in {-22,-11,0,11,22} {
                    \fill ($(source.center)+(\dotangle:1cm)$) circle[radius=0.6pt];
                }
                \draw[cut] (0,-0.75) -- (0,0.75);
                \node at (0,-1.25) {(\panel) $N_c=\cutlegs$};
            \end{scope}
        }
    \end{tikzpicture}
    \caption{The three two-factor cut topologies entering a generic two-loop form-factor \gls{rge}. Gray and blue blobs denote \gls{sm} amplitudes and source form factors, respectively, with loop orders indicated by superscripts. In the \gls{rge}, the \gls{sm} blob represents the appropriate matrix element of $\mathcal M^\dagger$ in \Eq{eq:rge-higher-loop-master}. The dashed lines denote the on-shell cuts. The two-particle cuts (a) and (b) contain a one-loop amplitude or form factor, whereas the three-particle cut (c) contains only tree-level factors. External-leg partitions and sums over allowed intermediate states are implicit.}
    \label{fig:two-loop-cut-topologies}
\end{figure}
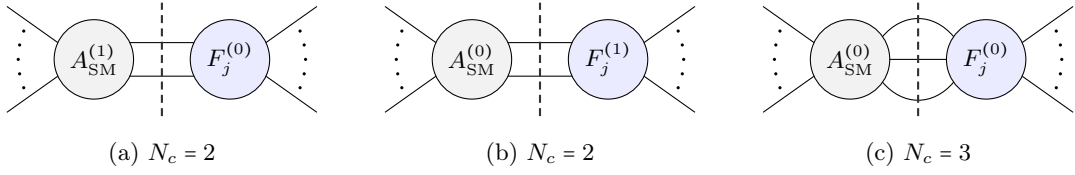

To extend the \gls{ssyt} argument of \Sec{sec:one-loop-ssyt-selection} beyond one loop, we must distinguish cuts built entirely from tree-level quantities from those containing loop amplitudes or form factors.
The channel projection applies to both, but the tree-level helicity bounds used above do not generally apply to the loop-level factors.
For a cut through $N_c\geq2$ states, with loop orders $L_L$ and $L_R$ on its two sides, loop counting gives
\begin{equation}
    L=L_L+L_R+N_c-1.
    \label{eq:cut-loop-counting}
\end{equation}
At two loops, this leaves the three possibilities shown in \Fig{fig:two-loop-cut-topologies}: a two-particle cut with a one-loop \gls{sm} amplitude, a two-particle cut with a one-loop source form factor, or a three-particle cut between two tree-level quantities~\cite{Bern:2019wie,Bern:2020ikv}.

The channel projection organizes all three topologies.
However, only the last one is determined entirely by tree-level quantities and their helicity structure.
The first two require loop amplitudes or form factors, including their rational terms and ultraviolet and infrared singularities.
Moreover, an $\mathcal O(\varepsilon)$ contribution can combine with a $1/\varepsilon$ pole to give a finite term, so a generic higher-loop anomalous dimension cannot be inferred from four-dimensional tree cuts alone~\cite{Bern:2020ikv,Baratella:2022nog}.

We call the cut with $N_c=L+1$ and $L_L=L_R=0$ the maximal tree cut.
It has the largest number of cut lines at fixed loop order among cuts separating the diagram into two connected pieces.
The question is therefore when these tree-level cuts suffice to determine the anomalous dimension.

The length selection rule identifies a case in which the sum of maximal tree cuts determines the full anomalous dimension.
Let a longer source operator $\mathcal O_l$ and a shorter target operator $\mathcal O_s$ have lengths $N_l$ and $N_s$, with $\delta N=N_l-N_s>0$ as in \Sec{sec:on-shell-rge-higher-loops}.
In a massless theory without relevant couplings, their anomalous dimension obeys~\cite{Bern:2019wie,Bern:2020ikv}
\begin{equation}
    \boxed{
    \gamma_{sl}^{(L)}=0
    \qquad\text{for}\qquad
    L\leq\delta N
    }.
    \label{eq:bern-length-selection}
\end{equation}
The first loop order not excluded by length alone is therefore
\begin{equation}
    L_\star=\delta N+1.
    \label{eq:first-length-allowed-order}
\end{equation}
At this order, only cuts through
\begin{equation}
    N_c=L_\star+1=\delta N+2
    \label{eq:bern-maximal-cut-number}
\end{equation}
states can contribute, and both factors in every such cut are tree-level quantities.
A contributing cut requires at least two uncut legs on the \gls{sm} side. With $N_c=N_l-N_s+2$, leg counting therefore forces the source form factor to have exactly $N_l$ legs and hence to be minimal.
It is therefore a local polynomial with only finitely many cut-channel components.
Only the corresponding \gls{sm} projections enter \Eq{eq:cut-coefficient-matrix-product}.

The length-counting and lower-loop form-factor arguments in \Secs{sec:on-shell-rge-higher-loops}{sec:on-shell-rge-infrared} exclude the iteration and infrared terms at this order.
The sum of maximal tree cuts therefore determines the full entry $\gamma_{sl}^{(L_\star)}$ via \Eq{eq:rge-maximal-tree-cut}.
The projection can then establish a non-renormalization theorem either because no cut-channel components match or because the contracted result cannot contain the target contact amplitude.

The length theorem identifies the relevant cut topologies, while the \gls{ssyt} analysis constrains which target shapes can occur after their contraction.
We next use the \gls{sm} helicity bounds to restrict these output shapes, following the logic of \Sec{sec:one-loop-ssyt-selection}.

\subsection{Selection rules for maximal tree cuts}
\label{sec:maximal-tree-cut-selection}

\begin{figure}[t]
    \centering
    \begin{tikzpicture}[
        every node/.style={font=\footnotesize},
        amplitude/.style={circle,draw,minimum size=1.25cm,align=center,inner sep=0pt,outer sep=0pt},
        cut/.style={densely dashed,black!80,line width=0.8pt}
    ]
        \path[use as bounding box] (-7.5,-2.0) rectangle (7.5,1.9);
        \node[amplitude,fill=gray!10] (sm) at (-2.1,0) {$A^{\mathrm{SM}}$};
        \node[amplitude,fill=blue!8] (source) at (2.1,0) {$A_l$};

        \draw (sm.35) .. controls (-0.9,1.0) and (0.9,1.0) .. (source.145);
        \draw (sm.10) .. controls (-0.9,0.35) and (0.9,0.35) .. (source.170);
        \draw (sm.325) .. controls (-0.9,-1.0) and (0.9,-1.0) .. (source.215);
        \foreach \yy in {-0.10,-0.25,-0.40}{
            \fill (-0.7,\yy) circle[radius=0.6pt];
            \fill (0.7,\yy) circle[radius=0.6pt];
        }
        \draw[cut] (0,-1.12) -- (0,1.12);
        \node at (0,1.48) {$N_c=L_\star+1$};

        \draw (sm.155) -- (-4.2,0.85);
        \draw (sm.205) -- (-4.2,-0.85);
        \draw (source.25) -- (4.2,0.85);
        \draw (source.335) -- (4.2,-0.85);
        \foreach \angle in {-18,-9,0,9,18}{
            \fill ($(source.center)+(\angle:1.6cm)$) circle[radius=0.6pt];
        }
        \node[align=center] at (-5.15,0) {$2$\\[3pt]uncut legs};
        \node[align=center] at (5.15,0) {$N_s-2$\\[3pt]uncut legs};
        \node[align=center] at (-2.1,-1.62) {SM tree amplitude\\[4pt]$N_c+2=N_l-N_s+4$};
        \node[align=center] at (2.1,-1.62) {Minimal source amplitude\\[4pt]$N_l=N_s+N_c-2$};
    \end{tikzpicture}
    \caption{Maximal tree cut for longer-to-shorter mixing at $L_\star=N_l-N_s+1$, evaluated on $N_s$ external particles. The gray \gls{sm} tree amplitude and the blue minimal source amplitude are sewn through $N_c=L_\star+1$ on-shell states, leaving two uncut external legs on the \gls{sm} side and $N_s-2$ on the source side. Sums over allowed intermediate states and external-leg assignments are implicit.}
    \label{fig:maximal-tree-cut}
\end{figure}
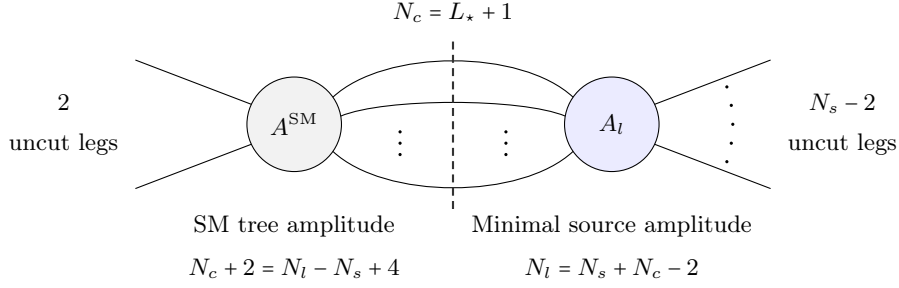

We first consider the longer-to-shorter mixing of \Sec{sec:higher-loop-cut-topologies} at its first length-allowed order $L_\star=\delta N+1$.
At this order, the contributing cuts contain the minimal source amplitude $A_l$ and an \gls{sm} tree amplitude sewn through $N_c=\delta N+2$ states, as shown in \Fig{fig:maximal-tree-cut}.
The \gls{sm} factor therefore has
\begin{equation}
    N_{\mathrm{SM}}
    =N_s-N_l+2N_c
    =\delta N+4
    =L_\star+3
    \label{eq:bern-sm-tree-multiplicity}
\end{equation}
legs.
The total helicity of the \gls{sm} factor is constrained by the bound in \Eq{eq:constraint_higher_point} discussed in \Sec{sec:sm-helicity-structure}.

We contract the matching cut-channel components according to \Eq{eq:cut-coefficient-matrix-product} and express the result in the Lorentz-invariant \gls{ssyt} basis using \Eq{eq:cut-in-li-ssyt-basis}.
We first illustrate how this restricts the target shapes at two loops and then give the general result.

We return to the dimension-six source component $\psi_-^2\phi^3$ of \Sec{sec:one-loop-ssyt-selection}, with shape $\Lambda_5(0,2)$ and total helicity $h_l=-1$.
We ask whether it can renormalize a target component $F_+^2\phi^2$, whose shape is $\Lambda_4(4,0)$ and whose total helicity is $h_s=+2$.
Here $\delta N=1$, so the first order not excluded by length is $L_\star=2$, and the relevant cut contains three intermediate states.

By \Eq{eq:bern-sm-tree-multiplicity}, the \gls{sm} factor has five legs, so the helicity bound allows for $h_{\mathrm{SM}}=\pm1$.
Evaluating the three-particle cut through the channel contraction in \Eq{eq:cut-coefficient-matrix-product} gives the following Lorentz-invariant \gls{ssyt} expansions.
We use the all-outgoing cut notation of \Sec{sec:one-loop-ssyt-selection}, now with three intermediate particles.
For $h_{\mathrm{SM}}=+1$, the cut has total helicity $h_{\mathcal C}=h_l+1=0$, giving
\begin{equation}
    \mathcal C
    =
    \sum_{\text{cut states}}\sigma_c
    \int d\Phi_3(P)\,
    A^{\mathrm{SM}}_{5,h=+1}\,
    A_l
    =
    \sum_k d_k^{(+)}(s)\,
    \mathcal Y_{\Lambda_4(2k,2k)}.
\label{eq:two-loop-example-plus}
\end{equation}
For $h_{\mathrm{SM}}=-1$, the corresponding cut instead has $h_{\mathcal C}=h_l-1=-2$, and
\begin{equation}
    \mathcal C
    =
    \sum_{\text{cut states}}\sigma_c
    \int d\Phi_3(P)\,
    A^{\mathrm{SM}}_{5,h=-1}\,
    A_l
    =
    \sum_k d_k^{(-)}(s)\,
    \mathcal Y_{\Lambda_4(2k,2k+4)}.
\label{eq:two-loop-example-minus}
\end{equation}
The coefficients $d_k^{(\pm)}(s)$ collect the contributions from the channel contraction after conversion to the Lorentz-invariant \gls{ssyt} basis.
As before, terms without fillings compatible with the external states are absent.

Neither expansion contains the target shape $\Lambda_4(4,0)$.
Producing it would require $h_{\mathrm{SM}}=h_s-h_l=3$, which is excluded by the nonexceptional helicity bound.
Under the completeness conditions stated in \Sec{sec:higher-loop-cut-topologies}, the absence of this target from the maximal tree cuts excludes the corresponding two-loop mixing.

By contrast, the conjugate target shape $\Lambda_4(0,4)$ is allowed by \Eq{eq:two-loop-example-minus}.
Determining whether its coefficient is nonzero requires the channel contraction and the remaining species and gauge information.

The \gls{ssyt} argument of \Sec{sec:one-loop-ssyt-selection} extends directly to maximal tree cuts through $N_c=L+1$ intermediate states.
Let the nonexceptional \gls{sm} tree amplitude $A_L$ have $N_{\mathrm{SM}}=N_L\geq4$ legs and the source tree amplitude $A_R=A_j^{\mathrm{tree}}$ have $N_R$ legs.
The cut leaves $N_{\mathcal C}=N_R+N_{\mathrm{SM}}-2(L+1)$ external legs, while the opposite helicities of the intermediate states cancel.
The \gls{sm} helicity bound therefore restricts the output tableaux to
\begin{equation}
    \left|
    (n_{\mathcal C}-\widetilde n_{\mathcal C})
    -(n_R-\widetilde n_R)
    \right|
    \leq2(N_{\mathrm{SM}}-4)
    =2\bigl[N_{\mathcal C}-N_R+2(L-1)\bigr].
    \label{eq:maximal-cut-output-ssyt-bound}
\end{equation}
As at one loop, a target violating this restriction is absent from the projected cut expansion, independently of the projection coefficients.

The source amplitude need not be minimal, but its weights satisfy $w(A_j^{\mathrm{tree}})\geq w_j$ and $\bar w(A_j^{\mathrm{tree}})\geq\bar w_j$.
At fixed operator dimension $\Delta$, \Eqs{eq:cut-weight-composition}{eq:ssyt-cheung-weight-dictionary} imply that the cut can contribute to a target component only if
\begin{equation}
    n_i\leq n_j+2(L-1),
    \qquad
    \widetilde n_i\leq\widetilde n_j+2(L-1).
    \label{eq:maximal-cut-ssyt-degree-selection}
\end{equation}
The target can therefore have at most $L-1$ more white columns and at most $L-1$ more barred blue columns than the source.
At one loop, \Eq{eq:cut-loop-counting} forces $N_c=2$ and $L_L=L_R=0$, so every two-factor cut is a maximal tree cut.
The $L=1$ limit therefore reproduces the one-loop result in \Eq{eq:one-loop-ssyt-degree-selection}.

These statements constrain only the maximal tree cuts. Contributions containing loop amplitudes, as well as iteration and infrared terms, must be considered separately before inferring a zero of the full anomalous dimension.
Denote by $\mathcal C_{ij}^{(L),\mathrm{max}}$ the contribution of maximal tree cuts with source $j$ to the chosen target contact component $i$.
In weight language, the exclusion reads
\begin{equation}
    \boxed{
    \mathcal C_{ij}^{(L),\mathrm{max}}=0
    \quad\text{if}\quad
    w_i+2(L-1)<w_j
    \quad\text{or}\quad
    \bar w_i+2(L-1)<\bar w_j
    }.
    \label{eq:maximal-cut-zero-weight}
\end{equation}
A target can receive a maximal-tree-cut contribution only if $|h_i-h_j|\leq N_i-N_j+2(L-1)$.
In \Figs{fig:dimension-six-ssyt-weight-lattice}{fig:dimension-eight-ssyt-weight-lattice}, this defines an upward-pointing cone whose vertex is shifted downward from the source by $2(L-1)$ units in $N$.
Targets outside this shifted cone have no maximal-tree-cut contribution, whereas targets inside it, including its boundaries, are not excluded.

For the longer-to-shorter mixing considered above, set $i=s$, $j=l$, and $L=L_\star=\delta N+1$.
At the same operator mass dimension, the target has $2\delta N$ more spinors than the source by \Eq{eq:dimension-helicity-ssyt-counts}.
Together with \Eq{eq:maximal-cut-ssyt-degree-selection}, this requires the increases in both spinor degrees to be non-negative.
The target therefore has $\delta N$ additional columns in total, distributed between the white and barred blue parts.
This compares the source and target column counts and is not a rule for multiplying the tableaux entering the cut.

Under the conditions of \Sec{sec:higher-loop-cut-topologies}, the maximal tree cuts determine the full entry at $L_\star=N_l-N_s+1$.
For fixed-helicity contact components, the equivalent weight and tableau exclusions can then be written as
\begin{equation}
    \boxed{
    \gamma_{sl}^{(L_\star)}=0
    \quad\text{if}\quad
    w_s+2(L_\star-1)<w_l
    \quad\text{or}\quad
    \bar w_s+2(L_\star-1)<\bar w_l
    }.
    \label{eq:bern-weight-zero}
\end{equation}
At the same operator dimension $\Delta$, this is equivalent to
\begin{equation}
    \boxed{
    \begin{gathered}
        \gamma_{sl}^{(L_\star)}=0
        \quad\text{if}\quad
        n_s>n_l+2(L_\star-1)
        \quad\text{or}\quad
        \widetilde n_s>\widetilde n_l+2(L_\star-1),\\
        \text{or, equivalently,}\quad
        N_s-N_l+2(L_\star-1)<\lvert h_s-h_l\rvert
    \end{gathered}
    }.
    \label{eq:bern-helicity-zero}
\end{equation}
To interpret \Eq{eq:bern-helicity-zero} in the same figures, fix the source and the loop order $L_\star$.
The length relation selects the target row $N_s=N_l-(L_\star-1)$, where the maximal tree cuts determine the full anomalous-dimension entry.
On this row, targets outside the shifted forward cone are excluded, while those inside or on its boundaries are not excluded by the non-renormalization theorem.
Targets below this row are already excluded by the length bound in \Eq{eq:bern-length-selection}.
For targets above it, the maximal-tree-cut test alone does not establish a zero of the full anomalous dimension.

In the two-loop example above, the target shapes at $(4,-2)$ and $(4,0)$ lie on the shifted cone's boundaries, while $(4,+2)$ lies outside and is excluded.
For operators containing several helicity components, the relevant source and target components must be examined separately.

Even for targets inside the cone, the channel projection can exclude mixing.
At $L_\star$ loops, $\gamma_{sl}^{(L_\star)}=0$ if, for every allowed maximal tree cut and intermediate state, the source and \gls{sm} amplitudes have no common $(J,\Lambda_c,T_c)$ component, as follows from \Eq{eq:cut-coefficient-matrix-product}.

If exceptional Yukawa contributions are present, the weaker bound $w_{\mathrm{SM}},\bar w_{\mathrm{SM}}\geq2$ replaces the bound used above~\cite{Cheung:2015aba}.
Using the same channel contraction, we therefore find that $\mathcal C_{ij}^{(L),\mathrm{max}}$ still vanishes if $w_i+2L<w_j$ or $\bar w_i+2L<\bar w_j$.
For contact components of equal mass dimension, this is equivalent to $\mathcal C_{ij}^{(L),\mathrm{max}}=0$ if $n_i>n_j+2L$ or $\widetilde n_i>\widetilde n_j+2L$.
This allows one additional column of either color compared with \Eq{eq:maximal-cut-ssyt-degree-selection}.
At $L_\star=\delta N+1$, this implies $\gamma_{sl}^{(L_\star)}=0$ if $|h_s-h_l|>\delta N+2$, or equivalently if $w_s+2L_\star<w_l$ or $\bar w_s+2L_\star<\bar w_l$.

\section{Conclusions} \label{sec:conclusions}

In the on-shell approach, the renormalization of effective operators is governed by the overlap of amplitudes on multiparticle phase space.
In this work, we have developed a representation-theoretic approach to these overlaps using Stiefel harmonics and their channel projections.

Our first main result concerns non-renormalization in the mixing of longer operators into shorter ones.
At the first loop order allowed by operator length, \(L_\star=\delta N+1\), with \(\delta N=N_l-N_s\), the anomalous dimension is determined entirely by the sum of maximal tree cuts.
In the massless setup considered here, length counting excludes the other cut topologies and the relevant infrared and iteration terms at this order.
The helicity bounds of the \gls{sm} tree amplitudes restrict which Lorentz-invariant \gls{ssyt} shapes can occur in the cut contractions.
We thereby establish higher-loop non-renormalization theorems for longer-to-shorter mixing, extending the one-loop helicity-weight restrictions to higher orders.

Our second main result is an explicit formalism for projecting massless on-shell amplitudes onto Stiefel harmonics.
By combining the \(U(N)\) representation theory of these harmonics with the physical partial-wave decomposition, we obtain a joint basis labeled by \((J,\Lambda,T)\), in which the Young-tableau labels resolve the kinematic degeneracies of multiparticle partial waves at fixed total angular momentum $J$. 
Orthonormality with respect to physical LIPS reduces the cut integrals to matrix products of projected amplitude coefficients, with the intermediate channel labels summed as in \Eq{eq:cut-coefficient-matrix-product}.
In particular, a local \gls{eft} contact amplitude has support on only finitely many channel harmonics, so that only a finite number of projections of the generally nonlocal \gls{sm} amplitude enter the cut.
We have illustrated this procedure through explicit one- and two-loop examples.

The channel decomposition also provides a criterion for vanishing cuts that goes beyond total-helicity counting.
For a cut to be nonvanishing, its two factors must share components with the same channel labels \((J,\Lambda,T)\) for an allowed intermediate state.
Systematically applying this criterion may reveal additional non-renormalization theorems beyond those obtained from helicity weights and operator length alone.

Several directions merit further investigation. An important technical challenge is the development of a systematic infrared-subtraction prescription for multiparticle harmonic projections, building on the treatment of infrared singularities in two-particle partial waves~\cite{Baratella:2020dvw}.
Such a prescription would extend the practical applicability of the projection formalism to more general amplitudes and cuts.
It would also be interesting to investigate whether the representation-theoretic structure leads to additional selection rules, including in extensions of \gls{smeft} involving dynamical gravity~\cite{Ruhdorfer:2019qmk,Baratella:2021guc}.
Along these lines, another natural direction is to extend the channel-harmonic analysis to nonlinear operator mixing, including the renormalization of dimension-eight operators by double insertions of dimension-six interactions, for which non-renormalization theorems have already been established~\cite{Cao:2023adc}.
Finally, the tableau-resolved partial-wave amplitudes provide a natural setting in which to explore constraints from unitarity, analyticity, and positivity. 
We believe that the framework developed here will be useful not only for understanding zeros in anomalous-dimension matrices, but also for uncovering further structure in multiparticle scattering amplitudes.

\paragraph{Note added.}
While this manuscript was in the final stages of preparation, \Ref{Bresciani:2026yuw} appeared on arXiv, independently deriving higher-loop non-renormalization theorems without employing an SSYT decomposition.
The helicity-based selection rules derived in \Sec{sec:maximal-tree-cut-selection} agree with the corresponding results of that work.

\acknowledgments

The work of NC ended in June 2026 and was supported in part by the U.S. Department of Energy under the grant DE-SC0011702.
MS is partially supported by the Swiss National Science Foundation under contract 200020-213104 and through the National Center of Competence in Research SwissMAP.
The work of MS was partially supported by the Collaborative Research Center SFB1258, the Munich Institute for Astro-, Particle and BioPhysics (MIAPbP), and the Excellence Cluster ORIGINS, which is funded by the Deutsche Forschungsgemeinschaft (DFG, German Research Foundation) under Germany’s Excellence Strategy – EXC-2094-390783311.
We used OpenAI Codex with GPT-6 Astra and Anthropic's Claude Code with Claude Opus 5.5 to assist with proofreading the manuscript for correctness, consistency, and typographical errors, as well as generating and editing equations and \LaTeX{} code for figures.
The authors take full responsibility for the final manuscript and its results.

\newpage

\appendix

\section{Channel bases and projection formulas} \label{app:basis-projection}

This appendix develops the joint \gls{ssyt} and partial-wave projection used in \Secs{sec:ssyt-decomposition}{sec:cut-integrals-channel-basis}. Starting from the Stiefel harmonics~\cite{Henning:2019mcv,Henning:2019enq} and the Poincaré partial-wave basis~\cite{Jiang:2020rwz,Shu:2021qlr}, we derive the projection and reconstruction formulas and explain how their labels combine. In particular, we show how Poincaré invariance contracts the open massive-spinor indices, leading to the channel functions $\mathcal B^J$, and how these functions are related to Lorentz-invariant \glspl{ssyt}. Throughout, we consider massless external particles at fixed timelike channel momentum, with species and helicities specified. Gauge, flavor, and permutation labels are suppressed.
All amplitude kernels and their projected coefficients in this appendix use the physical convention of \Sec{sec:on-shell-rge-one-loop}.
All-outgoing inputs must first be converted to that convention.

\subsection{Fixed-momentum kinematics and conventions}

We start by choosing $N\geq2$ spinors $\lambda_{i \alpha}$, with $\tilde{\lambda}_{i \dot{\alpha}} = (\lambda_{i \alpha})^*$, associated with massless momenta with fixed total massive momentum
\begin{equation}
  P^{\alpha\dot\alpha}=\sum_{i=1}^N \lambda_i^\alpha \tilde\lambda_i^{\dot\alpha},
  \qquad P^2>0, \qquad P^0>0.
  \label{eq:bp-fixedPintro}
\end{equation}
We expand each massless spinor in two basis spinors $x_{I\alpha}$ forming a spin frame for the timelike momentum $P$,
\begin{equation}
    \lambda_{i\alpha}=Z_i{}^I x_{I\alpha},
    \qquad I=1,2.
    \label{eq:bp-Lambda-Zx}
\end{equation}
We use the same massive spin frame for all initial, final, and cut configurations with total momentum $P$.
Each row is an ordinary two-component spinor and may be associated with a massless momentum, while together they reconstruct the massive momentum $P = x^\dagger x$. Therefore, $x^I_{\alpha}$ and $\tilde{x}_{I \dot{\alpha}}$ may be viewed as single massive spinors carrying the massive little-group index $I=\{1,2\}$. We choose the common phase of the frame such that $\det x=\sqrt{s}$, with $s=P^2$. Contracting the Lorentz indices of $T^{\alpha\ldots,\dot\alpha\ldots}$ with $x^I_\alpha$ and $\widetilde x_{I\dot\alpha}$ expresses the same tensor with open $\SU(2)_P$ indices. Since the massive spinors form a basis for timelike $P$, this change of basis is invertible and no information is lost.

We follow \Ref{Durieux:2019eor} for spinor contractions, index raising and lowering, and complex conjugation, with $\epsilon^{12}=-\epsilon_{12}=1$ for both Lorentz and massive little-group indices.
With $\overline Z_{iI}=(Z_i{}^I)^*$, the conjugate spinors satisfy
\begin{equation}
    \widetilde x_{I\dot\alpha}=(x^I_\alpha)^*,
    \qquad
    (x_{I\alpha})^*=-\widetilde x^I_{\dot\alpha},
    \qquad
    \widetilde\lambda_{i\dot\alpha}
    =-\overline Z_{iI}\widetilde x^I_{\dot\alpha}.
    \label{eq:bp-conjugate-spin-frame}
\end{equation}

Undotted and dotted Lorentz indices are converted into conjugate massive little-group indices. For $\SU(2)_P$, these representations are equivalent through the invariant tensor $\epsilon_{IJ}$, so tensors containing both kinds of index can be decomposed into definite spins $J$~\cite{Arkani-Hamed:2017jhn}. Contracting with the massive spinors establishes the conversion of indices. It does not by itself select a definite $J$.

Since the spin frame already reconstructs $P$, substituting \Eq{eq:bp-Lambda-Zx} into the fixed-momentum condition in \Eq{eq:bp-fixedPintro} gives
\begin{equation} \label{eq:bp-condZ}
    Z^\dagger Z = \mathbf{1}_2,
\end{equation}
and thus the spinor configurations satisfying \Eq{eq:bp-fixedPintro} form the complex Stiefel manifold~\cite{Henning:2019mcv}
\begin{equation}
  X_P\simeq V_2(\CC^N)
  =\{Z\in \CC^{N\times 2}:Z^\dagger Z=\mathbf{1}_2\}
  \simeq \frac{\U(N)}{\U(N-2)}.
  \label{eq:bp-stiefelintro}
\end{equation}
There are two commuting actions on $Z_{i}{}^I$ that leave \Eq{eq:bp-condZ} invariant\footnote{The reason why the Lorentz transformations acting within the fixed-$P$ space are restricted to $\SU(2)_P$ is that we fix a timelike momentum $P$ and therefore restrict ourselves to its massive little group. For $\U(N)$, fixing $P$ still leaves the freedom of rotating the massless spinors in a way that leaves $P$ unchanged. Choosing a particular spinor configuration singles out the stabilizer $\U(N-2)$, as discussed in \Ref{Henning:2019mcv}. This constrains which $\U(N)$ representations occur in the harmonics, but the space of allowed configurations still transforms under the full $\U(N)$.}
\begin{equation} \label{eq:bp-2trafos}
  Z\longmapsto UZ,
  \qquad U\in \U(N),
  \qquad\text{and}\qquad
  Z\longmapsto Zu,
  \qquad u\in \SU(2)_P.
\end{equation}
The Stiefel constraint also admits a right $\U(2)$ action, but its central $\U(1)$ is already contained in the left $\U(N)$ action. The first rotates particle labels. The second rotates the two spin-frame columns and is the physical massive little group of $P$ associated with the Poincaré group. Since the two actions commute, the Hilbert space can be decomposed simultaneously under both groups.

For a fixed helicity assignment $\mathbf h=\{h_1,\ldots,h_N\}$, the label $\alpha=\{p_1,h_1;\ldots;p_N,h_N\}$ specifies a physical momentum configuration, with tensor-product state
\begin{equation}
    \ket{\alpha}\equiv\ket{p_1,h_1;\ldots;p_N,h_N}.
\end{equation}
Before separating the total momentum of the multiparticle state, we use the full relativistic one-particle normalization
\begin{equation}
    \langle p',h'|p,h\rangle
    =(2\pi)^3\,2E_{\mathbf p}\,
    \delta^{(3)}(\mathbf p'-\mathbf p)\,
    \delta_{h'h}.
\end{equation}
This label does not contain a choice of spinor-helicity representatives, since independent changes of the spinor phases leave every momentum unchanged:
\begin{equation}
    \lambda_i\longmapsto e^{i\theta_i}\lambda_i,
    \qquad
    \widetilde\lambda_i\longmapsto e^{-i\theta_i}\widetilde\lambda_i.
\end{equation}
Writing $D_{\boldsymbol\theta}=\operatorname{diag}(e^{i\theta_1},\ldots,e^{i\theta_N})$, the points $Z$ and $D_{\boldsymbol\theta}Z$ consequently correspond to the same physical configuration $\alpha$.
It is therefore useful to distinguish the physical ket $\ket{\alpha}$ from $\ket{Z;\mathbf h}$, which denotes the same physical state expressed using the chosen spinor representatives $Z$.
For each momentum configuration $\alpha$, we denote the chosen spinor representative in the fixed massive spin frame by $Z(\alpha)$.
We choose the phase convention of the momentum states so that $\ket{\alpha}=\ket{Z(\alpha);\mathbf h}$.
Using the helicity convention of \Sec{sec:on-shell-rge-form-factors}, we set $q_i=2h_i$ and choose
\begin{equation}
    \ket{D_{\boldsymbol\theta}Z;\mathbf h}
    =e^{-i\mathbf q\cdot\boldsymbol\theta}\ket{Z;\mathbf h}.
    \label{eq:bp-lifted-ket-equivariance}
\end{equation}
It follows that a wavefunction $\Psi(Z)=\braket{Z;\mathbf h|\Psi}$ satisfies
\begin{equation}
    \Psi(D_{\boldsymbol\theta}Z)
    =e^{i\mathbf q\cdot\boldsymbol\theta}\Psi(Z).
    \label{eq:bp-wavefunction-equivariance}
\end{equation}

We keep the physical phase-space measure $d\Phi_N(P)$ defined in \Eq{eq:appPS_LIPS_measure_introduce}, and write its total volume as
\begin{equation}
    \mathcal N_N(s)=\int d\Phi_N(P).
    \label{eq:bp-normalized-phase-space}
\end{equation}
The Stiefel measure remains normalized by $\int d\mu_N=1$.
For integrands whose little-group phases cancel, the two measures obey $\int d\Phi_N\,F=\mathcal N_N(s)\int d\mu_N\,F$, as shown in \App{app:phase-space}.
The separation of the total momentum follows from
\begin{equation}
    \int\prod_{i=1}^{N}\frac{d^3\mathbf p_i}{(2\pi)^3\,2E_i}\,G(\alpha)
    =
    \int\frac{d^4P}{(2\pi)^4}
    \int d\Phi_N(P)\,G(\alpha),
\end{equation}
where $G$ is a test function and the total-momentum integration runs over the allowed momenta.
After separating this total-momentum integration, we use bra-ket notation at fixed total momentum $P$, suppressing the common factor $(2\pi)^4\delta^{(4)}(P'-P)$ unless explicitly displayed.
The fixed-$P$ resolution of the identity is therefore
\begin{equation}
    \mathbf1_{P,\mathbf h}
    =\int d\Phi_N(P)\,\ket{\alpha}\bra{\alpha}
    =\mathcal N_N(s)\int d\mu_N(Z)\,
    \ket{Z;\mathbf h}\bra{Z;\mathbf h}.
    \label{eq:bp-configuration-resolution}
\end{equation}
With the common total-momentum factor suppressed, the overlap of two configurations at fixed $P$ and external helicities is
\begin{equation}
    \langle\alpha'|\alpha\rangle
    =\delta_{\Phi_N}(\alpha',\alpha).
    \label{eq:bp-configuration-inner-product}
\end{equation}
Here $\delta_{\Phi_N}$ is the delta function with respect to the physical phase-space measure. Integrating $\delta_{\Phi_N}(\alpha',\alpha)f(\alpha)$ over $d\Phi_N(P;\alpha)$ gives $f(\alpha')$, with the same phase convention used for both momentum states.
Fixing $P$ does not fix the individual particle momenta, so the configuration delta function remains.

\subsection{Poincaré partial-wave projection}

At fixed total momentum $P$ and helicities $\mathbf h$, the multiparticle states $\ket{\alpha}$ can be decomposed into states $\ket{P;J,M,a}$ of definite total angular momentum.\footnote{For two-particle states, this is the standard helicity partial-wave construction of \Ref{Jacob:1959at}.}
The label $a$ distinguishes independent copies of the spin-$J$ representation~\cite{Shu:2021qlr}, for which we choose an orthonormal basis.
Before suppressing the total-momentum delta function, the normalization and the Poincaré Clebsch--Gordan coefficients (CGCs) are specified by
\begin{align}
    \langle P';J',M',a'|P;J,M,a\rangle
    &=(2\pi)^4\delta^{(4)}(P'-P)\,
    \delta_{J'J}\delta_{M'M}\delta_{a'a},\\
    \langle\alpha|P;J,M,a\rangle
    &=(2\pi)^4\delta^{(4)}\!\left(P-\sum_i p_i\right)
    C^{JMa}(\alpha;P).
\end{align}
In the fixed-$P$ notation introduced above, the latter becomes
\begin{equation}
  C^{JMa}(\alpha;P)=\braket{\alpha|P;J,M,a}.
  \label{eq:bp-PoincareCGC}
\end{equation}
The index $M$ denotes a component in the spin-$J$ representation of $\SU(2)_P$.
It can be written as $M=-J,\ldots,J$ or represented by completely symmetric massive-spinor indices $(I_1,\ldots,I_{2J})$, as in \Ref{Shu:2021qlr}.
Complex conjugation gives the dual massive little-group tensor, for example $(Y^*)_{I_1\ldots I_r}\equiv(Y^{I_1\ldots I_r})^*$.
The CGCs are therefore wavefunctions on multiparticle phase space with an open spin-$J$ index.
The complete orthonormal partial-wave basis provides the resolution of the identity
\begin{align}
  \mathbf{1}_{P,\mathbf{h}}
  &= \sum_{JMa} \ket{P;J,M,a}\bra{P;J,M,a}.
\end{align}
Together with the integral resolution in \Eq{eq:bp-configuration-resolution}, this gives the orthogonality and completeness relations
\begin{align}
  \int d\Phi_N(P)\,
  C^{JMa}(\alpha;P)^*
  C^{J'M'a'}(\alpha;P)
  &=\delta_{JJ'}\delta_{MM'}\delta_{aa'},
  \label{eq:bp-Poincareorth}\\
  \sum_{J,M,a}
  C^{JMa}(\alpha;P)
  C^{JMa}(\beta;P)^*
  &=\delta_{\Phi_N}(\alpha,\beta).
  \label{eq:bp-Poincarecomplete}
\end{align}

Given a scattering amplitude $A(\alpha_f;\alpha_i)$ with fixed initial- and final-state configurations $\alpha_i$ and $\alpha_f$, each with total momentum $P$
\begin{equation}
    A(\alpha_f;\alpha_i) \equiv \langle \alpha_f | \mathcal M | \alpha_i \rangle,
\end{equation}
we can project onto its partial-wave coefficients $a_{a_f a_i}^J(s)$,
\begin{equation}
  {a}^{J}_{a_f a_i}(s)
  =
  \int d\Phi_{N_f}(P)\,d\Phi_{N_i}(P)\,
  C^{J M a_f}(\alpha_f;P)^{*}\,
  A(\alpha_f;\alpha_i)\,
  C^{J M a_i}(\alpha_i;P).
\end{equation}
The partial-wave coefficients depend on the channel invariant $s$, the total spin $J$, and the labels $a_f,a_i$.
Rotational invariance makes them independent of $M$, so we use the same fixed value in both CGCs without summing over it.
Conversely, the amplitude $A(\alpha_f;\alpha_i)$ can be written as
\begin{equation}
  A(\alpha_f;\alpha_i)
  =
  \sum_{J,M,a_f,a_i}
  C^{JMa_f}_{f}(\alpha_f;P)
  a^{J}_{a_f a_i}(s)
  C^{JMa_i}_{i}(\alpha_i;P)^*.
\end{equation}

Using the spinor representatives introduced above, the same wavefunctions can be written as
\begin{equation}
  C^{JMa}(Z;P)
  =
  \braket{Z;\mathbf h|P;J,M,a}.
\end{equation}
For a chosen representative $Z(\alpha)$, this gives $C^{JMa}(\alpha;P)=C^{JMa}(Z(\alpha);P)$.

\subsection{Projection onto Stiefel harmonics}

The \gls{ssyt} construction of \Ref{Henning:2019mcv} provides Stiefel harmonics on the same fixed-$P$ kinematic space.
They are labeled by a $\U(N)$ representation $\Lambda$ and a Young tableau filling $T$,
\begin{equation}
    Y_{\Lambda,T}(Z) = \langle Z;\mathbf{h}|P;\Lambda,T\rangle.
\end{equation}
This overlap uses the same fixed-total-momentum convention as the Poincaré CGCs.\footnote{Before suppressing the common total-momentum factor, the overlap is $\langle\alpha|P;\Lambda,T\rangle=(2\pi)^4\delta^{(4)}(P-\sum_i p_i)\,Y_{\Lambda,T}(Z(\alpha))$, with the massive-spinor indices left implicit.}
For a given shape $\Lambda$, the filling $T$ fixes the individual helicities, so we do not include $\mathbf h$ in the ket.
The full $\U(N)$ construction includes different helicity assignments. 
For fixed external helicities, we keep only the compatible tableau fillings.
Once the shape and filling are specified, no additional $\U(N)$ degeneracy indices are needed.
The harmonic $Y_{\Lambda,T}(Z)$ is nevertheless tensor-valued, with open right-$\SU(2)_P$ indices, as illustrated in \Eq{eq:covariant-ssyt-example}.

The Stiefel harmonics of \Ref{Henning:2019mcv} form a complete basis for wavefunctions at fixed total momentum $P$ and external helicities when all independent spin components are included.
Expressing these components in terms of $J,M$ is a change of basis and preserves completeness.
We orthonormalize the harmonics using physical LIPS.
Relative to unit normalization with $d\mu_N$, this includes a factor $\mathcal N_N(s)^{-1/2}$.
We suppress the fixed momentum $P$ in the argument $Y_{\Lambda,T}(Z)$, including the resulting normalization dependence on $s$.
Orthonormalization can mix \gls{ssyt} fillings with the same $\Lambda$ and external helicities and may require different normalization factors for their spin-$J$ components. We use $T$ to label the resulting orthonormal basis.
Throughout this subsection, the open massive-spinor indices of the harmonics and states are left implicit.
We resolve them into components of definite $J,M$ only when introducing the joint basis below.

The complete orthonormal basis of Stiefel harmonics provides the resolution of the identity
\begin{align}
  \mathbf{1}_{P,\mathbf{h}}
  &=
  \sum_{\Lambda,T}\ket{P;\Lambda,T}\bra{P;\Lambda,T}.
\end{align}
Together with the integral resolution in \Eq{eq:bp-configuration-resolution}, this gives the orthogonality and completeness relations
\begin{align}
    \mathcal N_N(s)\int d\mu_N(Z)\,
    Y_{\Lambda,T}(Z)^*Y_{\Lambda',T'}(Z)
    &=\delta_{\Lambda\Lambda'}\delta_{TT'},
    \label{eq:bp-SSYT-orthogonality}\\
    \sum_{\Lambda,T}
    Y_{\Lambda,T}(Z)Y_{\Lambda,T}(W)^*
    &=\frac{\delta_{\mu_N;\mathbf h}(Z,W)}
    {\mathcal N_N(s)}.
\end{align}
The identity in the suppressed massive-spinor indices is understood in the orthogonality relation.
The Stiefel orthogonality relation also holds without restricting the helicities, since harmonics with different helicity assignments are orthogonal under the little-group phase integrations.
In the resolution of the identity and the completeness relation, the corresponding spin indices are contracted implicitly.
The kernel $\delta_{\mu_N;\mathbf h}$ plays the role of a delta function for wavefunctions with the chosen external helicities.
For any such wavefunction $F$, whose spinor-phase dependence is given by \Eq{eq:bp-wavefunction-equivariance}, integration gives the original wavefunction
\begin{equation}
    \int d\mu_{N}(W) \,
  \delta_{\mu_N;\mathbf{h}}(Z,W)\,F(W)=F(Z).
\end{equation}

We can now project a scattering amplitude onto the Stiefel harmonics in direct analogy with the partial-wave projection above. Consider fixed initial and final helicity assignments and define the amplitude in spinor-helicity variables by
\begin{equation} \label{eq:bp-AZZ}
  A(Z_f;Z_i)
  \equiv
  \bra{Z_f;\mathbf h_f}\mathcal M\ket{Z_i;\mathbf h_i}.
\end{equation}
Here, $Z_i\in V_2(\mathbb C^{N_i})$ and $Z_f\in V_2(\mathbb C^{N_f})$ describe the initial and final fixed-$P$ spinor configurations, respectively. The matrix elements of $\mathcal M$ in the basis of Stiefel harmonics are defined by
\begin{equation}
  b_{\Lambda_fT_f;\Lambda_iT_i}(s)
  \equiv
  \bra{P;\Lambda_f,T_f}
  \mathcal M
  \ket{P;\Lambda_i,T_i}.
  \label{eq:bp-SSYT-amplitude-coefficient}
\end{equation}
The projection leaves the initial and final massive-spinor indices open, so $b$ is a tensor despite these indices being suppressed.
In the reconstruction, its indices are contracted with the corresponding indices of the two harmonics.
Inserting \Eq{eq:bp-configuration-resolution} on both sides of $\mathcal M$ gives one factor of $\mathcal N_N(s)$ for each Stiefel integration.
\begin{equation}
    b_{\Lambda_fT_f;\Lambda_iT_i}(s)
    =
    \mathcal N_{N_f}(s)\mathcal N_{N_i}(s)
    \int d\mu_{N_f}(Z_f)\,d\mu_{N_i}(Z_i)
    Y^{(f)}_{\Lambda_f,T_f}(Z_f)^*
    A(Z_f;Z_i)
    Y^{(i)}_{\Lambda_i,T_i}(Z_i).
    \label{eq:bp-SSYT-amplitude-projection}
\end{equation}
Conversely, completeness of the Stiefel harmonics gives
\begin{equation}
  A(Z_f;Z_i)
  =
  \sum_{\Lambda_f,T_f}
  \sum_{\Lambda_i,T_i}
  Y^{(f)}_{\Lambda_f,T_f}(Z_f)\,
  b_{\Lambda_fT_f;\Lambda_iT_i}(s)
  Y^{(i)}_{\Lambda_i,T_i}(Z_i)^*.
  \label{eq:bp-SSYT-amplitude-expansion}
\end{equation}
Since the auxiliary $\U(N)$ is not a symmetry of the $S$-matrix, the coefficient tensor $b$ need not be diagonal in the tableau labels $(\Lambda_f,T_f)$ and $(\Lambda_i,T_i)$.

Since we keep only harmonics with the same initial- and final-state helicity assignments as the amplitude, the little-group phases cancel in the projection integrand $Y^{(f)*}A Y^{(i)}$ in \Eq{eq:bp-SSYT-amplitude-projection}.
Using \Eq{eq:appPS_LIPS_to_stiefel_measures}, the normalized phase integrations therefore give unity, and the factors $\mathcal N_{N_f}(s)\mathcal N_{N_i}(s)$ convert the Stiefel measures into physical LIPS.
Choosing the same spinor representatives for the harmonics and the amplitude gives
\small
\begin{equation}
b_{\Lambda_fT_f;\Lambda_iT_i}(s)
=
\int d\Phi_{N_f}(P)\,d\Phi_{N_i}(P)
Y^{(f)}_{\Lambda_f,T_f}
(Z_f(\alpha_f))^*
A(\alpha_f;\alpha_i)\,
Y^{(i)}_{\Lambda_i,T_i}
(Z_i(\alpha_i)).
\label{eq:bp-SSYT-amplitude-projection-LIPS}
\end{equation}
\normalsize

\subsection{Joint \texorpdfstring{\glsentryshort{ssyt}}{SSYT} and partial-wave projection} \label{appsec:joint_basis}

The joint basis combines the \gls{ssyt} labels of the Stiefel harmonics with the physical partial-wave labels.
Since the left-$\U(N)$ and right-$\SU(2)_P$ actions in \Eq{eq:bp-2trafos} commute, the spin decomposition can be performed separately for each covariant Stiefel harmonic $Y_{\Lambda,T}$.

The conformal-helicity duality of \Ref{Henning:2019mcv} relates the allowed $\U(N)$ representations to conformal representations of $\SU(2,2)$, with quantum numbers $(\Delta,j_1,j_2)$ and a total helicity $h$.
The two Lorentz spins $j_1$ and $j_2$ combine into total-spin representations of the massive little group $\SU(2)_P$ according to
\begin{equation}
  V_{j_1}\otimes V_{j_2}
  =\bigoplus_{J=\lvert j_1-j_2\rvert}^{j_1+j_2}V_J.
\end{equation}
A fixed $\U(N)$ representation therefore generally contains several total spins $J$, rather than selecting a single partial wave.\footnote{For $N=2$, each $\U(2)$ representation instead carries a single physical spin $J$, as shown in \App{app:two-particle-casimirs}.}

We now decompose each harmonic $Y_{\Lambda,T}(Z)$ into components of definite total spin $J$ and spin projection $M$ under $\SU(2)_P$.
The corresponding channel harmonic is given by
\begin{equation}
  Y_{\Lambda,T;J,M}(Z)
  =
  \braket{Z;\mathbf h|P;\Lambda,T;J,M}.
  \label{eq:bp-joint-harmonics}
\end{equation}
This decomposition makes the spin content explicit while leaving the tableau labels unchanged.
For each $(\Lambda,T)$, the angular-momentum decomposition above contains each allowed $J$ once.
Thus, at fixed $J$, the partial-wave degeneracy label $a$ can be fully specified by $(\Lambda,T)$.
Independent species, gauge, flavor, and permutation labels remain suppressed.

At fixed helicities, the complete orthonormal joint basis provides the resolution of the identity
\begin{align}
\mathbf{1}_{P,\mathbf h}
&=
\sum_{\Lambda T}
\sum_{JM}
\ket{P;\Lambda,T;J,M}
\bra{P;\Lambda,T;J,M}.
\label{eq:bp-joint-resolution-identity}
\end{align}
Together with the integral resolution in \Eq{eq:bp-configuration-resolution}, this gives the orthogonality relation
\begin{equation}
\mathcal N_N(s)\int d\mu_N(Z)\,
Y_{\Lambda,T;J,M}(Z)^*
Y_{\Lambda',T';J',M'}(Z)
=
\delta_{\Lambda\Lambda'}
\delta_{TT'}
\delta_{JJ'}
\delta_{MM'},
\label{eq:bp-joint-orthogonality}
\end{equation}
where we use the orthonormal harmonics introduced above.

Inserting the resolution of the identity in \Eq{eq:bp-configuration-resolution} on both sides of $\mathcal M$ gives the fully resolved projection coefficients,
\begin{equation}
\begin{aligned}
& c^{J_fM_f;J_iM_i}_{\Lambda_fT_f;\Lambda_iT_i}(s)
=
\bra{P;\Lambda_f,T_f;J_f,M_f}
\mathcal M
\ket{P;\Lambda_i,T_i;J_i,M_i}\\
&=
\mathcal N_{N_f}(s)\mathcal N_{N_i}(s)
\int d\mu_{N_f}(Z_f)\,d\mu_{N_i}(Z_i)\,
\braket{P;\Lambda_f,T_f;J_f,M_f|Z_f;\mathbf h_f}\\
&\qquad\times
\bra{Z_f;\mathbf h_f}\mathcal M\ket{Z_i;\mathbf h_i}\,
\braket{Z_i;\mathbf h_i|P;\Lambda_i,T_i;J_i,M_i}\\
&=
\mathcal N_{N_f}(s)\mathcal N_{N_i}(s)
\int d\mu_{N_f}(Z_f)\,d\mu_{N_i}(Z_i)\,
Y^{(f)}_{\Lambda_f,T_f;J_f,M_f}(Z_f)^*
A(Z_f;Z_i)
Y^{(i)}_{\Lambda_i,T_i;J_i,M_i}(Z_i).
\end{aligned}
\label{eq:bp-joint-amplitude-projection}
\end{equation}
These are the components of the coefficient tensor $b$ in \Eq{eq:bp-SSYT-amplitude-coefficient} after expressing its initial and final massive-spinor indices in the $J,M$ bases.
Completeness then gives the reconstruction
\begin{equation}
\begin{aligned}
A(Z_f;Z_i)
={}&
\sum_{\Lambda_f,T_f}
\sum_{\Lambda_i,T_i}
\sum_{J_f,M_f}
\sum_{J_i,M_i}
Y^{(f)}_{\Lambda_f,T_f;J_f,M_f}(Z_f)
c^{J_fM_f;J_iM_i}_{\Lambda_fT_f;\Lambda_iT_i}(s)
Y^{(i)}_{\Lambda_i,T_i;J_i,M_i}(Z_i)^*.
\label{eq:bp-joint-amplitude-expansion-general}
\end{aligned}
\end{equation}

At fixed timelike $P$, Poincaré invariance requires $S$, and hence $\mathcal M$, to commute with the physical $\SU(2)_P$ rotations.
For fixed $(\Lambda_f,T_f)$ and $(\Lambda_i,T_i)$, Schur's lemma therefore gives
\begin{equation}
c^{J_fM_f;J_iM_i}_{\Lambda_fT_f;\Lambda_iT_i}(s)
=
\delta_{J_fJ_i}\,
\delta_{M_fM_i}\,
c^J_{\Lambda_fT_f;\Lambda_iT_i}(s),
\label{eq:bp-joint-amplitude-schur}
\end{equation}
with $J=J_f=J_i$ in the nonzero blocks.
Matrix elements between different total spins vanish, while each equal-spin block is proportional to the identity in its spin indices.
The reduced coefficient $c^J$ is independent of $M$, but need not be diagonal in the tableau labels because the auxiliary $\U(N)$ is not a symmetry of the $S$-matrix.
Setting $J_f=J_i=J$ and $M_f=M_i=M$ in \Eq{eq:bp-joint-amplitude-projection} gives this coefficient for any fixed $M$, without summing over $M$.
The two complete projections integrate out the fixed-$P$ configurations, and Lorentz invariance leaves only $s=P^2$ as a continuous kinematic argument.
Substituting \Eq{eq:bp-joint-amplitude-schur} into \Eq{eq:bp-joint-amplitude-expansion-general} gives
\begin{equation}
A(Z_f;Z_i)
=
\sum_{J,M}
\sum_{\Lambda_f,T_f}
\sum_{\Lambda_i,T_i}
Y^{(f)}_{\Lambda_f,T_f;J,M}(Z_f)
c^J_{\Lambda_fT_f;\Lambda_iT_i}(s)
Y^{(i)}_{\Lambda_i,T_i;J,M}(Z_i)^*.
\label{eq:bp-joint-amplitude-expansion}
\end{equation}
The reduced coefficient $c^J$ carries no open massive-spinor indices. Thus the identity in \Eq{eq:bp-joint-amplitude-schur} directly contracts those of the two harmonics through the remaining sum over $M$.
Defining the channel function
\begin{equation}
\mathcal B^J_{\Lambda_f,T_f;\Lambda_i,T_i}(Z_f,Z_i) \equiv \sum_M Y^{(f)}_{\Lambda_f,T_f;J,M}(Z_f)Y^{(i)*}_{\Lambda_i,T_i;J,M}(Z_i),
\end{equation}
we obtain the expansion used in \Sec{sec:ssyt-decomposition},
\begin{equation} \label{eq:bp-A_in_BJ}
A(Z_f;Z_i)
=
\sum_{J}
\sum_{\Lambda_f,T_f}
\sum_{\Lambda_i,T_i}
c^J_{\Lambda_f,T_f;\Lambda_i,T_i}(s)\,
\mathcal B^J_{\Lambda_f,T_f;\Lambda_i,T_i}(Z_f,Z_i).
\end{equation}

\subsection{Relation to Lorentz-invariant \texorpdfstring{\glsentryshortpl{ssyt}}{SSYTs}}

The same argument applies to an individual Lorentz-invariant \gls{ssyt} of \Ref{Henning:2019enq}, using the notation introduced in \Sec{sec:ssyt-decomposition}.
Choose a partition of its $N=N_i+N_f$ external legs into initial and final channels and denote the crossed polynomial by $\mathcal Y^{(N)}_{\Lambda,T}(Z_f,Z_i)$.
The incoming legs are crossed using \Eq{eq:rge-crossing-convention}, with reversed helicities and conjugate particle quantum numbers as specified in \Sec{sec:on-shell-rge-one-loop}.
These are crossed polynomial values, distinct from physical amplitude kernels as explained in \Sec{sec:ssyt-decomposition}.
Since the crossed polynomial is Lorentz invariant, the argument of \Eq{eq:bp-joint-amplitude-schur} also applies here, giving the reduced projection coefficients
\small
\begin{equation}
y^{J;\Lambda,T}_{\Lambda_fT_f;\Lambda_iT_i}(s)
=
\mathcal N_{N_f}(s)\mathcal N_{N_i}(s)
\int d\mu_{N_f}(Z_f)\,d\mu_{N_i}(Z_i)
Y^{(f)}_{\Lambda_f,T_f;J,M}(Z_f)^*
\mathcal Y^{(N)}_{\Lambda,T}(Z_f,Z_i)
Y^{(i)}_{\Lambda_i,T_i;J,M}(Z_i).
\label{eq:bp-scalar-ssyt-projection}
\end{equation}
\normalsize
Rotational invariance makes these coefficients independent of $M$, so we use the same fixed value in both harmonics without summing over it.
Completeness gives the expansion directly in terms of the channel functions $\mathcal B^J$,
\begin{equation} \label{eq:bp-decomp_LI_SSYT}
\mathcal Y^{(N)}_{\Lambda,T}(Z_f,Z_i) = \sum_{J}
\sum_{\Lambda_f,T_f}
\sum_{\Lambda_i,T_i}
y^{J;\Lambda,T}_{\Lambda_fT_f;\Lambda_iT_i}(s) \mathcal B^J_{\Lambda_f,T_f;\Lambda_i,T_i}(Z_f,Z_i).
\end{equation}
Thus every crossed Lorentz-invariant \gls{ssyt} is a sum of pairwise contractions of covariant initial- and final-channel harmonics.
For a polynomial \gls{ssyt}, this sum is finite since at fixed $P$ it is a polynomial of bounded degree in $Z_i,Z_f$ and their conjugates, which spans only finitely many harmonic representations.

Unlike the channel wavefunctions $Y$, whose normalization is inherited by their contractions $\mathcal B^J$, the Lorentz-invariant spinor polynomials $\mathcal Y^{(N)}_{\Lambda,T}$ are not normalized with respect to physical LIPS.
Conversely, after compensating the powers of $s$ introduced by the spin-frame construction and the physical-LIPS normalization, each channel function $\mathcal B^J$ constructed from polynomial \glspl{ssyt} becomes a Lorentz-invariant polynomial in the external spinors.
Since $\mathcal B^J$ has mass dimension $4-N$ and fixed helicities, the polynomial $s^p\mathcal B^J$ has a single shape $\Lambda=\Lambda_N(n,\widetilde n)$, with $n-\widetilde n=2h$ as in \Eq{eq:helicity-ssyt-degree} and $n+\widetilde n=4p-2N+8$. By \Eq{eq:dimension-helicity-ssyt-counts}, this is the shape of the contact amplitudes of operators of mass dimension $\Delta=2p+4$.
Completeness of the Lorentz-invariant \gls{ssyt} basis then gives the finite expansion
\begin{equation}
    \mathcal B^J_{\Lambda_f,T_f;\Lambda_i,T_i}(Z_f,Z_i)
    =s^{-p}\sum_{T}
    K^{J;\Lambda,T}_{\Lambda_fT_f;\Lambda_iT_i}\,
    \mathcal Y^{(N)}_{\Lambda,T}(Z_f,Z_i).
    \label{eq:bp-finite-channel-LI-expansion}
\end{equation}
Here the $K$ are numerical coefficients for the chosen channel and basis conventions.
The two expansions use different tableau labels. The channel harmonics are constructed from separate initial- and final-state tableaux, whereas the Lorentz-invariant \glspl{ssyt} describe all external legs together.
Expressing one set in terms of the other therefore does not contradict the linear independence of the Lorentz-invariant \glspl{ssyt}.

Substituting the finite reverse expansion into \Eq{eq:bp-A_in_BJ} gives the schematic form used in \Eq{eq:sm-amplitude-in-li-ssyt},
\begin{equation}
A(Z_f;Z_i)
=
\sum_{\Lambda,T}
d_{\Lambda,T}(s)\,
\mathcal Y^{(N)}_{\Lambda,T}(Z_f,Z_i).
\end{equation}
A finite expansion can be chosen for a local polynomial amplitude. Exchange or loop amplitudes generally require an infinite expansion.
The harmonic reconstruction formulas above apply to square-integrable kernels.
Soft or collinear singularities can require a regulator before projection, as discussed in \Sec{sec:on-shell-rge-infrared}.
In particular, the finite local expansion used for anomalous dimensions applies to the UV counterterm after the required cut sum and IR subtraction, not to an arbitrary individual cut.

\subsection{One-sided projection and cut contractions}
\label{appsec:one-sided-projection}

The cut integral in \Eq{eq:physical-cut-kernel} can also be evaluated by projecting only the cut legs of each amplitude, leaving the uncut external configurations fixed.
This follows the one-sided partial-wave projection of \Ref{Shu:2021qlr}, here formulated in the joint \gls{ssyt} and partial-wave basis.
We use the same amplitudes $A_L(Z_f;Z_c)$ and $A_R(Z_c;Z_i)$ as in \Sec{sec:cut-integrals-channel-basis}, with timelike total cut momentum $P$ and $a_c=(\Lambda_c,T_c)$.
Only the cut channel needs to admit the fixed-$P$ harmonic expansion, so the same construction applies to form factors with $q\neq0$, keeping the operator momentum and the uncut momenta fixed.
Divergent cut-channel overlaps still require regularization or subtraction, as discussed in \Sec{sec:on-shell-rge-infrared}.
For each choice of intermediate states, define the one-sided coefficients
\begin{equation}
\begin{aligned}
    f^{JM}_{L;a_c}(Z_f;s)
    & =
    \int d\Phi_{N_c}(P)\,
    A_L(Z_f;Z_c)\,
    Y^{(c)}_{a_c;J,M}(Z_c),\\
    f^{JM}_{R;a_c}(Z_i;s)
    & =
    \int d\Phi_{N_c}(P)\,
    Y^{(c)}_{a_c;J,M}(Z_c)^*\,
    A_R(Z_c;Z_i).
\end{aligned}
\label{eq:bp-one-sided-projection}
\end{equation}
The cut states are incoming for $A_L$ and outgoing for $A_R$, which fixes the different conjugations.
There is no sum over $M$ in either projection.
Only the cut configuration has been integrated over, so the coefficients still depend on the unprojected external states and carry an open spin-$J$ index, represented here by $M$.
Unlike the reduced coefficients $c^J(s)$, they need not be independent of $M$ or depend only on $s$.
Completeness gives the reconstruction formulas
\begin{equation}
\begin{aligned}
    A_L(Z_f;Z_c)
    & =
    \sum_{J,M,a_c}
    f^{JM}_{L;a_c}(Z_f;s)\,
    Y^{(c)}_{a_c;J,M}(Z_c)^*,\\
    A_R(Z_c;Z_i)
    & =
    \sum_{J,M,a_c}
    Y^{(c)}_{a_c;J,M}(Z_c)\,
    f^{JM}_{R;a_c}(Z_i;s).
\end{aligned}
\label{eq:bp-one-sided-reconstruction}
\end{equation}
Inserting these expressions into the cut integral and using \Eq{eq:bp-joint-orthogonality} yields
\begin{equation}
    \mathcal C_{i\to f}(Z_f;Z_i)
    =
    \sum_{\text{cut states}}
    \sum_{J,M,a_c}
    f^{JM}_{L;a_c}(Z_f;s)\,
    f^{JM}_{R;a_c}(Z_i;s).
\label{eq:bp-one-sided-cut-contraction}
\end{equation}
The phase-space integral pairs equal cut-channel labels and contracts the remaining spin indices through the sum over $M$.
No additional phase-space factor appears because the harmonics are normalized with the same physical measure as the cut.
Thus the cut vanishes if the two factors have no common $(J,a_c)$ component for any allowed intermediate state, without requiring a projection of the uncut legs.

For scattering amplitudes admitting the complete two-sided projection above, expansion of the remaining external configurations gives
\begin{equation}
\begin{aligned}
    f^{JM}_{L;a_c}(Z_f;s)
    & =
    \sum_{a_f}
    Y^{(f)}_{a_f;J,M}(Z_f)\,
    c^J_{L;a_fa_c}(s),\\
    f^{JM}_{R;a_c}(Z_i;s)
    & =
    \sum_{a_i}
    c^J_{R;a_ca_i}(s)\,
    Y^{(i)}_{a_i;J,M}(Z_i)^*.
\end{aligned}
\label{eq:bp-one-sided-two-sided-relation}
\end{equation}
Here $a_f=(\Lambda_f,T_f)$ and $a_i=(\Lambda_i,T_i)$, as in \Sec{sec:cut-integrals-channel-basis}.
Substituting these expressions into \Eq{eq:bp-one-sided-cut-contraction} reproduces \Eq{eq:cut-channel-composition}, with the sum over $M$ combining the external harmonics into $\mathcal B^J$.
If the source is polynomial in the cut spinors, its one-sided expansion at fixed spectator kinematics contains only finitely many cut-channel components.

\section{Phase-space measures and integrals}\label{app:phase-space}%

This appendix records the precise relation between the usual Lorentz-invariant phase space (LIPS) measure \(d\Phi_N(P)\) and the Stiefel measure \(d\mu_N(Z)\). We begin by recalling how the standard phase-space measures are expressed in terms of spinor variables. This naturally leads to the relation between the LIPS and Stiefel measures. As a cross-check, we independently verify the standard phase-space normalization formulas using spinor variables. We additionally present an explicit set of coordinates on the Stiefel manifold, allowing an explicit parameterization of the measure \(d\mu_N(Z)\). Throughout, an eye is kept on the various integrals encountered throughout the main text.

Related spinor-helicity parameterizations of massless multiparticle phase space are discussed in \Refs{Larkoski:2020thc,Degrande:2025uil}.

The single-particle phase-space measure for a massless particle of momentum \(p\) (the measure on the forward light cone) is given by
\begin{equation}\label{eq:appPS_single_particle_measure}
  d\Pi_p = \frac{d^4p}{(2\pi)^3}\delta(p^2)\theta(p^0) = \frac{d^3\mathbf p}{(2\pi)^32E_p}, ~~~~E_p = |\mathbf{p}|.
\end{equation}
For positive-energy momenta we introduce a spinor variable \(\lambda\), with complex conjugate \(\widetilde{\lambda} = \lambda^*\), such that \(p_{\alpha\dot\alpha} = \lambda_\alpha \widetilde{\lambda}_{\dot \alpha}\). Since this parameterization trivializes the \(\delta(p^2)\theta(p^0)\), intuitively one expects the measure on the light cone to be proportional to the flat spinor measure,
\[
\frac{d^3\mathbf p}{2E_p} \propto d^4\lambda,
\]
where \(d^4\lambda\) stands for the four real integration variables in the complex \(2\)-spinor \(\lambda\). This is indeed the case. One only needs to pay attention to the fact that the spinors are a redundant parameterization of the momentum: the net spinor phase, which is identified with the little-group phase, does not contribute to the momentum. We must therefore mod out by the volume of this \(\U(1)\) little-group factor,
\[
\frac{d^3\mathbf p}{2E_p} = \frac{d^4\lambda}{\operatorname{vol}\U(1)} = \frac{d^4\lambda}{2\pi}.
\]
This equality is understood for little-group-invariant integrands. Thus, in terms of spinor variables, the single-particle measure is simply expressed as
\begin{equation}\label{eq:appPS_single_particle_measure_spinors}
  d\Pi_p = \frac{d^3\mathbf p}{(2\pi)^32E_p} = \frac{d^4\lambda}{(2\pi)^4}.
\end{equation}

Note that one can explicitly verify this by picking a spinor parameterization. Given a (positive-energy, massless) momentum \(p\), let \(\lambda(p)\) denote the map to spinor variables. This map is, of course, ambiguous up to the net phase of \(\lambda\). We may parameterize a general spinor as \(\lambda = e^{i\alpha}\lambda(p)\), explicitly exhibiting the little-group phase. The expectation is that \(d^4\lambda \propto d\alpha \,d\Pi_p\). Indeed, choosing the parameterization
\[
\def\arraystetch{1.5}
\lambda = e^{i\alpha} \times \sqrt{2E}\begin{pmatrix} \cos(\theta/2) \\ \sin(\theta/2)\,e^{i\phi} \end{pmatrix},
\]
one readily finds
\[
d^4\lambda = \frac{E}{2}\,dE\,d\Omega_2\,d\alpha.
\]
For a massless momentum,
\[
\frac{d^3\mathbf p}{2E} = \frac{E}{2}\,dE\,d\Omega_2.
\]
Therefore
\[
d^4\lambda = \frac{d^3\mathbf p}{2E}\,d\alpha,
\]
exactly as anticipated. When the physics is independent of the little-group phase (which is the case when an integral over \(d\Pi_p\) arises), the \(\alpha\) integral simply contributes a \((2\pi)\) factor, verifying \Eq{eq:appPS_single_particle_measure_spinors}.

Before fixing the total momentum, the multiparticle phase space is the Cartesian product of the single-particle phase spaces, so its measure is the product measure,\footnote{The measures are defined for labeled momenta. Symmetry factors for identical particles are included in the cut-state sums, as specified in \Sec{sec:on-shell-rge-one-loop}.}
\[
d\Pi_N = d\Pi_{p_1}\cdots d\Pi_{p_N} = \frac{d^{4N}\lambda}{(2\pi)^{4N}}.
\]
Finally, the LIPS measure \(d\Phi_N(P)\) is obtained by requiring that the \(N\) particles have momentum \(P\),
\begin{equation}\label{eq:appPS_LIPS_measure_introduce}
  d\Phi_N(P) = d\Pi_N\,(2\pi)^4\delta^{(4)}\big(P - \sum_ip_i\big) = \frac{d^{4N}\lambda}{(2\pi)^{4N-4}}\,\delta^{(4)}(P - \lambda^\dagger\lambda).
\end{equation}
In the last equality we use the matrix notation \(\lambda = \{\lambda_{i\alpha}, ~i =1,\dots, N\}\), so that \(\lambda\) is an \(N\times 2\) matrix and \(\lambda^\dagger \lambda = \sum_i \lambda_{i\alpha}\widetilde{\lambda}_{i\dot\alpha}\) is the momentum carried by the \(N\) particles.\footnote{Note that the delta function is defined with respect to the four real momentum components \(P^\mu\). An additional Jacobian factor is needed if one wants to express the delta function with respect to the matrix components \(P_{\alpha\dot\alpha}\).} Equation~\eqref{eq:appPS_LIPS_measure_introduce} gives the basic way of writing an integral over the \(N\)-body Lorentz-invariant phase space in terms of spinors. This also highlights a major advantage of spinors: the \(N\)-body phase space simply has the flat measure \(d^{4N}\lambda\), with LIPS obtained by imposing the quadratic constraint \(\lambda^\dagger \lambda = P\).

We now relate the phase-space measure to the Stiefel manifold. We will show that \(\int d^{4N}\lambda\) naturally factorizes into an integral over the momentum carried by the particles times an integral over the Stiefel manifold. The momentum integral is trivially carried out in the LIPS measure by using the momentum delta function. Let us call the total momentum carried by the spinors \(Q\),
\[
Q = \lambda^\dagger \lambda.
\]
The delta function in \Eq{eq:appPS_LIPS_measure_introduce} will set \(Q = P\), but we leave it independent for now. For \(N\ge 2\), \(Q\) is generically a positive-definite Hermitian \(2\times 2\) matrix (i.e.\ \(Q^\mu\) is a positive-energy timelike momentum). We may therefore perform the polar decomposition\footnote{This is completely analogous to the usual polar decomposition of a vector, \(x = \hat{n}\,r = \hat{n}\,\sqrt{x^Tx}\), where \(\hat{n}\) is a unit vector, \(\hat{n}^T\hat{n} = 1\).}
\begin{equation}\label{eq:appPS_polar_decomp}
  \lambda = Z\,x, ~~~~x = \sqrt{Q}, ~~~~ Z^\dagger Z =\mathbf{1}_2.
\end{equation}
Because \(Z\) is an \(N\times 2\) complex matrix satisfying \(Z^\dagger Z = \mathbf{1}_2\), it belongs to the Stiefel manifold,
\[
Z \in V_2(\mathbb{C}^N) \simeq \frac{\U(N)}{\U(N-2)}.
\]
To fix \(x\) uniquely in the polar decomposition we choose it to be the positive-definite Hermitian square root of \(Q\).\footnote{In the main text we frequently consider \(x\) to be an arbitrary \(2\times 2\) complex matrix, in which case there is an additional \(\U(2)\) redundancy \(x \to g\,x\), \(Z \to Z\,g^\dagger\), where the \(\SU(2)\subset \U(2)\) is identified with the massive little group of the corresponding momentum. Here we use the \(\U(2)\) freedom to gauge-fix \(x\) to be the positive-definite Hermitian square root \(\sqrt{Q}\).}

The polar decomposition gives us the change of variables \(\lambda \to (Q,Z)\), where \(Q \in \operatorname{Herm}_2^+\), \(Z \in V_2(\mathbb{C}^N)\).
The corresponding measure must have the general form
\[
d^{4N}\lambda = \mathcal{J}_N(Q)\, d^4Q\,d\mu_N(Z),
\]
where \(d\mu_N(Z)\) is the \(\U(N)\)-invariant (Haar) measure on the Stiefel manifold, normalized by
\begin{equation}\label{eq:appPS_stiefel_normalization}
  \int_{V_2(\mathbb{C}^N)} d\mu_N(Z) = 1,
\end{equation}
while \(\mathcal{J}_N(Q)\) is a Jacobian factor. This Jacobian is fixed, up to a normalization constant, by Lorentz invariance and dimensional analysis. Namely, since \(d^{4N}\lambda\) is Lorentz invariant, the Jacobian can only depend on \(\det Q\). Dimensional analysis then implies
\begin{equation}\label{eq:appPS_dlambda_to_dQ_dZ_constant}
  d^{4N}\lambda = C_N\,(\det Q)^{N-2}\,d^4Q\,d\mu_N(Z),
\end{equation}
for some to-be-determined constant \(C_N\).

To determine \(C_N\) we consider the Gaussian integral,
\[
I_N = \int_{\mathbb{C}^{N\times 2}}d^{4N}\lambda\,e^{-\operatorname{tr}(\lambda^\dagger\lambda)},
\]
and evaluate it two different ways. In the first way we simply perform the Gaussian integrals, where the \(2N\) independent complex Gaussian integrations each give \(\pi\), and hence
\[
I_N = \pi^{2N}.
\]
Next we evaluate \(I_N\) using the polar decomposition,
\[
\begin{gathered}
  I_N = C_N \int_{Q>0}d^4Q\,(\det Q)^{N-2} e^{-\operatorname{tr}Q} \int d\mu_N(Z) \vspace{.15cm} \\
  \Rightarrow ~~\pi^{2N} = C_N \int_{Q>0}d^4Q\,(\det Q)^{N-2} e^{-\operatorname{tr}Q} .
\end{gathered}
\]
To perform the integral over \(Q\) one explicitly parameterizes a positive-definite Hermitian \(2\times 2\) matrix and carries out the integration to find\footnote{Parameterize
\[
\begin{gathered}
  Q = Q^\mu \sigma_\mu = \begin{pmatrix} a & z \\ z^* & b \end{pmatrix}, ~~~~a>0,~~b>0,~~ab> |z|^2 \vspace{.15cm} \\
  d^4Q = \frac{1}{2} da\,db\,d\operatorname{Re}z\,d\operatorname{Im}z,
\end{gathered}
\]
where the factor of \(1/2\) arises in the Jacobian from \(a = Q^0 + Q^3\), \(b = Q^0 - Q^3\). Introduce \(t = b - |z|^2/a\), so that the positivity conditions are \(a>0\), \(t>0\). Upon changing variables from \(b\) to \(t\), the integral becomes expressed in terms of elementary integrals and the result follows.
}
\[
\int_{Q>0}d^4Q\,(\det Q)^{N-2} e^{-\operatorname{tr}Q} = \frac{\pi}{2}\,\Gamma(N)\,\Gamma(N-1).
\]
Hence
\begin{equation}\label{eq:appPS_C_N}
  C_N = \frac{2\pi^{2N-1}}{\Gamma(N)\,\Gamma(N-1)},
\end{equation}
and therefore
\begin{equation}\label{eq:appPS_dlambda_to_dQ_dZ_answer}
  \boxed{~
    d^{4N}\lambda = \frac{2\pi^{2N-1}}{\Gamma(N)\,\Gamma(N-1)}\,(\det Q)^{N-2}\,d^4Q\,d\mu_N(Z), ~~~Q = \lambda^\dagger\lambda.
    ~}
\end{equation}

Returning to the LIPS measure, we substitute \Eq{eq:appPS_dlambda_to_dQ_dZ_answer} into the spinor representation of LIPS in \Eq{eq:appPS_LIPS_measure_introduce} to obtain
\[
\int d\Phi_N(P) = \frac{C_N}{(2\pi)^{4N-4}}\int_{Q>0}d^4Q\,(\det Q)^{N-2}\,\delta^4(P-Q) \int_{V_2(\mathbb{C}^N)}d\mu_N(Z).
\]
The delta function trivializes the \(Q\) integral, setting \(Q = P\). Since \(\det P = P^2 = s\) we obtain
\begin{equation}\label{eq:appPS_LIPS_to_stiefel_measures}
  \boxed{~
    \int d\Phi_N(P) = \frac{C_N}{(2\pi)^{4N-4}}\,s^{N-2}\,\int d\mu_N(Z).
    ~}
\end{equation}
In this way we see that an integration over LIPS is precisely an integral over the Stiefel manifold when expressed in terms of spinor variables.

We note that our calculation also informs us of the overall phase-space normalization \(\mathcal{N}_N(s) \equiv \int d\Phi_N(P)\). Using the normalization of the Stiefel measure and substituting \(C_N\),
\begin{equation}\label{eq:appPS_LIPS_normalization}
  \begin{gathered}
    \int d\Phi_N(P) = \mathcal{N}_N(s) \int d\mu_N(Z), \vspace{.15cm} \\
    \mathcal{N}_N(s) = \int d\Phi_N(P) = \frac{C_N}{(2\pi)^{4N-4}}\,s^{N-2} = \frac{s^{N-2}}{2(4\pi)^{2N-3}\,\Gamma(N)\,\Gamma(N-1)}.
  \end{gathered}
\end{equation}

A word of caution, which we hope is already clear: The equality of integrals in \Eq{eq:appPS_LIPS_to_stiefel_measures} holds when the integrand is little-group independent. In other words, whenever a phase-space integral \(\int d\Phi_N(P)\) arises, \Eq{eq:appPS_LIPS_to_stiefel_measures} tells us how to rewrite the integral in terms of spinors. The converse is not true: an integral that starts over \(d^{4N}\lambda\) or over \(d\mu_N(Z)\) is not necessarily equivalent to an integral over LIPS, unless the integrand is little-group independent.

In this regard, one may wonder if there is a way to factor out the \(\U(1)^N\) little-group phases in the Stiefel measure \(d\mu_N(Z)\). This is indeed possible, and in \Sec{sub:appPS_little_group_phases} we show how it happens for our chosen parameterization of the Stiefel manifold. The upshot is that, in our chosen coordinates, the measure factorizes
\[
d\mu_N(Z) = \left(\prod_{i=1}^N\frac{d\phi_i}{2\pi}\right) \times d\nu_N(\theta,\psi,\chi),
\]
where \(\phi_i\) are the little-group angles and \((\theta,\psi,\chi)\) are \(3N-4\) other angles that explicitly parameterize LIPS. That is, we have the full equality of measures
\[
d\Phi_N(P) = \mathcal{N}_N(s)\,d\nu_N(\theta,\psi,\chi).
\]

\subsection{Explicit coordinates on the Stiefel manifold}\label{appsub:PS_explicit_stiefel_coords}

This section constructs explicit coordinates on the Stiefel manifold, i.e.\ we solve the constraint \(Z^\dagger Z = \mathbf{1}_2\). The corresponding measure is constructed in \Sec{sub:appPS_invariant_measure}.

It will be easiest to think about this as constructing a general two-frame in \(\mathbb{C}^N\), i.e.\ a pair of mutually orthogonal unit vectors. In this language we view the \(N\times 2\) matrix \(Z\) as comprised of two \(N\)-vectors,
\begin{equation}\label{eq:appPS_Z_uv}
  Z = \begin{pmatrix} u & v \end{pmatrix}.
\end{equation}
The constraint supplies the geometry,
\[
\def\arraystretch{1.2}
Z^\dagger Z = \mathbf{1}_2 ~~~\Rightarrow~~~ \begin{pmatrix} |u|^2 & u^\dagger v \\ v^\dagger u & |v|^2 \end{pmatrix} = \begin{pmatrix} 1 & \\ & 1 \end{pmatrix},
\]
which tells us that \(u\) and \(v\) are mutually orthogonal (complex) unit vectors. The constraint \(|u|^2 = 1\) says that \(u \in S^{2N-1}\); \(v\) also spans a sphere, but the constraint \(u^\dagger v = v^\dagger u = 0\) tells us that this sphere lies in the tangent space to \(S^{2N-1}\). Thus, the Stiefel manifold has the {\it local} geometry\footnote{The precise statement is that the Stiefel manifold is a sphere bundle over the sphere, where we have the fibration
\[
S^{2N-3} \longrightarrow V_2(\mathbb{C}^N) \overset{\pi}{\longrightarrow} S^{2N-1},
\]
with projection \(\pi(u,v) = u\). The fiber above a particular \(u\) consists of unit vectors in the complex orthogonal complement \(u_{\mathbb{C}}^\perp\), which is indeed \(S^{2N-3}\).
}
\[
V_2(\mathbb{C}^N) \sim S^{2N-1}\times S^{2N-3}.
\]
Note that this is consistent with the homogeneous space picture of the Stiefel manifold,
\[
V_2(\mathbb{C}^N) \simeq \frac{\U(N)}{\U(N-2)} \sim \frac{\U(N)}{\U(N-1)} \times \frac{\U(N-1)}{\U(N-2)} \sim S^{2N-1}\times S^{2N-3}.
\]
In this way we can anticipate that the measure will factorize into the product measure of the corresponding spheres.

\subsubsection{Coordinates on the real Stiefel manifold}

As a warm-up, we first construct coordinates on the {\it real} Stiefel manifold. Our algorithm is most transparent here. Moreover, the real results are recycled for the complex case. The real Stiefel manifold of ordered orthonormal two-frames in \(\mathbb{R}^N\) is\footnote{This is the relevant manifold for spinors in \(2+1\) dimensions~\cite{Henning:2019mcv}.}
\[
V_2(\mathbb{R}^N) \simeq \frac{\OO(N)}{\OO(N-2)} \sim S^{N-1} \times S^{N-2},
\]
described by a {\it real} \(N \times 2\) matrix \(Z\) obeying \(Z^TZ = \mathbf{1}_2\). To construct the two-frame, as in \Eq{eq:appPS_Z_uv} we parameterize \(Z = (u,v)\), where the constraint \(Z^T Z = \mathbf{1}\) implies \(u^2 = v^2 = 1\), \(u\cdot v  = 0\).

Let us give a brief, but technical, description of our algorithm to construct a general two-frame. It is elaborated below. Initialize \(u \in S^{N-1}\), e.g.\ by taking standard hyperspherical coordinates. Since \(u\cdot v = 0\), we have \(v \in T_{u}S^{N-1}\). We can construct a frame \(e_a\), \(a=1,\dots, N-1\), for the tangent space \(T_uS^{N-1}\) by orthonormalizing the tangent vectors \(\widetilde{e}_a = \partial_{\theta_a}u \in T_uS^{N-1}\), which are automatically orthogonal to \(u\) since \(u\) is a unit vector, \(u\cdot \widetilde{e}_a = u\cdot \partial_{\theta_a}u = 0\). A general vector \(w\) in the tangent space may then be decomposed according to \(w \in T_uS^{N-1} ~\leftrightarrow~ w = \undertilde{w}^ae_a\) where \(\undertilde{w} = (w_1,\dots ,w_{N-1}) \in \mathbb{R}^{N-1}\). Therefore we initialize \(\undertilde{v} \in S^{N-2}\), i.e.\ \(\undertilde{v}\) is a \((N-1)\)-dimensional unit vector, and take \(v = \undertilde{v}^a e_a\). This gives the two-frame \((u,v)\).

In more detail, we begin by taking \(u \in S^{N-1}\), choosing the usual hyperspherical coordinates
\begin{equation}\label{eq:appPS_real_hypersphere_coords}
  \def\arraystretch{1.1}
  u = \begin{pmatrix} c_1 \\ s_1\,c_2 \\ s_1\,s_2\,c_3 \\ \vdots \\ s_1\cdots s_{N-2}\,c_{N-1} \\ s_1\cdots s_{N-2}\,s_{N-1} \end{pmatrix}, ~~~~~ \begin{aligned} c_a &\equiv \cos\theta_a, \\[0.8ex] s_a &\equiv \sin\theta_a . \end{aligned}
\end{equation}
Because \(u\) is a unit vector, it is orthogonal to its derivatives, \(u^T\partial_{\theta_a} u = 0\). We use this to construct an orthogonal frame \(u^\perp\). Specifically, the \(N-1\) vectors \(\widetilde{e}_a = \partial_{\theta_a}u\) are automatically orthogonal to \(u\). We can orthonormalize the set using the Gram--Schmidt procedure,\footnote{For the standard parameterization in \Eq{eq:appPS_real_hypersphere_coords} the \(\widetilde{e}_a\) come out automatically orthogonal, and only need to be normalized. We discuss and prove this below. \label{foot:appPS_orthogonality_e_a}} creating a set \(\{e_a, ~a=1,\dots,N-1\}\) obeying \(u\cdot e_a = 0\) and \(e_a\cdot e_b = \delta_{ab}\). These \(e_a = e_a(\theta)\) depend only on the angles in \(u\). To construct a general \(v\), we begin by initializing a general \(\undertilde{v} \in S^{N-2}\) (\(\undertilde{v}\) is a \((N-1)\)-dimensional unit vector),
\begin{equation}\label{eq:appPS_utv_coords}
  \def\arraystretch{1.1}
  \undertilde{v} = \begin{pmatrix} c_1' \\ s_1'\,c_2' \\ s_1'\,s_2'\,c_3' \\ \vdots \\ s_1'\cdots s_{N-2}' \end{pmatrix}, ~~~~~ \begin{aligned} c_b' &\equiv \cos\psi_b, \\[0.8ex] s_b' &\equiv \sin\psi_b, \end{aligned}
\end{equation}
which depends on the \(N-2\) angles \(\psi_b\). We then construct \(v\) according to
\begin{equation}\label{eq:appPS_v_construct}
  v = \undertilde{v}^a e_a.
\end{equation}
Because \(u\cdot e_a = 0\) we immediately have \(u\cdot v = 0\). Moreover, \(v^2 = 1\),
\[
v^2 = \undertilde{v}^a\undertilde{v}^b (e_a\cdot e_b) = \undertilde{v}^a \undertilde{v}^a = 1.
\]

In footnote~\ref{foot:appPS_orthogonality_e_a} we noted that the \(\widetilde{e}_a = \partial_{\theta_a}u\) come out automatically orthogonal when using the standard hyperspherical coordinates in \Eq{eq:appPS_real_hypersphere_coords}. This follows from the recursive construction of these coordinates. To see this, begin with the real unit vector
\[
r_N(\theta_1,\dots,\theta_{N-1}) \in S^{N-1}.
\]
We can write the hyperspherical parameterization recursively as
\[
r_N = \begin{pmatrix} \cos\theta_1 \\ \sin\theta_1\,r_{N-1}(\theta_2,\dots,\theta_{N-1}) \end{pmatrix},
\]
where \(r^T_{N-1}r_{N-1} = 1\). It is easy to see that \(\partial_{\theta_1}r_N\) is orthogonal to \(\partial_{\theta_a}r_N\) for any \(a>1\),
\[
(\partial_{\theta_1}r_N)^T (\partial_{\theta_a}r_N) = \cos\theta_1\sin\theta_1\,r_{N-1}^T\partial_{\theta_a}r_{N-1} = 0,
\]
since \(r_{N-1}\) has unit norm. Therefore the first angular derivative is orthogonal to every remaining angular derivative. What about two derivatives with \(a,b>1\)? We have
\[
(\partial_{\theta_a}r_N)^T(\partial_{\theta_b}r_N) = \sin^2\theta_1(\partial_{\theta_a}r_{N-1})^T(\partial_{\theta_b}r_{N-1}).
\]
Thus, if the derivatives are mutually orthogonal for \(r_{N-1}\), they are orthogonal for \(r_N\). Since the claim is trivially true for \(N=2\), induction proves it for every \(N\).

In fact, induction also gives the norms:
\[
\begin{gathered}
  (\partial_{\theta_a}r_N)^T(\partial_{\theta_b}r_N) = \delta_{ab} \prod_{j=1}^{a-1}\sin^2\theta_j, \vspace{.15cm} \\
  e.g.~~|\partial_{\theta_1}r_N|^2 = 1, ~~~|\partial_{\theta_2}r_N|^2 = \sin^2\theta_1, ~~~|\partial_{\theta_3}r_N|^2 = \sin^2\theta_1\sin^2\theta_2, ~~\dots.
\end{gathered}
\]
This allows us to write a closed formula for the orthonormal basis,
\begin{equation}\label{eq:appPS_e_a_closed_formula}
  e_a(u) = \frac{1}{\prod_{j=1}^{a-1}\sin\theta_j}\,\frac{\partial u}{\partial \theta_a}, ~~~~a = 1,\dots, N-1.
\end{equation}
These coordinates are valid away from the usual hyperspherical-coordinate singularities. The resulting formulas for the invariant measure hold globally in the sense of integration, since the excluded coordinate boundaries have measure zero.

\subsubsection{Coordinates on the complex Stiefel manifold}

The complex case follows the same logic, this time applied to the complex unit vector \(u \in S^{2N-1} \subset \mathbb{C}^N\), where \(u\) is an \(N\)-dimensional complex unit vector satisfying \(u^\dagger u = 1\). The condition \(u^\dagger v = 0\) tells us \(v\) lies in the complex orthogonal complement to \(u\), \(v\in u_{\mathbb{C}}^\perp\). A general parameterization of \(u\) is given by standard complex hyperspherical coordinates,
\begin{equation}\label{eq:appPS_complex_hypersphere_coords}
  \begin{gathered}
    u = \begin{pmatrix} e^{i\phi_1}c_1 \\ e^{i\phi_2} s_1\,c_2 \\ e^{i\phi_3} s_1\,s_2\,c_3 \\ \vdots \\ e^{i\phi_{N-1}}s_1\cdots s_{N-2}\,c_{N-1} \\ e^{i\phi_N} s_1\cdots s_{N-2}\,s_{N-1} \end{pmatrix},
    ~~~~~\begin{aligned} c_a &\equiv \cos\theta_a, \\ s_a &\equiv \sin\theta_a, \end{aligned}
    ~~~~~\begin{aligned} \theta_a &\in [0,\pi/2], \qquad a=1,\ldots,N-1, \\ \phi_i &\in [0,2\pi), \qquad i=1,\ldots,N. \end{aligned}
  \end{gathered}
\end{equation}
Note that this takes the form
\begin{equation}\label{eq:appPS_u_complex_decompose}
  u = D_\phi \,r_N(\theta), ~~~~D_\phi = \operatorname{diag}(e^{i\phi_1},\dots,e^{i\phi_N}),
\end{equation}
with \(r_N(\theta)\) a real unit vector defined in \Eq{eq:appPS_real_hypersphere_coords}. It follows immediately that the \(\theta_a\) derivatives of \(u\) lie in the complex orthogonal complement \(u_{\mathbb{C}}^\perp\), allowing us to construct a basis for this space. In fact, a basis for \(u_{\mathbb{C}}^\perp\) is given by the \(e_a\) of \Eq{eq:appPS_e_a_closed_formula} (using \Eq{eq:appPS_complex_hypersphere_coords} for \(u\)). Therefore, to construct \(v\) we initialize a complex \(\undertilde{v} \in S^{2N-3} \subset \mathbb{C}^{N-1}\),
\begin{equation}\label{eq:appPS_utv_coords_complex}
  \begin{gathered}
    \undertilde{v} = \begin{pmatrix} e^{i\chi_1}c_1' \\ e^{i\chi_2}s_1'\,c_2' \\ e^{i\chi_3}s_1'\,s_2'\,c_3' \\ \vdots \\ e^{i\chi_{N-1}} s_1'\cdots s_{N-2}' \end{pmatrix},
    ~~~~~\begin{aligned} c'_b &\equiv \cos\psi_b, \\ s'_b &\equiv \sin\psi_b, \end{aligned}
    ~~~~~\begin{aligned} \psi_b &\in [0,\pi/2], \qquad b=1,\ldots,N-2, \\ \chi_a &\in [0,2\pi), \qquad a=1,\ldots,N-1. \end{aligned}
    \vspace{.15cm} \\
    \undertilde{v} = \undertilde{D}_\chi r_{N-1}(\psi), ~~~~\undertilde{D}_\chi = \operatorname{diag}(e^{i\chi_1},\dots,e^{i\chi_{N-1}}),
  \end{gathered}
\end{equation}
and form \(v\) by
\begin{equation}\label{eq:appPS_v_construct_complex}
  v = \undertilde{v}^ae_a, ~~~~~ e_a(u) = \frac{1}{\prod_{j=1}^{a-1}\sin\theta_j} \frac{\partial u}{\partial \theta_a}.
\end{equation}

As a concrete example, consider the \(N=2\) case where we parameterize \(u\) and \(v\) as
\begin{equation}\label{eq:appPS_complex_stiefel_N=2}
  u = \begin{pmatrix} e^{i\phi_1}\,c_\theta \\ e^{i\phi_2}\,s_\theta \end{pmatrix}, ~~~~ v = e^{i\chi} \begin{pmatrix} -e^{i\phi_1}\,s_\theta \\ e^{i\phi_2}\,c_\theta \end{pmatrix}.
\end{equation}
Note that \(v = \undertilde{v}e_1\) where \(\undertilde{v} = e^{i\chi}\) and \(e_1 = \widetilde{e}_1 = \partial_\theta u\). As a second example, for \(N=3\) we find
\begin{equation}\label{eq:appPS_complex_stiefel_N=3}
  \begin{gathered}
    u = \begin{pmatrix} e^{i\phi_1}\,c_1 \\ e^{i\phi_2}\,s_1\,c_2 \\ e^{i\phi_3}\,s_1\,s_2 \end{pmatrix}, ~~~~e_1 = \partial_{\theta_1}u = \begin{pmatrix} -e^{i\phi_1}\,s_1 \\ e^{i\phi_2}\,c_1\,c_2 \\ e^{i\phi_3}\,c_1\,s_2 \end{pmatrix}, ~~~~e_2 = \frac{1}{s_1}\partial_{\theta_2}u = \begin{pmatrix} 0 \\ -e^{i\phi_2}\,s_2 \\ e^{i\phi_3}\,c_2 \end{pmatrix}
    \vspace{.15cm} \\
    \undertilde{v} = \begin{pmatrix} e^{i\chi_1}\cos\psi \\ e^{i\chi_2}\sin\psi \end{pmatrix} ~~~\Rightarrow~~~ v = \undertilde{v}^ae_a = e^{i\chi_1}\cos\psi\,e_1 + e^{i\chi_2}\sin\psi\,e_2.
  \end{gathered}
\end{equation}

\subsection{The invariant Stiefel measure}\label{sub:appPS_invariant_measure}

Next we turn to computing the Stiefel measure \(d\mu_N(Z)\) using the parameterization \(Z = (u,v)\) worked out above. We saw above that the Stiefel manifold is locally the product of two spheres, \(V_2(\mathbb{C}^N) \sim S^{2N-1}\times S^{2N-3}\), where the first sphere is controlled by the \(N\)-dimensional complex unit vector \(u\), while the second sphere is controlled by the \((N-1)\)-dimensional complex unit vector \(\undertilde{v}\). It should not be surprising, therefore, that the measure factorizes
\begin{equation}\label{eq:appPS_stiefel_measure_spheres}
  \begin{aligned}
    d\mu_N(u,v) &= \frac{d\Omega_{2N-1}(u)}{\operatorname{vol}(S^{2N-1})}\,\frac{d\Omega_{2N-3}(\undertilde{v})}{\operatorname{vol}(S^{2N-3})} \\
    &= \frac{\Gamma(N)\Gamma(N-1)}{4\pi^{2N-1}}\,d\Omega_{2N-1}(u)\,d\Omega_{2N-3}(\undertilde{v}),
  \end{aligned}
\end{equation}
where \(v = \undertilde{v}^ae_a\), \(d\Omega_k\) is the standard spherical measure on \(S^k\), and we used the volume of the sphere,
\[
\operatorname{vol}(S^{2m-1}) = \frac{2\pi^m}{\Gamma(m)}.
\]

To see that \Eq{eq:appPS_stiefel_measure_spheres} gives the invariant Stiefel measure, recall that \(V_2(\mathbb C^N)=\U(N)/\U(N-2)\) is a homogeneous space, and therefore admits a unique normalized \(\U(N)\)-invariant measure. We construct such a measure by first integrating the unit vector \(u\) over \(S^{2N-1}\) with its normalized spherical measure, and then, for each fixed \(u\), integrating the second unit vector \(v\) over the sphere \(S^{2N-3}\subset u^\perp_{\mathbb C}\), again with its normalized spherical measure. This construction is \(\U(N)\)-invariant: any \(U\in \U(N)\) preserves the spherical measure on \(S^{2N-1}\), and maps the unit sphere in \(u^\perp_{\mathbb C}\) isometrically onto the unit sphere in \((Uu)^\perp_{\mathbb C}\), preserving its spherical measure as well. Thus, the resulting measure is normalized and \(\U(N)\)-invariant, and by uniqueness must be the invariant Stiefel measure.

Consider a unit vector \(z\in S^{2m-1} \subset \mathbb{C}^m\), parameterized using the standard coordinates in \Eq{eq:appPS_complex_hypersphere_coords}. Then the ordinary spherical element is
\begin{equation}\label{eq:appPS_spherical_volume_element}
  d\Omega_{2m-1} = \left[\prod_{a=1}^{m-1}(\sin\theta_a)^{2(m-a)-1}\cos\theta_a d\theta_a\right]\prod_{i=1}^m d\phi_i.
\end{equation}
From
\[
\int_0^{\pi/2}\sin^{2(m-a)-1}\theta_a\,\cos\theta_a\,d\theta_a = \frac{1}{2(m-a)},
\]
the full normalized Stiefel measure in \Eq{eq:appPS_stiefel_measure_spheres} is given by
\begin{equation}\label{eq:appPS_stiefel_measure_full}
  \boxed{
  \begin{aligned}
    d\mu_N =& \prod_{a=1}^{N-1}\Big[2(N-a)(\sin\theta_a)^{2(N-a)-1}\cos\theta_a\,d\theta_a\Big]
    \times \prod_{i=1}^N \frac{d\phi_i}{2\pi} \\
    &~\times \prod_{b=1}^{N-2}\Big[2(N-1-b)(\sin\psi_b)^{2(N-1-b)-1}\cos\psi_b\,d\psi_b\Big]
    \times \prod_{a=1}^{N-1}\frac{d\chi_a}{2\pi}
  \end{aligned}
  }
\end{equation}

\subsection{Separating the little-group phases}\label{sub:appPS_little_group_phases}

There's a nice feature about our parameterization of the Stiefel manifold: the \(N\) independent little-group phases are exactly the angles \(\phi_i\) appearing in the first vector \(u\) in \Eq{eq:appPS_complex_hypersphere_coords}. This is not immediately obvious, but it follows from how we constructed the orthonormal frame.

Recall that \(u = D_\phi r_N(\theta)\), with \(D_\phi = \operatorname{diag}(e^{i\phi_1},\dots,e^{i\phi_N})\) and \(r_N(\theta)\) a real unit vector. We showed that the orthonormal vectors \(e_a\) are given, up to normalization, by the \(\theta_a\) derivatives of \(u\), \Eq{eq:appPS_v_construct_complex}. Since \(D_\phi\) doesn't depend on polar angles,
\[
e_a = D_\phi\,e_a^{(0)}(\theta),
\]
where the vectors \(e_a^{(0)}(\theta)\) are real. Define
\[
E_0(\theta) = \begin{pmatrix} e_1^{(0)} & \cdots & e_{N-1}^{(0)} \end{pmatrix},
\]
in terms of which the second Stiefel vector is
\[
v = D_\phi E_0(\theta)\,\undertilde{v}(\psi,\chi).
\]
Therefore the complete \(N\times 2\) Stiefel matrix is
\begin{equation}\label{eq:appPS_Z_explicit_matrix_form}
  Z = \begin{pmatrix} u & v \end{pmatrix} = D_\phi\begin{pmatrix} r_N(\theta) & E_0(\theta)\,\undertilde{v}(\psi,\chi) \end{pmatrix}.
\end{equation}

Now consider an individual little-group transformation,
\[
Z \to D_\alpha Z, ~~~~D_\alpha = \operatorname{diag}(e^{i\alpha_1},\dots,e^{i\alpha_N}).
\]
Using \Eq{eq:appPS_Z_explicit_matrix_form},
\[
D_\alpha Z = D_{\phi+\alpha}\begin{pmatrix} r_N(\theta) & E_0(\theta)\,\undertilde{v}(\psi,\chi) \end{pmatrix}.
\]
This shows that the \(\phi_i\) shift under the little-group transformation, while the other angles are unchanged,
\[
\begin{gathered}
  \phi_i \to \phi_i + \alpha_i \vspace{.15cm} \\
  \theta_a,~\psi_b,~\chi_a~\text{unchanged}.
\end{gathered}
\]
In other words, our coordinates separate the little-group fibers from the physical phase-space coordinates!

\subsubsection{Physical phase-space coordinates and the LIPS measure}

The complete Stiefel manifold has \(4N-4\) real dimensions. Of these, the \(N\) angles \(\phi_i\) correspond to the little-group phases, as we just saw. The remaining physical coordinates are
\[
\underbrace{\theta_1,\dots,\theta_{N-1}}_{N-1},~~~ \underbrace{\psi_1,\dots,\psi_{N-2}}_{N-2},~~~ \underbrace{\chi_1,\dots,\chi_{N-1}}_{N-1},
\]
giving \(3N-4\) coordinates, exactly the dimension of physical \(N\)-particle LIPS. We can therefore write the normalized Stiefel measure in the form
\begin{equation}\label{eq:appPS_stiefel_measure_extract_LG_phases}
  d\mu_N(Z) = \left(\prod_{i=1}^N\frac{d\phi_i}{2\pi}\right)\,d\nu_N(\theta,\psi,\chi),
\end{equation}
where \(d\nu_N\) is the normalized measure on physical phase space, whose explicit expression is obtained from \Eq{eq:appPS_stiefel_measure_full} by removing the \(\phi_i\) integrations.

Using our previous LIPS normalization, i.e.\ \(\mathcal{N}_N(s)\) from \Eq{eq:appPS_LIPS_normalization}, we obtain
\begin{equation}\label{eq:appPS_LIPS_measure_dnu_N}
  \boxed{~d\Phi_N(P) = \mathcal{N}_N(s)\,d\nu_N(\theta,\psi,\chi).~}
\end{equation}
This is a particularly concrete realization of the LIPS-Stiefel relationship discussed earlier.

Let us check the \(N=2\) case, where \(u,v\) are given in \Eq{eq:appPS_complex_stiefel_N=2}. The normalized Stiefel measure is
\[
d\mu_2 = 2\sin\theta\cos\theta\,d\theta\,\frac{d\phi_1}{2\pi}\,\frac{d\phi_2}{2\pi}\,\frac{d\chi}{2\pi}.
\]
Removing the little-group phases,
\[
d\nu_2 = 2\sin\theta\cos\theta\,d\theta\,\frac{d\chi}{2\pi}.
\]
Defining \(\Theta = 2\theta\) this gives
\[
d\nu_2 = \frac{\sin\Theta\,d\Theta\,d\chi}{4\pi} = \frac{d\Omega_2}{4\pi},
\]
exactly the normalized angular measure on the ordinary two-particle phase space \(S^2\). Recalling the normalization \(\mathcal{N}_2(s) = 1/(8\pi)\) we find
\[
d\Phi_2(P) = \mathcal{N}_2(s)\,d\nu_2 = \frac{d\Omega_2}{32\pi^2},
\]
precisely the conventionally normalized 2-particle LIPS measure.
Up to little-group phase choices and angle relabelings, \Eqs{eq:appPS_complex_stiefel_N=2}{eq:appPS_complex_stiefel_N=3} agree with the two- and three-particle spinor parameterizations of \Ref{Caron-Huot:2016cwu}.
The two-particle parameterization is also used in \Refs{Baratella:2020lzz,Bern:2020ikv}, with different phase-space normalization conventions.
We use the physical LIPS normalization in \Eq{eq:appPS_LIPS_normalization}, keeping identical-particle symmetry factors separate.

\subsection{Phase space in simplex coordinates}\label{sub:appPS_simplified_measure}

Above we expressed the complex unit vectors \(u\) and \(\undertilde{v}\) using hyperspherical coordinates. This is natural for constructing the frame. However, for writing the integration measure, it is sometimes more convenient to replace the polar angles by the squared component magnitudes. In this spirit, define
\[
t_i = |u_i|^2.
\]
Since \(u^\dagger u = 1\),
\[
t_i \ge 0, ~~~\sum_{i=1}^N t_i = 1.
\]
Thus the \(t_i\) belong to the standard \((N-1)\)-simplex \(\Delta_{N-1}\). We can write
\begin{equation}
  u_i = \sqrt{t_i}\, e^{i\phi_i}.
\end{equation}
Similarly, for \(\undertilde{v} \in S^{2N-3} \subset \mathbb{C}^{N-1}\), define
\[
\rho_a = |\undertilde{v}_a|^2,
\]
with
\[
\rho_a \ge 0, ~~~\sum_{a=1}^{N-1}\rho_a = 1,
\]
so that the \(\rho_a\) belong to the simplex \(\Delta_{N-2}\). Then
\begin{equation}
  \undertilde{v}_a = \sqrt{\rho_a}\, e^{i\chi_a}.
\end{equation}

This causes a remarkable simplification to the measure: for a uniformly distributed complex unit vector \(u \in S^{2N-1}\), the squared magnitudes \(t_i\) are uniformly distributed on the simplex. Explicitly
\begin{equation}
  \frac{d\Omega_{2N-1}(u)}{\operatorname{vol}(S^{2N-1})} = (N-1)!\left(\prod_{i=1}^{N-1}dt_i\right)\left(\prod_{i=1}^N\frac{d\phi_i}{2\pi}\right),
\end{equation}
where \(t_N = 1 - \sum_{i=1}^{N-1}t_i\) and the integration domain is
\[
t_i \ge 0, ~~~\sum_{i=1}^{N-1}t_i \le 1.
\]
The factor \((N-1)!\) normalizes the simplex measure, since
\[
\operatorname{vol}(\Delta_{N-1}) = \frac{1}{(N-1)!}.
\]

Exactly the same argument gives
\begin{equation}
  \frac{d\Omega_{2N-3}(\undertilde{v})}{\operatorname{vol}(S^{2N-3})} = (N-2)!\left(\prod_{a=1}^{N-2}d\rho_a\right)\left(\prod_{a=1}^{N-1}\frac{d\chi_a}{2\pi}\right).
\end{equation}
Therefore the full normalized Stiefel measure is simply
\begin{equation}
  d\mu_N(Z) = (N-1)!(N-2)!\left(\prod_{i=1}^{N-1}dt_i\right)\left(\prod_{a=1}^{N-2}d\rho_a\right)\left(\prod_{i=1}^N\frac{d\phi_i}{2\pi}\right)\left(\prod_{a=1}^{N-1}\frac{d\chi_a}{2\pi}\right).
\end{equation}
The resulting measure contains no trigonometric factors. In particular, after quotienting by the little group
\begin{equation}
  \boxed{~ d\Phi_N(P) %
    = \mathcal{N}_N(s)\,(N-1)!(N-2)!\left(\prod_{i=1}^{N-1}dt_i\right)\left(\prod_{a=1}^{N-2}d\rho_a\right)\left(\prod_{a=1}^{N-1}\frac{d\chi_a}{2\pi}\right), ~}
\end{equation}
where the normalization \(\mathcal{N}_N(s)\) is given in \Eq{eq:appPS_LIPS_normalization} and \(t\) and \(\rho\) are integrated over their respective simplices. The above is a remarkably simple formula for the physical phase-space measure of massless particles.

\FloatBarrier
\section{Examples of channel projections and cut integrals} \label{app:examples}

In this appendix we provide three concrete examples of projection onto Stiefel harmonics as well as the joint \gls{ssyt} and partial-wave projection of \App{app:basis-projection}.
In \Apps{app:examples-dipole}{app:examples-cuh}, we evaluate the two- and three-particle cuts contributing to one-loop dipole mixing and two-loop $\mathcal O_{uH}\to\mathcal O_{HG}$ mixing using the projection method and compare with the direct phase-space integration of the cut integrals. 
In \App{app:examples-3to3}, we project a six-gluon amplitude onto the lowest-degree channel harmonics with three particles on each side, illustrating how the \gls{ssyt} labels distinguish independent channel components at fixed total angular momentum.

\FloatBarrier
\subsection{One-loop mixing of dipole operators}
\label{app:examples-dipole}
\suppressfloats[t]

\begin{figure}[tb]
    \centering
    \begin{subfigure}[t]{0.48\textwidth}
        \centering
        \begin{tikzpicture}[every node/.style={font=\footnotesize}, amplitude/.style={circle,draw,minimum size=1.05cm,align=center,inner sep=0pt,outer sep=0pt}]
            \path[use as bounding box] (-3.15,-1.15) rectangle (3.15,1.15);
            \node[amplitude,fill=gray!10] (sm) at (-1.25,0) {$A_L^{(\mathrm{a})}$};
            \node[amplitude,fill=blue!8] (eft) at (1.25,0) {$A_R^{(\mathrm{a})}$};
            \draw[vector] (sm.25) -- (eft.155);
            \draw[dashed] (sm.335) -- (eft.205);
            \node[anchor=south east,inner sep=0pt] at (-0.12,0.68) {$(-\ell_1)_{B_-}$};
            \node[anchor=south west,inner sep=0pt] at (0.12,0.68) {$(\ell_1)_{B_+}$};
            \node[anchor=north east,inner sep=0pt] at (-0.12,-0.68) {$(-\ell_2)_{H_k}$};
            \node[anchor=north west,inner sep=0pt] at (0.12,-0.68) {$(\ell_2)_{H^\dagger_k}$};
            \draw[vector] (sm.155) -- (-2.3,0.85) node[left] {$3_{W^a_+}$};
            \draw[dashed] (sm.205) -- (-2.3,-0.85) node[left] {$4_{H^\dagger_i}$};
            \draw (eft.25) -- (2.3,0.85) node[right] {$1_e$};
            \draw (eft.335) -- (2.3,-0.85) node[right] {$2_{l_j}$};
            \draw[densely dashed,black!80,line width=0.8pt] (0,-0.55) -- (0,0.55);
        \end{tikzpicture}
        \caption{Gauge boson and scalar.}
        \label{fig:ex-dipole-bh-cut}
    \end{subfigure}\hfill
    \begin{subfigure}[t]{0.48\textwidth}
        \centering
        \begin{tikzpicture}[every node/.style={font=\footnotesize}, amplitude/.style={circle,draw,minimum size=1.05cm,align=center,inner sep=0pt,outer sep=0pt}]
            \path[use as bounding box] (-3.15,-1.15) rectangle (3.15,1.15);
            \node[amplitude,fill=gray!10] (sm) at (-1.25,0) {$A_L^{(\mathrm{b})}$};
            \node[amplitude,fill=blue!8] (eft) at (1.25,0) {$A_R^{(\mathrm{b})}$};
            \draw[vector] (sm.25) -- (eft.155);
            \draw (sm.335) -- (eft.205);
            \node[anchor=south east,inner sep=0pt] at (-0.12,0.68) {$(-\ell_1)_{B_-}$};
            \node[anchor=south west,inner sep=0pt] at (0.12,0.68) {$(\ell_1)_{B_+}$};
            \node[anchor=north east,inner sep=0pt] at (-0.12,-0.68) {$(-\ell_2)_{\bar l_k}$};
            \node[anchor=north west,inner sep=0pt] at (0.12,-0.68) {$(\ell_2)_{l_k}$};
            \draw[vector] (sm.155) -- (-2.3,0.85) node[left] {$3_{W^a_+}$};
            \draw (sm.205) -- (-2.3,-0.85) node[left] {$2_{l_j}$};
            \draw (eft.25) -- (2.3,0.85) node[right] {$1_e$};
            \draw[dashed] (eft.335) -- (2.3,-0.85) node[right] {$4_{H^\dagger_i}$};
            \draw[densely dashed,black!80,line width=0.8pt] (0,-0.55) -- (0,0.55);
        \end{tikzpicture}
        \caption{Gauge boson and lepton.}
        \label{fig:ex-dipole-lb-cut}
    \end{subfigure}
    \caption{The two cuts contributing to $\mathcal O_{BHle}\to\mathcal O_{WHle}$ mixing. The left and right blobs denote the tree-level \gls{sm} amplitudes and source form factors, respectively. All state labels use the all-outgoing convention. Helicity signs follow the operator convention of \Sec{sec:on-shell-rge-form-factors}. Here the fermions $l$ and $e$ have helicity $+1/2$, while $\bar l$ has helicity $-1/2$. The internal weak index $k$ is summed.}
    \label{fig:ex-dipole-cuts}
\end{figure}
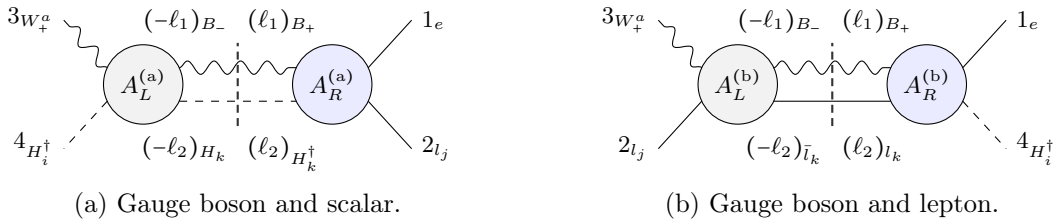

We consider the one-loop mixing of the source operator $\mathcal O_{BHle}$ into the target $\mathcal O_{WHle}$ through the two cuts shown in \Fig{fig:ex-dipole-cuts}.
We use the tree-level amplitudes and normalization conventions of \Ref{Baratella:2020lzz}, with the factor $C_j/\Lambda^2$ removed from each dimension-six operator form factor as in \Eq{eq:rge-form-factor-definition}. The \gls{sm} amplitudes for panels (a) and (b) are obtained from Eqs.~(95) and (96) of that reference, using the crossing convention of \Sec{sec:on-shell-rge-one-loop}.
In \Apps{app:examples-dipole}{app:examples-cuh} the displayed spinor expressions $A_L$ use all-outgoing legs, with the crossed cut legs ordered as in \Sec{sec:on-shell-rge-one-loop}, while $A_R$ is the source form factor with its stated overall factors removed.
The direct cut formulas therefore contain $\sigma_c$.
The coefficients $f_L,f_R$ instead use the physical definitions of \App{appsec:one-sided-projection}, so the conversion phase is included in $f_L$ before contracting the coefficients.
The target form factor is given by
\begin{equation}
    F_{WHle}(1_e,2_{l_j},3_{W^a_+},4_{H^\dagger_i})
    =\langle31\rangle\langle32\rangle\,(T^a)_{ij}.
    \label{eq:ex-dipole-target}
\end{equation}
Here $i,j,k$ label weak-doublet components and $a,b$ label weak-adjoint components. The generators obey $\operatorname{tr}(T^aT^b)=\delta^{ab}/2$.

The tree-level source form factor vanishes on the target external states, which contain a $W$ instead of a $B$, so the infrared terms in \Eq{eq:rge-one-loop-uv} vanish.
The minimal form factor is also independent of the \gls{sm} couplings, so the coupling-derivative term is absent.
Equation~\eqref{eq:rge-one-loop-ir-free} then gives
\begin{equation}
    \gamma^{(1)}_{WHle,BHle}\,F_{WHle}
    =-\frac{(T^a)_{ij}}{\pi}
    \sum_{c\in\{\mathrm{a},\mathrm{b}\}}\sigma_c
    \int d\Phi_2(P_c)\,A_L^{(c)}A_R^{(c)}.
    \label{eq:ex-dipole-rge}
\end{equation}
Here $c=\mathrm{a},\mathrm{b}$ labels the corresponding panels of \Fig{fig:ex-dipole-cuts}, with $P_{\mathrm{a}}=p_3+p_4$ and $P_{\mathrm{b}}=p_2+p_3$.
The tree-level \gls{sm} amplitudes $A_L^{(c)}$ and source form factors $A_R^{(c)}$ have their gauge group structures removed to simplify the notation. Both contractions give the common tensor $(T^a)_{ij}$ displayed explicitly in \Eq{eq:ex-dipole-rge}.
We denote the contribution of each summand by $\gamma^{(1,c)}_{WHle,BHle}$.
After canceling $(T^a)_{ij}$, the target polynomial to be matched is $\langle31\rangle\langle32\rangle$.

We first evaluate the contribution $\gamma^{(1,\mathrm{a})}_{WHle,BHle}$ from \Fig{fig:ex-dipole-cuts}(a).
The source form factor $A_R^{(\mathrm{a})}$ has outgoing legs $1_e$, $2_{l_j}$, $(\ell_1)_{B_+}$ and $(\ell_2)_{H^\dagger_k}$.
The cut momenta $\ell_1,\ell_2$ have positive energy, and these two intermediate states enter $A_L^{(\mathrm{a})}$ as incoming states, scattering into $3_{W^a_+}$ and $4_{H^\dagger_i}$.
In the all-outgoing representation used in the figure, the corresponding \gls{sm} cut legs are $(-\ell_1)_{B_-}$ and $(-\ell_2)_{H_k}$.
Momentum conservation on the \gls{sm} side gives $-\ell_1-\ell_2+p_3+p_4=0$, so $P=P_{\mathrm{a}}=\ell_1+\ell_2=p_3+p_4$, with $s=P^2>0$.
The two tree-level factors are given by
\begin{equation}
    A_R^{(\mathrm{a})}=F_{BHle}=\langle\ell_11\rangle\langle\ell_12\rangle,
    \qquad
    A_L^{(\mathrm{a})}=g_1g_2Y_H\,
    \frac{\langle\ell_23\rangle\langle43\rangle}{\langle\ell_2\ell_1\rangle\langle\ell_14\rangle}.
    \label{eq:ex-dipole-a-factors}
\end{equation}
Here $Y_H$ is the hypercharge of $H$.
The source and target have $h=+2$, whereas $A_L^{(\mathrm{a})}$ has total all-outgoing helicity zero.
The cut contraction $\sum_k(T^a)_{ik}\delta_{kj}=(T^a)_{ij}$ reproduces the target tensor.

To evaluate the contribution from panel (a) in \Eq{eq:ex-dipole-rge}, we first identify the cut-channel component of the source.
It depends on the cut spinors only through $\lambda_{\ell_1}^2$.
The cut legs therefore have operator helicities $(h_{\ell_1},h_{\ell_2})=(+1,0)$, and the cut-spinor degrees are $(n,\widetilde n)=(2,0)$.
The unique \gls{ssyt} at these degrees and helicities is a white row filled twice with $\ell_1$.
We denote the cut and outgoing harmonics for panel (a) by $Y^{(c,\mathrm a)}$ and $Y^{(f,\mathrm a)}$, respectively, and use the label $\mathrm b$ for panel (b).
Using the spinors defined in \Eqs{eq:bp-Lambda-Zx}{eq:bp-conjugate-spin-frame}, its physical-LIPS-normalized harmonic is
\begin{equation}
    \begin{array}{@{}c@{}}
        \begin{ytableau}
            \ell_1 & \ell_1
        \end{ytableau}
    \end{array}
    :\qquad
    Y^{(c,\mathrm a)\,I_1I_2}(Z_c)
    =\sqrt{\frac{3}{\mathcal N_2}}\,
    \frac{\langle x^{I_1}\ell_1\rangle\langle x^{I_2}\ell_1\rangle}{s}
    =\sqrt{\frac{3}{\mathcal N_2}}\,Z_{\ell_1}{}^{I_1}Z_{\ell_1}{}^{I_2}.
    \label{eq:ex-dipole-a-harmonic}
\end{equation}
Here $\mathcal N_2$ is the two-particle phase-space volume defined in \Eq{eq:appPS_LIPS_normalization}.
We use symmetric spinor indices as below \Eq{eq:bp-PoincareCGC} and suppress the fixed labels $(a_c,J)$.

For the projections, we use the $N=2$ parameterization in \Eq{eq:appPS_complex_stiefel_N=2} and the measure in \Eq{eq:appPS_stiefel_measure_full}:
\begin{equation}
\begin{aligned}
    Z_{\ell_1}{}^I&=e^{i\phi_1}\bigl(\cos\theta,-e^{i\chi}\sin\theta\bigr)^I,\\
    Z_{\ell_2}{}^I&=e^{i\phi_2}\bigl(\sin\theta,e^{i\chi}\cos\theta\bigr)^I,\\
    d\mu_2&=2\sin\theta\cos\theta\,d\theta\,
    \frac{d\phi_1}{2\pi}\frac{d\phi_2}{2\pi}\frac{d\chi}{2\pi},
    \qquad \mathcal N_2=\frac{1}{8\pi}.
\end{aligned}
\label{eq:ex-two-particle-measure}
\end{equation}
The angle $\theta$ ranges over $[0,\pi/2]$, and all phases range over $[0,2\pi)$.
The little-group phases cancel in the cut and projection integrands, allowing the use of \Eq{eq:appPS_LIPS_normalization}, \ie, $\int d\Phi_2=\mathcal N_2\int d\mu_2$.

The projection of the source form factor needed to evaluate the integral in \Eq{eq:ex-dipole-rge} follows from \Eq{eq:bp-one-sided-projection}:
\begin{equation}
\begin{aligned}
    f_{R,I_1I_2}
    &=\mathcal N_2\int d\mu_2\,
    \bigl(Y^{(c,\mathrm a)*}\bigr)_{I_1I_2}\,A_R^{(\mathrm{a})}\\
    &=\sqrt{3\mathcal N_2}\,
    \langle x_{K_1}1\rangle\langle x_{K_2}2\rangle
    \int d\mu_2\,
    \overline Z_{\ell_1 I_1}\overline Z_{\ell_1 I_2}
    Z_{\ell_1}{}^{K_1}Z_{\ell_1}{}^{K_2}\\
    &=\sqrt{\frac{\mathcal N_2}{3}}\,
    \langle x_{(I_1}1\rangle\langle x_{I_2)}2\rangle.
\end{aligned}
\label{eq:ex-dipole-a-fR}
\end{equation}
Here parentheses indicate symmetrization, with a factor of $1/n!$ for $n$ indices.
The result can also be read off directly by inserting \Eq{eq:bp-Lambda-Zx} into $A_R^{(\mathrm{a})}$ and comparing with \Eq{eq:ex-dipole-a-harmonic}.
The reconstruction formula \Eq{eq:bp-one-sided-reconstruction} therefore gives
\begin{equation}
    A_R^{(\mathrm{a})}=Y^{(c,\mathrm a)\,I_1I_2}f_{R,I_1I_2},
    \label{eq:ex-dipole-a-reconstruction}
\end{equation}
so the source lies entirely in this single channel component.

The orthogonality relation \Eq{eq:bp-joint-orthogonality} therefore implies that only the \gls{sm} component matching this cut-channel harmonic contributes to the cut.
The physical incoming and outgoing two-particle configurations of $A_L^{(\mathrm{a})}$ both have operator helicities $(+1,0)$.
As seen from the projection below, the outgoing harmonic corresponds to the same two-box white row, filled twice with leg $3$.
Using \Eq{eq:bp-one-sided-projection} we find the coefficient $f_L$ to be given by
\begin{equation}
\begin{aligned}
    f_L^{I_1I_2}
    &=\mathcal N_2\int d\mu_2\,A_L^{(\mathrm{a})}\,Y^{(c,\mathrm a)\,I_1I_2}
    =-\frac{\mathcal N_2}{2}\,g_1g_2Y_H\,Y^{(f,\mathrm a)\,I_1I_2}
    ,\\
    Y^{(f,\mathrm a)\,I_1I_2}
    &=\sqrt{\frac{3}{\mathcal N_2}}\,
    \frac{\langle x^{I_1}3\rangle\langle x^{I_2}3\rangle}{s}.
\end{aligned}
\label{eq:ex-dipole-a-fL}
\end{equation}
To obtain this result, we used the common frame aligned with the final momenta, $\lambda_3=x_1$ and $\lambda_4=x_2$.

The intermediate particles in panel (a) are distinct bosons, so the conversion phase and counting weight are both unity.
Equation~\eqref{eq:bp-one-sided-cut-contraction} then gives
\begin{equation}
\begin{aligned}
    \gamma^{(1,\mathrm{a})}_{WHle,BHle}\,\langle31\rangle\langle32\rangle
    &=-\frac{1}{\pi}\,f_L^{I_1I_2}f_{R,I_1I_2}\\
    &=\frac{\mathcal N_2\,g_1g_2Y_H}{2\pi s}\,
    \langle x^{I_1}3\rangle\langle x^{I_2}3\rangle
    \langle x_{(I_1}1\rangle\langle x_{I_2)}2\rangle \\
    &=\frac{\mathcal N_2}{2\pi}\,g_1g_2Y_H\,
    \langle31\rangle\langle32\rangle.
\end{aligned}
\label{eq:ex-dipole-a-cut}
\end{equation}
Matching the target polynomial and inserting $\mathcal N_2=1/(8\pi)$ gives
\begin{equation}
    \gamma^{(1,\mathrm{a})}_{WHle,BHle}
    =\frac{g_1g_2Y_H}{16\pi^2}.
    \label{eq:ex-dipole-a-result}
\end{equation}
Direct integration of the same term in \Eq{eq:ex-dipole-rge} gives
\begin{equation}
\begin{aligned}
    \gamma^{(1,\mathrm{a})}_{WHle,BHle}\,\langle31\rangle\langle32\rangle
    &=
    -\frac{1}{\pi}\int d\Phi_2(P)\,A_L^{(\mathrm{a})}A_R^{(\mathrm{a})} \\
    &=-\frac{\mathcal N_2}{\pi}\int d\mu_2\,A_L^{(\mathrm{a})}A_R^{(\mathrm{a})}
    =\frac{g_1g_2Y_H}{16\pi^2}\,
    \langle31\rangle\langle32\rangle,
    \label{eq:ex-dipole-a-direct}
\end{aligned}
\end{equation}
in agreement with \Eq{eq:ex-dipole-a-result}.

We next consider panel (b) of \Fig{fig:ex-dipole-cuts}, with intermediate states $(\ell_1)_{B_+}$ and $(\ell_2)_{l_k}$ and total momentum $P=P_{\mathrm b}=\ell_1+\ell_2=p_2+p_3$, where $s=P^2$.
The source and crossed \gls{sm} factors are
\begin{equation}
    A_R^{(\mathrm b)}=\langle\ell_11\rangle\langle\ell_1\ell_2\rangle,
    \qquad
    A_L^{(\mathrm b)}=-i\,g_1g_2Y_l\,
    \frac{\langle23\rangle^2}{\langle2\ell_1\rangle\langle\ell_1\ell_2\rangle},
    \qquad \sigma_{\mathrm b}=i.
    \label{eq:ex-dipole-b-factors}
\end{equation}

The cut legs have operator helicities $(+1,+1/2)$ and cut-spinor degrees $(n,\widetilde n)=(3,0)$.
The source operator has only one nonzero cut-channel component, with normalized harmonic
\begin{equation}
    \begin{array}{@{}c@{}}
        \begin{ytableau}
            \ell_1 & \ell_1 \\
            \ell_2
        \end{ytableau}
    \end{array}
    :\qquad
    Y^{(c,\mathrm b)\,I}(Z_c)
    =\sqrt{\frac{2}{\mathcal N_2}}\,
    \frac{\langle\ell_1\ell_2\rangle\langle x^I\ell_1\rangle}{s}
    =\sqrt{\frac{2}{\mathcal N_2}}\,
    \epsilon_{K_1K_2}Z_{\ell_1}{}^{K_1}Z_{\ell_2}{}^{K_2} Z_{\ell_1}{}^I.
    \label{eq:ex-dipole-b-harmonic}
\end{equation}
As before, by \Eq{eq:bp-joint-orthogonality}, only the matching \gls{sm} component contributes to the cut.
Using \Eq{eq:bp-one-sided-projection} and the parameterization in \Eq{eq:ex-two-particle-measure}, now with $\lambda_3=x_1$ and $\lambda_2=x_2$, gives
\begin{equation}
\begin{aligned}
    f_{R,I}
    &=\mathcal N_2\int d\mu_2\,
    \bigl(Y^{(c,\mathrm b)*}\bigr)_I A_R^{(\mathrm b)}
    =\sqrt{\frac{s\mathcal N_2}{2}}\,\langle x_I1\rangle,\\
    f_L^I
    &=i\mathcal N_2\int d\mu_2\,
    A_L^{(\mathrm b)}Y^{(c,\mathrm b)\,I}
    =\sqrt{2\mathcal N_2}\,g_1g_2Y_l\,
    \frac{\langle23\rangle\langle x^I3\rangle}{s}.
\end{aligned}
\label{eq:ex-dipole-b-projections}
\end{equation}
The factor $i$ converts the all-outgoing input of \Eq{eq:ex-dipole-b-factors} to the physical \gls{sm} matrix element.
Contracting the two projection coefficients according to \Eq{eq:bp-one-sided-cut-contraction} gives
\begin{equation}
\begin{aligned}
    \gamma^{(1,\mathrm b)}_{WHle,BHle}\,
    \langle31\rangle\langle32\rangle
    &=-\frac{1}{\pi}\,f_L^If_{R,I}\\
    &=\frac{g_1g_2Y_l}{8\pi^2}\,
    \langle31\rangle\langle32\rangle.
\end{aligned}
\label{eq:ex-dipole-b-cut}
\end{equation}
Direct integration gives the same result:
\begin{equation}
\begin{aligned}
    -\frac{\sigma_{\mathrm b}}{\pi}\int d\Phi_2(P)\,
    A_L^{(\mathrm b)}A_R^{(\mathrm b)}
    &=-\frac{\mathcal N_2g_1g_2Y_l}{\pi}\,
    \langle23\rangle^2
    \int d\mu_2\,\frac{\langle\ell_11\rangle}{\langle2\ell_1\rangle}\\
    &=\frac{g_1g_2Y_l}{8\pi^2}\,
    \langle31\rangle\langle32\rangle.
\end{aligned}
\label{eq:ex-dipole-b-direct}
\end{equation}

Adding the two contributions and using $Y_H=Y_l+Y_e$ yields
\begin{equation}
\begin{aligned}
    \gamma^{(1)}_{WHle,BHle}
    &=\gamma^{(1,\mathrm a)}_{WHle,BHle}
    +\gamma^{(1,\mathrm b)}_{WHle,BHle}\\
    &=\frac{g_1g_2}{16\pi^2}\,(Y_H+2Y_l)
    =\frac{g_1g_2}{16\pi^2}\,(3Y_l+Y_e).
\end{aligned}
\label{eq:ex-dipole-result}
\end{equation}
The contributions from panels (a) and (b) reproduce Eqs.~(38) and (39) of \Ref{Baratella:2020lzz}, while their sum agrees with its Eq.~(40), after extracting the source Wilson coefficient.

\FloatBarrier
\subsection{Two-loop mixing of \texorpdfstring{$\mathcal O_{uH}$}{O\_uH} into \texorpdfstring{$\mathcal O_{HG}$}{O\_HG}}
\label{app:examples-cuh}

\begin{figure}[t]
    \centering
    \begin{tikzpicture}[every node/.style={font=\footnotesize}, amplitude/.style={circle,draw,minimum size=1.10cm,align=center,inner sep=0pt,outer sep=0pt}]
        \path[use as bounding box] (-3.8,-1.55) rectangle (3.8,1.55);
        \node[amplitude,fill=gray!10] (sm) at (-1.8,0) {$A_L$};
        \node[amplitude,fill=blue!8] (eft) at (1.8,0) {$A_R$};
        \draw (sm.35) .. controls (-0.85,0.95) and (0.85,0.95) .. (eft.145);
        \draw (sm.0) -- (eft.180);
        \draw[dashed] (sm.325) .. controls (-0.85,-0.95) and (0.85,-0.95) .. (eft.215);
        \node[anchor=south east,inner sep=0pt] at (-0.12,1.08) {$(-\ell_1)_{\bar q_r}$};
        \node[anchor=south west,inner sep=0pt] at (0.12,1.08) {$(\ell_1)_{q^r}$};
        \node[anchor=south east,inner sep=0pt] at (-0.12,0.16) {$(-\ell_2)_{\bar u}$};
        \node[anchor=south west,inner sep=0pt] at (0.12,0.16) {$(\ell_2)_u$};
        \node[anchor=north east,inner sep=0pt] at (-0.12,-1.08) {$(-\ell_3)_{H^{\dagger k}}$};
        \node[anchor=north west,inner sep=0pt] at (0.12,-1.08) {$(\ell_3)_{H_k}$};
        \draw[vector] (sm.155) -- (-3,0.85) node[left] {$3_{g^A_+}$};
        \draw[vector] (sm.205) -- (-3,-0.85) node[left] {$4_{g^B_+}$};
        \draw[dashed] (eft.25) -- (3,0.85) node[right] {$1_{H_i}$};
        \draw[dashed] (eft.335) -- (3,-0.85) node[right] {$2_{H^{\dagger j}}$};
        \draw[densely dashed,black!80,line width=0.8pt] (0,-0.98) -- (0,0.98);
    \end{tikzpicture}
    \caption{The three-particle cut contributing to the two-loop mixing $\mathcal O_{uH}\to\mathcal O_{HG}$. The left and right blobs denote the tree-level \gls{sm} amplitude and source form factor, respectively. All state labels use the all-outgoing convention. In the operator-helicity convention of \Sec{sec:on-shell-rge-form-factors}, $q$ and $u$ have helicity $+1/2$, while $\bar q$ and $\bar u$ have helicity $-1/2$. Internal weak and color indices are summed.}
    \label{fig:ex-cuh-cut}
\end{figure}
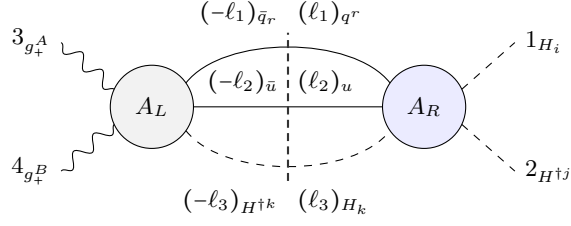

We next consider the mixing of $\mathcal O_{uH}$ into $\mathcal O_{HG}$, taking one quark flavor and real Yukawa and Wilson coefficients.
The source and target have lengths five and four, so \Eq{eq:bern-length-selection} excludes one-loop mixing.
The first allowed contribution is the three-particle cut in \Fig{fig:ex-cuh-cut}.
We evaluate the source and target form-factor components with total operator helicities $+1$ and $+2$, respectively.
We use the form-factor and \gls{sm}-amplitude normalizations of Eqs.~(2.17), (2.19) and (2.37) of \Ref{EliasMiro:2020tdv}, with the source Wilson coefficient removed.

\suppressfloats[t]

The target form factor is
\begin{equation}
    F_{HG}(1_{H_i},2_{H^{\dagger j}},3_{g^A_+},4_{g^B_+})
    =-2\,\delta_i^j\delta^{AB}\,\langle34\rangle^2.
    \label{eq:ex-cuh-target}
\end{equation}
Here $i,j,k,r$ label weak-doublet components and $A,B$ label color-adjoint components.

The intermediate states on the source side are $(\ell_1)_{q^r}$, $(\ell_2)_u$ and $(\ell_3)_{H_k}$, with $P=\ell_1+\ell_2+\ell_3=p_3+p_4$ and $s=P^2>0$.
The source has weak tensor $\delta_i^j\delta_k^r+\delta_k^j\delta_i^r$ and a color identity between the two fermions.
Contracting the weak indices gives $3\delta_i^j$, while each \gls{sm} color ordering gives $\operatorname{tr}(T^AT^B)=\delta^{AB}/2$.
We remove these group structures and include both color orderings in $A_L$.
Equation~\eqref{eq:rge-maximal-tree-cut} then becomes
\begin{equation}
    \gamma^{(2)}_{HG,uH}\,F_{HG}
    =-\frac{3\delta_i^j\delta^{AB}}{2\pi}\,\sigma_c
    \int d\Phi_3(P)\,A_LA_R.
    \label{eq:ex-cuh-rge}
\end{equation}
The crossed fermions in $A_L$ are ordered as $(-\ell_2)_{\bar u}$ and $(-\ell_1)_{\bar q_r}$, opposite to their order in $A_R$.
Using the crossing convention of \Sec{sec:on-shell-rge-one-loop}, the two factors expressed in positive-energy cut spinors are
\begin{equation}
\begin{aligned}
    A_R&=\langle\ell_1\ell_2\rangle,\\
    A_L&=-2y\,g_s^2\left(
    \frac{[\ell_1\ell_2]^2}{[\ell_13][34][4\ell_2]}
    +\frac{[\ell_1\ell_2]^2}{[\ell_14][43][3\ell_2]}
    \right),
    \qquad \sigma_c=i^2=-1.
\end{aligned}
\label{eq:ex-cuh-factors}
\end{equation}
This phase belongs to the direct all-outgoing cut.
It is included in the physical left projection below, so it is not inserted again when contracting $f_L$ with $f_R$.
Here $y$ is the Yukawa coupling and $g_s$ is the strong gauge coupling.

The source has cut-leg operator helicities $(+1/2,+1/2,0)$ and cut-spinor degrees $(n,\widetilde n)=(2,0)$.
At these degrees and helicities, the two candidate harmonics are
\begin{equation}
\begin{aligned}
    \begin{array}{@{}c@{}}
        \begin{ytableau}
            \ell_1 & \ell_2
        \end{ytableau}
    \end{array}
    &:\qquad
    Y_{\mathrm{row}}^{(c)\,I_1I_2}(Z_c)
    =\sqrt{\frac{12}{\mathcal N_3}}\,
    \frac{\langle x^{(I_1}\ell_1\rangle\langle x^{I_2)}\ell_2\rangle}{s}
    =\sqrt{\frac{12}{\mathcal N_3}}\,
    Z_{\ell_1}{}^{(I_1}Z_{\ell_2}{}^{I_2)},\\
    \begin{array}{@{}c@{}}
        \begin{ytableau}
            \ell_1 \\
            \ell_2
        \end{ytableau}
    \end{array}
    &:\qquad
    Y^{(c)}(Z_c)
    =\sqrt{\frac{3}{\mathcal N_3}}\,
    \frac{\langle\ell_1\ell_2\rangle}{\sqrt{s}}
    =\sqrt{\frac{3}{\mathcal N_3}}\,
    \epsilon_{K_1K_2}Z_{\ell_1}{}^{K_1}Z_{\ell_2}{}^{K_2}.
\end{aligned}
\label{eq:ex-cuh-harmonic}
\end{equation}

To evaluate the projections, we choose the common spin frame of \Eq{eq:bp-Lambda-Zx} such that $\lambda_3=x_1$ and $\lambda_4=x_2$.
The parameterization and unit-normalized measure in \Eqs{eq:appPS_complex_stiefel_N=3}{eq:appPS_stiefel_measure_full} give
\begin{equation}
\begin{aligned}
    Z_{\ell_1}{}^I
    &=e^{i\phi_1}\bigl(\cos\theta_1,-e^{i\chi_1}\sin\theta_1\cos\psi\bigr)^I,\\
    Z_{\ell_2}{}^I
    &=e^{i\phi_2}\bigl(\sin\theta_1\cos\theta_2,e^{i\chi_1}\cos\theta_1\cos\theta_2\cos\psi-e^{i\chi_2}\sin\theta_2\sin\psi\bigr)^I,\\
    Z_{\ell_3}{}^I
    &=e^{i\phi_3}\bigl(\sin\theta_1\sin\theta_2,e^{i\chi_1}\cos\theta_1\sin\theta_2\cos\psi+e^{i\chi_2}\cos\theta_2\sin\psi\bigr)^I,\\
    d\mu_3
    &=16\sin^3\theta_1\cos\theta_1\sin\theta_2\cos\theta_2\sin\psi\cos\psi\,d\theta_1\,d\theta_2\,d\psi \frac{d\phi_1}{2\pi}\frac{d\phi_2}{2\pi}\frac{d\phi_3}{2\pi}\frac{d\chi_1}{2\pi}\frac{d\chi_2}{2\pi}, \\
    \mathcal N_3 & =\frac{s}{256\pi^3}.
\end{aligned}
\label{eq:ex-three-particle-parametrization}
\end{equation}
The angles $\theta_1,\theta_2,\psi$ range over $[0,\pi/2]$, and all phases range over $[0,2\pi)$.
We can convert this measure to the physical LIPS using \Eq{eq:appPS_LIPS_normalization}.

Using \Eq{eq:bp-one-sided-projection}, the source projection onto the row \gls{ssyt} gives
\begin{equation}
\begin{aligned}
    f_{R,\mathrm{row},I_1I_2}
    &=\mathcal N_3\int d\mu_3\,
    \bigl(Y_{\mathrm{row}}^{(c)*}\bigr)_{I_1I_2}A_R\\
    &=\sqrt{12s\mathcal N_3}\int d\mu_3\,
    \overline Z_{\ell_1(I_1}\overline Z_{\ell_2 I_2)}
    \epsilon_{K_1K_2}Z_{\ell_1}{}^{K_1}Z_{\ell_2}{}^{K_2}
    =0.
    \label{eq:ex-cuh-row-projection}
\end{aligned}
\end{equation}
The integral vanishes because the symmetric row harmonic carries $J=1$, while the antisymmetric spinor contraction in $A_R$ is a $J=0$ scalar.
The source therefore lies entirely in the column component.
By \Eq{eq:bp-joint-orthogonality}, only the matching \gls{sm} projection contributes to the cut.
The remaining projections give
\begin{equation}
\begin{aligned}
    f_R
    &=\mathcal N_3\int d\mu_3\,Y^{(c)*}A_R
    =\sqrt{\frac{s\mathcal N_3}{3}},\\
    f_L
    &=-\mathcal N_3\int d\mu_3\,A_LY^{(c)}
    =4y\,g_s^2\sqrt{\frac{3\mathcal N_3}{s}}\,
    \frac{\langle34\rangle^2}{s}.
\end{aligned}
\label{eq:ex-cuh-projections}
\end{equation}
The two color orderings give equal contributions to $f_L$.
Contracting the coefficients according to \Eq{eq:bp-one-sided-cut-contraction} therefore yields
\begin{equation}
    \gamma^{(2)}_{HG,uH}\,F_{HG}
    =-\frac{3\delta_i^j\delta^{AB}}{2\pi}\,f_Lf_R = -\frac{6y\,g_s^2\mathcal N_3}{\pi s}\,
    \delta_i^j\delta^{AB}\,\langle34\rangle^2.
\label{eq:ex-cuh-result}
\end{equation}
Direct integration of \Eq{eq:ex-cuh-rge} gives the same result,
\begin{equation}
\begin{aligned}
    -\frac{3\delta_i^j\delta^{AB}}{2\pi}\,\sigma_c\int d\Phi_3(P)\,A_LA_R
    &=-\frac{6y\,g_s^2\mathcal N_3}{\pi}\,\delta_i^j\delta^{AB}
    \int d\mu_3\,
    \frac{[\ell_1\ell_2]^2\langle\ell_1\ell_2\rangle}
    {[\ell_13][34][4\ell_2]}\\
    &=-\frac{6y\,g_s^2\mathcal N_3}{\pi s}\,
    \delta_i^j\delta^{AB}\,\langle34\rangle^2.
\end{aligned}
\label{eq:ex-cuh-direct}
\end{equation}
Matching the result in \Eq{eq:ex-cuh-result} to the target form factor in \Eq{eq:ex-cuh-target} gives
\begin{equation}
    \gamma^{(2)}_{HG,uH}
    =\frac{3y\,g_s^2}{(16\pi^2)^2}.
    \label{eq:ex-cuh-anomalous-dimension}
\end{equation}
For real couplings, \Eq{eq:ex-cuh-anomalous-dimension} agrees with \Ref{Born:2026xkr} as well as the corrected two-loop result of \Ref{DiNoi:2024ajj}.\footnote{The opposite sign in Eq.~(4.22) of \Ref{EliasMiro:2020tdv} can be traced to its Eq.~(4.21), where the crossed amplitude is evaluated at the positive-energy spinors of the cut quarks. Applying the continuation in \Eq{eq:rge-crossing-convention} to the two cut-quark spinors in the \gls{sm} amplitude gives a factor $i^2=-1$.}

In both examples above, the Stiefel harmonics selected by the source form factors carry a single total spin $J$, so the calculations coincide with the one-sided partial-wave projections described in \App{appsec:one-sided-projection}.
We next consider a channel with three particles on each side, where several independent components contribute at fixed $J$ and are distinguished by their \gls{ssyt} labels.

\FloatBarrier
\subsection{Two-sided projection of a six-gluon amplitude}
\label{app:examples-3to3}
\suppressfloats[t]

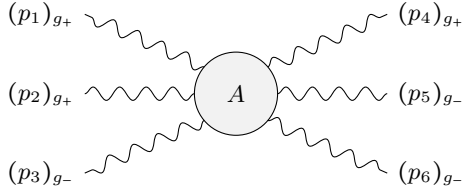
\begin{figure}[t]
    \centering
    \begin{tikzpicture}[every node/.style={font=\footnotesize}, amplitude/.style={circle,draw,minimum size=1.10cm,align=center,inner sep=0pt,outer sep=0pt}]
        \path[use as bounding box] (-3.1,-1.35) rectangle (3.1,1.35);
        \node[amplitude,fill=gray!10] (sm) at (0,0) {$A$};
        \draw[vector] (sm.140) -- (-2,1.05) node[left] {$(p_1)_{g_+}$};
        \draw[vector] (sm.180) -- (-2,0) node[left] {$(p_2)_{g_+}$};
        \draw[vector] (sm.220) -- (-2,-1.05) node[left] {$(p_3)_{g_-}$};
        \draw[vector] (sm.40) -- (2,1.05) node[right] {$(p_4)_{g_+}$};
        \draw[vector] (sm.0) -- (2,0) node[right] {$(p_5)_{g_-}$};
        \draw[vector] (sm.320) -- (2,-1.05) node[right] {$(p_6)_{g_-}$};
    \end{tikzpicture}
    \caption{The color-ordered tree-level six-gluon amplitude $A$ in \Eq{eq:ex-mhv6-crossed}, with incoming gluons $1,2,3$ on the left and outgoing gluons $4,5,6$ on the right. All momenta have positive energy, and helicity signs follow the operator convention of \Sec{sec:on-shell-rge-form-factors}.}
    \label{fig:ex-mhv6-channel}
\end{figure}

In \Apps{app:examples-dipole}{app:examples-cuh}, the \gls{sm} amplitude has a two-particle final state, with a unique final-state harmonic for each allowed $J$ at fixed helicities, as discussed in \App{app:two-particle-casimirs}. By \Eq{eq:bp-one-sided-two-sided-relation}, each $f^{JM}_{L;a_c}$ is therefore proportional to this harmonic. The reduced projection coefficients $c^J$ defined in \Eq{eq:bp-joint-amplitude-schur} can thus be read off without an additional integration over the final state. The analogous simplification holds for $f_R$ when the initial state contains two particles. With three particles on each side, several initial- and final-state harmonics can carry the same $J$, and $c^J$ forms a matrix whose rows label the final-state harmonics and whose columns label the initial-state harmonics. 
We illustrate this with the six-gluon amplitude in \Fig{fig:ex-mhv6-channel}, first projecting onto a representative pair of Stiefel harmonics and then onto the channel harmonics in \Eqs{eq:ex-mhv6-joint-explicit-final}{eq:ex-mhv6-joint-explicit-initial}.

The process in \Fig{fig:ex-mhv6-channel} has incoming and outgoing operator helicities $(+1,+1,-1)$ and $(+1,-1,-1)$, respectively.
We use positive-energy momenta for all six particles, with $P=p_1+p_2+p_3=p_4+p_5+p_6$ and $s=P^2>0$.
Using the crossing convention of \Eq{eq:rge-crossing-convention}, the color-ordered tree-level MHV amplitude~\cite{Parke:1986gb,Dixon:1996wi}, with the gauge coupling and color factor removed, is
\begin{equation}
    A(Z_f;Z_i)
    =-\frac{\langle34\rangle^3}
    {\langle12\rangle\langle23\rangle\langle45\rangle\langle56\rangle\langle61\rangle}.
    \label{eq:ex-mhv6-crossed}
\end{equation}
There is no fermionic conversion phase in this example.
When the incoming momenta are crossed, the color ordering remains $(1,2,3,4,5,6)$.
In the all-outgoing representation, the incoming helicities are reversed, giving total operator helicity $h=-2$, consistent with the helicity constraint in \Eq{eq:constraint_higher_point}.
The amplitude has only collinear poles, which give finite projection integrals below.

The helicities fix the minimal spinor content to $\lambda_1^2\lambda_2^2\widetilde\lambda_3^2$ on the initial side and $\lambda_4^2\widetilde\lambda_5^2\widetilde\lambda_6^2$ on the final side. 
At these minimal spinor degrees, the initial and final helicities allow the following three \glspl{ssyt} on each side:
\begin{equation}
    \begin{array}{l@{\qquad}c@{\qquad}c@{\qquad}c}
        \text{final:}
        &
        \begin{array}[t]{@{}c@{}}
            \begin{ytableau}
                *(blue!40) 4 & *(blue!40) 4 & *(blue!40) 4 & *(blue!40) 4 & 4 & 4 \\
                *(blue!40) 5 & *(blue!40) 5 & *(blue!40) 6 & *(blue!40) 6
            \end{ytableau}
        \end{array}
        &
        \begin{array}[t]{@{}c@{}}
            \begin{ytableau}
                *(blue!40) 4 & *(blue!40) 4 & *(blue!40) 4 & 4 & 4 \\
                *(blue!40) 5 & *(blue!40) 6
            \end{ytableau}
        \end{array}
        &
        \begin{array}[t]{@{}c@{}}
            \begin{ytableau}
                *(blue!40) 4 & *(blue!40) 4 & 4 & 4
            \end{ytableau}
        \end{array}
        \\[3.5ex]
        & \Lambda_f=(6,4) & \Lambda_f=(5,2) & \Lambda_f=(4,0)
        \\[3ex]
        \text{initial:}
        &
        \begin{array}[t]{@{}c@{}}
            \begin{ytableau}
                *(blue!40) 1 & *(blue!40) 1 & 1 & 1 & 2 & 2 \\
                *(blue!40) 2 & *(blue!40) 2
            \end{ytableau}
        \end{array}
        &
        \begin{array}[t]{@{}c@{}}
            \begin{ytableau}
                *(blue!40) 1 & *(blue!40) 1 & 1 & 1 & 2 \\
                *(blue!40) 2 & *(blue!40) 2 & 2
            \end{ytableau}
        \end{array}
        &
        \begin{array}[t]{@{}c@{}}
            \begin{ytableau}
                *(blue!40) 1 & *(blue!40) 1 & 1 & 1 \\
                *(blue!40) 2 & *(blue!40) 2 & 2 & 2
            \end{ytableau}
        \end{array}
        \\[3.5ex]
        & \Lambda_i=(6,2) & \Lambda_i=(5,3) & \Lambda_i=(4,4)
    \end{array}
    \label{eq:ex-mhv6-tableaux}
\end{equation}
The labels $\Lambda_f$ and $\Lambda_i$ give the row lengths of the combined white and blue diagrams, drawn as in \Eq{eq:covariant-ssyt-example}.
Each displayed shape has one filling compatible with the chosen helicities, so we suppress $T$.
Using the common spin frame of \Eqs{eq:bp-Lambda-Zx}{eq:bp-conjugate-spin-frame}, the dimensionless \gls{ssyt} polynomials before normalization are
\begin{equation}
\begin{aligned}
    \widehat Y^{(f)}_{(6,4)}
    &=s^{-3}\,\langle x^{I_1}4\rangle\langle x^{I_2}4\rangle\,
    [5\widetilde x_{(J_1}][5\widetilde x_{J_2}][6\widetilde x_{J_3}][6\widetilde x_{J_4)}],
    &&J\in\{1,2,3\},\\
    \widehat Y^{(f)}_{(5,2)}
    &=s^{-5/2}\,\langle x^{I_1}4\rangle\langle x^{I_2}4\rangle\,[56]\,
    [5\widetilde x_{(J_1}][6\widetilde x_{J_2)}],
    &&J\in\{0,1,2\},\\
    \widehat Y^{(f)}_{(4,0)}
    &=s^{-2}\,\langle x^{I_1}4\rangle\langle x^{I_2}4\rangle\,[56]^2,
    &&J=1,\\[1ex]
    \widehat Y^{(i)}_{(6,2)}
    &=s^{-3}\,\langle x^{(K_1}1\rangle\langle x^{K_2}1\rangle\langle x^{K_3}2\rangle\langle x^{K_4)}2\rangle\,
    [3\widetilde x_{L_1}][3\widetilde x_{L_2}],
    &&J\in\{1,2,3\},\\
    \widehat Y^{(i)}_{(5,3)}
    &=s^{-5/2}\,\langle12\rangle\,\langle x^{(K_1}1\rangle\langle x^{K_2)}2\rangle\,
    [3\widetilde x_{L_1}][3\widetilde x_{L_2}],
    &&J\in\{0,1,2\},\\
    \widehat Y^{(i)}_{(4,4)}
    &=s^{-2}\,\langle12\rangle^2\,[3\widetilde x_{L_1}][3\widetilde x_{L_2}],
    &&J=1,
\end{aligned}
\label{eq:ex-mhv6-ssyt-polynomials}
\end{equation}
where we state the spin components $J$ of each \gls{ssyt} explicitly.
Here the hats distinguish the unnormalized polynomials from the normalized harmonics of \Eq{eq:bp-SSYT-orthogonality}.
The displayed spins $J$ follow from coupling the Lorentz spins $j_1$ and $j_2$, given by half the numbers of white and blue open indices, respectively, as discussed in \App{appsec:joint_basis}.

The shapes $(4,0)$ and $(4,4)$ carry only white or only blue open indices and contain only $J=1$.
Their normalization follows from \Eq{eq:bp-SSYT-orthogonality}
\begin{equation}
    \int d\Phi_3(P)\,
    \bigl(Y^{(f)*}_{(4,0)}\bigr)_{I_1I_2}\,
    Y^{(f)\,K_1K_2}_{(4,0)}
    =\delta^{(K_1}_{I_1}\delta^{K_2)}_{I_2},
    \label{eq:ex-mhv6-orthonormality}
\end{equation}
and analogously for $Y^{(i)}_{(4,4)}$. Hence, the normalized harmonics are given by
\begin{equation}
    Y^{(f)\,I_1I_2}_{(4,0)}
    =\sqrt{\frac{90}{\mathcal N_3}}\,
    \frac{\langle x^{I_1}4\rangle\langle x^{I_2}4\rangle[56]^2}{s^2},
    \quad
    Y^{(i)}_{(4,4),L_1L_2}
    =\sqrt{\frac{90}{\mathcal N_3}}\,
    \frac{\langle12\rangle^2[3\widetilde x_{L_1}][3\widetilde x_{L_2}]}{s^2},
\label{eq:ex-mhv6-harmonics-simple}
\end{equation}
with the phase-space volume $\mathcal N_3$ defined in \Eq{eq:ex-three-particle-parametrization}.

For the other four shapes, which contain several spins, we use the channel harmonics defined in \Eq{eq:bp-joint-harmonics} and normalize each spin-$J$ component separately according to \Eq{eq:bp-joint-orthogonality}.
Following \App{appsec:joint_basis}, we decompose the tensors in \Eq{eq:ex-mhv6-ssyt-polynomials} into components of definite total spin $J$ and spin projection $M$ by raising the lower indices using $\epsilon^{IJ}$ and applying the appropriate symmetrizations and antisymmetrizations.
The channel harmonics satisfy the physical-LIPS orthogonality condition of \Eq{eq:bp-joint-orthogonality},
\begin{equation}
    \int d\Phi_3(P)\,
    Y_{\Lambda;J,M}^*Y_{\Lambda';J',M'}
    =\delta_{\Lambda\Lambda'}\delta_{JJ'}\delta_{MM'},
    \label{eq:ex-mhv6-joint-orthogonality}
\end{equation}
separately on the initial and final sides.
After separating off the invariant $\epsilon$ tensors associated with antisymmetric index pairs, we represent each spin-$J$ component by a completely symmetric tensor with $2J$ upper massive-spinor indices, whose components in the standard orthonormal spin-$J$ basis are the channel harmonics $Y_{\Lambda;J,M}$.
The final-state harmonics read
\begin{equation}
\begin{aligned}
    Y^{(f)}_{(6,4);3}{}^{K_1\ldots K_6}
    &=\sqrt{\tfrac{360}{\mathcal N_3}}\,s^{-3}\,
    \langle x^{(K_1}4\rangle\langle x^{K_2}4\rangle[5\widetilde x^{K_3}][5\widetilde x^{K_4}][6\widetilde x^{K_5}][6\widetilde x^{K_6)}],\\
    Y^{(f)}_{(6,4);2}{}^{K_1\ldots K_4}
    &=-\sqrt{\tfrac{840}{\mathcal N_3}}\,s^{-3}\,
    \bigl(\langle4|P|5]\,\langle x^{(K_1}4\rangle[5\widetilde x^{K_2}][6\widetilde x^{K_3}][6\widetilde x^{K_4)}]\\
    &\qquad+\langle4|P|6]\,\langle x^{(K_1}4\rangle[5\widetilde x^{K_2}][5\widetilde x^{K_3}][6\widetilde x^{K_4)}]\bigr),\\
    Y^{(f)}_{(6,4);1}{}^{K_1K_2}
    &=\sqrt{\tfrac{126}{\mathcal N_3}}\,s^{-3}\,
    \bigl(\langle4|P|5]^2[6\widetilde x^{K_1}][6\widetilde x^{K_2}]\\
    &\qquad+4\langle4|P|5]\langle4|P|6]\,[5\widetilde x^{(K_1}][6\widetilde x^{K_2)}]
    +\langle4|P|6]^2[5\widetilde x^{K_1}][5\widetilde x^{K_2}]\bigr),\\
    Y^{(f)}_{(5,2);2}{}^{K_1\ldots K_4}
    &=\sqrt{\tfrac{252}{\mathcal N_3}}\,s^{-5/2}\,[56]\,
    \langle x^{(K_1}4\rangle\langle x^{K_2}4\rangle[5\widetilde x^{K_3}][6\widetilde x^{K_4)}],\\
    Y^{(f)}_{(5,2);1}{}^{K_1K_2}
    &=-\sqrt{\tfrac{189}{\mathcal N_3}}\,s^{-5/2}\,[56]\,
    \bigl(\langle4|P|6]\,\langle x^{(K_1}4\rangle[5\widetilde x^{K_2)}]
    +\langle4|P|5]\,\langle x^{(K_1}4\rangle[6\widetilde x^{K_2)}]\bigr),\\
    Y^{(f)}_{(5,2);0}
    &=\sqrt{\tfrac{420}{\mathcal N_3}}\,s^{-5/2}\,[56]\,\langle4|P|5]\langle4|P|6],\\
    Y^{(f)}_{(4,0);1}{}^{K_1K_2}
    &=\sqrt{\tfrac{90}{\mathcal N_3}}\,s^{-2}\,[56]^2\,\langle x^{K_1}4\rangle\langle x^{K_2}4\rangle,
\end{aligned}
\label{eq:ex-mhv6-joint-explicit-final}
\end{equation}
and the initial-state harmonics are
\begin{equation}
\begin{aligned}
    Y^{(i)}_{(6,2);3}{}^{K_1\ldots K_6}
    &=\sqrt{\tfrac{360}{\mathcal N_3}}\,s^{-3}\,
    \langle x^{(K_1}1\rangle\langle x^{K_2}1\rangle\langle x^{K_3}2\rangle\langle x^{K_4}2\rangle[3\widetilde x^{K_5}][3\widetilde x^{K_6)}],\\
    Y^{(i)}_{(6,2);2}{}^{K_1\ldots K_4}
    &=-\sqrt{\tfrac{840}{\mathcal N_3}}\,s^{-3}\,
    \bigl(\langle1|P|3]\,\langle x^{(K_1}1\rangle\langle x^{K_2}2\rangle\langle x^{K_3}2\rangle[3\widetilde x^{K_4)}]\\
    &\qquad+\langle2|P|3]\,\langle x^{(K_1}1\rangle\langle x^{K_2}1\rangle\langle x^{K_3}2\rangle[3\widetilde x^{K_4)}]\bigr),\\
    Y^{(i)}_{(6,2);1}{}^{K_1K_2}
    &=\sqrt{\tfrac{126}{\mathcal N_3}}\,s^{-3}\,
    \bigl(\langle1|P|3]^2\langle x^{K_1}2\rangle\langle x^{K_2}2\rangle\\
    &\qquad+4\langle1|P|3]\langle2|P|3]\,\langle x^{(K_1}1\rangle\langle x^{K_2)}2\rangle
    +\langle2|P|3]^2\langle x^{K_1}1\rangle\langle x^{K_2}1\rangle\bigr),\\
    Y^{(i)}_{(5,3);2}{}^{K_1\ldots K_4}
    &=\sqrt{\tfrac{252}{\mathcal N_3}}\,s^{-5/2}\,\langle12\rangle\,
    \langle x^{(K_1}1\rangle\langle x^{K_2}2\rangle[3\widetilde x^{K_3}][3\widetilde x^{K_4)}],\\
    Y^{(i)}_{(5,3);1}{}^{K_1K_2}
    &=-\sqrt{\tfrac{189}{\mathcal N_3}}\,s^{-5/2}\,\langle12\rangle\,
    \bigl(\langle1|P|3]\,\langle x^{(K_1}2\rangle[3\widetilde x^{K_2)}]
    +\langle2|P|3]\,\langle x^{(K_1}1\rangle[3\widetilde x^{K_2)}]\bigr),\\
    Y^{(i)}_{(5,3);0}
    &=\sqrt{\tfrac{420}{\mathcal N_3}}\,s^{-5/2}\,\langle12\rangle\,\langle1|P|3]\langle2|P|3],\\
    Y^{(i)}_{(4,4);1}{}^{K_1K_2}
    &=\sqrt{\tfrac{90}{\mathcal N_3}}\,s^{-2}\,\langle12\rangle^2\,[3\widetilde x^{K_1}][3\widetilde x^{K_2}].
\end{aligned}
\label{eq:ex-mhv6-joint-explicit-initial}
\end{equation}

We first evaluate the projection onto the pair of Stiefel harmonics $(4,0)\times(4,4)$.
Using \Eqs{eq:bp-SSYT-amplitude-projection}{eq:bp-SSYT-amplitude-projection-LIPS} gives
\begin{equation}
\begin{aligned}
    b_{I_1I_2;L_1L_2}
    &=\mathcal N_3^2\int d\mu_3(Z_f)\,d\mu_3(Z_i)\,
    \bigl(Y^{(f)*}_{(4,0)}\bigr)_{I_1I_2}
    A(Z_f;Z_i)Y^{(i)}_{(4,4),L_1L_2}\\
    &=\int d\Phi_3^{(f)}(P)\,d\Phi_3^{(i)}(P)\,
    \bigl(Y^{(f)*}_{(4,0)}\bigr)_{I_1I_2}
    A(Z_f;Z_i)Y^{(i)}_{(4,4),L_1L_2}\\
    &=\frac{\beta}{2}\bigl(\epsilon_{I_1L_1}\epsilon_{I_2L_2}+\epsilon_{I_1L_2}\epsilon_{I_2L_1}\bigr).
\end{aligned}
\label{eq:ex-mhv6-b-def}
\end{equation}
Since both harmonics carry only $J=1$, rotational invariance fixes $b$ up to a single coefficient $\beta$, as in \Eq{eq:bp-joint-amplitude-schur}.
In practice, we evaluate $\beta$ by contracting $b$ with the invariant $\epsilon^{I_1L_1}\epsilon^{I_2L_2}$ already before performing the integrals.
On the right-hand side of \Eq{eq:ex-mhv6-b-def}, this contraction gives $3\beta$.
Under the integral, we use $\langle x^I i\rangle^*=[i\widetilde x_I]$, $[ij]^*=-\langle ij\rangle$ and $[i\widetilde x_I][j\widetilde x^I]=\sqrt s\,[ij]$, so that the two harmonics contract into $(90/\mathcal N_3)\langle12\rangle^2\langle56\rangle^2[34]^2/s^3$.
The contraction therefore gives
\begin{equation}
\begin{aligned}
    \beta
    &=\frac{30}{\mathcal N_3\,s^3}\int d\Phi_3^{(f)}(P)\,d\Phi_3^{(i)}(P)\,
    A\,\langle12\rangle^2\langle56\rangle^2[34]^2\\
    &=-\frac{30}{\mathcal N_3\,s^3}\int d\Phi_3^{(f)}(P)\,
    \frac{\langle56\rangle}{\langle45\rangle}\,I(Z_f),
\end{aligned}
\label{eq:ex-mhv6-beta}
\end{equation}
where
\begin{equation}
\begin{aligned}
    I(Z_f)
    &=\int d\Phi_3^{(i)}(P)\,
    \frac{\langle34\rangle^3[34]^2\langle12\rangle}{\langle23\rangle\langle61\rangle}\\
    &=\frac{\mathcal N_3}{24}\,\langle4|P|6]\,
    \left(
    \frac{\langle4|P|4]^2}{\langle6|P|6]}
    +\frac{\langle4|P|4]\,\langle4|P|6]\langle6|P|4]}{\langle6|P|6]^2}
    +\frac{\langle4|P|6]^2\langle6|P|4]^2}{\langle6|P|6]^3}
    \right),
\end{aligned}
\label{eq:ex-mhv6-inner-def}
\end{equation}
which depends only on the final-state kinematics.
To evaluate this integral, we use the parameterization and measure in \Eq{eq:ex-three-particle-parametrization}, with $1,2,3$ in place of $\ell_1,\ell_2,\ell_3$.
For this integral, we expand the initial spinors in an auxiliary spin frame $x'$ aligned with leg $6$, such that $\langle x'^16\rangle=\langle6|P|6]^{1/2}$ and $\langle x'^26\rangle=0$, while the harmonics keep their components in the common frame $x$.
In these coordinates, the two poles in the denominator of the integrand in \Eq{eq:ex-mhv6-inner-def} are of the form
\begin{equation}
    \langle23\rangle=-\sqrt s\,e^{i(\phi_2+\phi_3+\chi_2)}\sin\theta_1\sin\psi,
    \qquad
    \langle61\rangle=\langle6|P|6]^{1/2}\,e^{i(\phi_1+\chi_1)}\sin\theta_1\cos\psi.
    \label{eq:ex-mhv6-monomial-poles}
\end{equation}
This makes sure that there is no sum of terms with different phases in the denominator and crucially simplifies the calculation since the phase integrals select the terms without net phase.
For the remaining integral over the final states in \Eq{eq:ex-mhv6-beta}, we use \Eq{eq:ex-three-particle-parametrization} with $(\ell_1,\ell_2,\ell_3)=(6,4,5)$, such that $\langle6|P|6]$ is independent of the phases. After inserting \Eq{eq:ex-mhv6-inner-def}, using momentum conservation, and integrating over the phases and $\theta_2$, we obtain
\begin{equation}
\begin{aligned}
    \beta
    &=\frac{5\mathcal N_3}{48s}
    \int_0^{\pi/2}4\sin^3\theta_1\cos\theta_1\,d\theta_1
    \int_0^{\pi/2}2\sin\psi\cos\psi\,d\psi
    \bigl(1+3\sin^2\theta_1\sin^2\psi
    +9\sin^4\theta_1\sin^4\psi\bigr) \\
    &=\frac{5\mathcal N_3}{48s}\left(1+1+\frac32\right)
    =\frac{35}{96}\,\frac{\mathcal N_3}{s}.
\end{aligned}
\label{eq:ex-mhv6-final-integrand}
\end{equation}
Substituting this result into \Eq{eq:ex-mhv6-b-def} gives the projection tensor
\begin{equation}
    b_{I_1I_2;L_1L_2}
    =\frac{35}{192}\,\frac{\mathcal N_3}{s}
    \bigl(\epsilon_{I_1L_1}\epsilon_{I_2L_2}+\epsilon_{I_1L_2}\epsilon_{I_2L_1}\bigr).
    \label{eq:ex-mhv6-b-result}
\end{equation}

We now apply the joint projection of \Eq{eq:bp-joint-amplitude-projection} to all pairs of the displayed harmonics.
Conservation of total angular momentum restricts the nonzero projections to $J_f=J_i$ and $M_f=M_i$, as in \Eq{eq:bp-joint-amplitude-schur}.
The reduced coefficients are therefore given by
\begin{equation}
    c^J_{\Lambda_f;\Lambda_i}
    =\mathcal N_3^2\int d\mu_3(Z_f)\,d\mu_3(Z_i)\,
    Y^{(f)*}_{\Lambda_f;J,M}(Z_f)
    A(Z_f;Z_i)Y^{(i)}_{\Lambda_i;J,M}(Z_i).
\label{eq:ex-mhv6-joint-projection}
\end{equation}
Here $M$ is arbitrary but fixed with no sum, as discussed below \Eq{eq:bp-joint-amplitude-schur}.
We insert the normalized harmonics of \Eqs{eq:ex-mhv6-joint-explicit-final}{eq:ex-mhv6-joint-explicit-initial}, parameterizing both configurations such that the remaining poles are phase monomials.
As above, we integrate over the initial states first, using the auxiliary frame $x'$ aligned with leg $6$, and then over the final states.
In each integral, we perform the phase integrations before the remaining angular integrations.

For the pair $(4,0)\times(4,4)$, both harmonics contain only $J=1$, so \Eq{eq:ex-mhv6-b-result} already gives
\begin{equation}
    c^1_{(4,0);(4,4)}=\beta=\frac{35}{96}\,\frac{\mathcal N_3}{s}.
    \label{eq:ex-mhv6-single-coefficient}
\end{equation}
The remaining pairs require separate projections for each spin $J$ present on both sides.
Ordering the rows by $\Lambda_f=(6,4),(5,2),(4,0)$ and the columns by $\Lambda_i=(6,2),(5,3),(4,4)$, we find
\begin{equation}
\begin{aligned}
    c^0
    &=-\frac{\mathcal N_3}{s}\,\frac{35}{288},
    &
    c^1
    &=\frac{\mathcal N_3}{s}\begin{pmatrix}
        0 & \frac{7\sqrt6}{288} & \frac{\sqrt{35}}{48}\\[0.5ex]
        -\frac{7\sqrt6}{288} & 0 & \frac{5\sqrt{210}}{288}\\[0.5ex]
        \frac{\sqrt{35}}{48} & -\frac{5\sqrt{210}}{288} & \frac{35}{96}
    \end{pmatrix},
    \\[1ex]
    c^2
    &=\frac{\mathcal N_3}{s}\begin{pmatrix}
        \frac{7}{24} & -\frac{91\sqrt{30}}{1440}\\[0.5ex]
        \frac{91\sqrt{30}}{1440} & \frac{2051}{720}
    \end{pmatrix},
    &
    c^3
    &=-\frac{\mathcal N_3}{s}\,\frac{93}{112}.
\end{aligned}
\label{eq:ex-mhv6-blocks}
\end{equation}
The coefficients $c^0$ and $c^3$ correspond to the pairs $(5,2)\times(5,3)$ and $(6,4)\times(6,2)$, respectively, while $c^2$ uses the first two shapes on each side.
Within the displayed set, $c^1$ and $c^2$ have ranks three and two, respectively, demonstrating that several independent channel components contribute at fixed $J$.
These matrices give the joint projection onto the harmonics in \Eqs{eq:ex-mhv6-joint-explicit-final}{eq:ex-mhv6-joint-explicit-initial}.

To reconstruct the full amplitude, inserting these coefficients into \Eq{eq:bp-joint-amplitude-expansion} gives the following terms in the channel expansion of \Eq{eq:bp-A_in_BJ}
\begin{equation}
    A(Z_f;Z_i)
    =\sum_{J=0}^{3}\sum_{\Lambda_f,\Lambda_i}^{\text{lowest degree}}
    c^J_{\Lambda_f;\Lambda_i}
    \sum_{M}
    Y^{(f)}_{\Lambda_f;J,M}(Z_f)
    Y^{(i)*}_{\Lambda_i;J,M}(Z_i)
    +\ldots.
    \label{eq:ex-mhv6-reconstruction}
\end{equation}
The tableau sums run over the shapes at the lowest spinor degrees allowed by the chosen helicities and containing the indicated spin $J$. The ellipsis denotes contributions involving higher-degree harmonics.
The sum over $M$ forms the channel function $\mathcal B^J_{\Lambda_f;\Lambda_i}$ of \Eq{eq:bp-A_in_BJ}.
For example, the pair evaluated above contributes
\begin{equation}
    c^1_{(4,0);(4,4)}\,\mathcal B^1_{(4,0);(4,4)}(Z_f,Z_i)
    =\frac{525}{16}\,\frac{\langle34\rangle^2[12]^2[56]^2}{s^4}.
    \label{eq:ex-mhv6-term}
\end{equation}
In \Eq{eq:ex-mhv6-reconstruction}, the \gls{ssyt} labels $\Lambda_f,\Lambda_i$ explicitly resolve the kinematic degeneracy indices at fixed $J$ in the multiparticle partial-wave expansion of \Ref{Shu:2021qlr}.
The calculated coefficients give the exact projection onto the lowest-degree harmonics, while reconstructing the full amplitude requires higher-degree contributions.
\FloatBarrier

\section{Conformal and auxiliary Casimirs}
\label{app:two-particle-casimirs}
\label{app:smeft-bases}

The conformal group $\SU(2,2)$ and the auxiliary group $\U(N)$ act on general polynomials in the spinors $\lambda_i^\alpha$ and $\widetilde\lambda_i^{\dot\alpha}$ of $N$ massless particles.
The $\U(N)$ action commutes with all conformal generators~\cite{Henning:2019mcv}.
We first collect the generators and quadratic Casimirs of both groups in spinor variables, then derive the relation between the Casimirs and its connection to the Pauli--Lubanski operator for $N=2$.
Throughout the calculation, $\lambda$ and $\widetilde\lambda$ are independent variables. The fixed-$P$ conditions of \Eq{eq:bp-fixedPintro} are imposed after evaluating the differential operators.

\subsection{Relation between conformal and auxiliary Casimirs}

We use the oscillator realization of \Ref{Henning:2019mcv}, with factors of $i$ absorbed into $\mathcal D$ and $E^i{}_j$. Writing $\partial_{i\alpha}=\partial/\partial\lambda_i^\alpha$ and $\widetilde\partial_{i\dot\alpha}=\partial/\partial\widetilde\lambda_i^{\dot\alpha}$, we denote the spinor-degree operators by
\begin{equation}
    \widehat n=\sum_i\lambda_i^\alpha\partial_{i\alpha},
    \qquad
    \widehat{\widetilde n}=\sum_i\widetilde\lambda_i^{\dot\alpha}\widetilde\partial_{i\dot\alpha}.
\label{eq:app-casimir-degree-operators}
\end{equation}
The conformal generators are the momentum $P$, the Lorentz generators $M$ and $\widetilde M$, the dilatation generator $\mathcal D$, and the special conformal generator $K$, which are given by
\begin{equation}
\begin{gathered}
    P_{\alpha\dot\alpha}
    =\sum_i\lambda_{i\alpha}\widetilde\lambda_{i\dot\alpha},
    \qquad
    K_{\alpha\dot\alpha}
    =-\sum_i\partial_{i\alpha}\widetilde\partial_{i\dot\alpha},\\[2pt]
    M_{\alpha\beta}
    =\frac{i}{2}\sum_i\bigl(\lambda_{i\alpha}\partial_{i\beta}+\lambda_{i\beta}\partial_{i\alpha}\bigr),
    \qquad
    \widetilde M_{\dot\alpha\dot\beta}
    =\frac{i}{2}\sum_i\bigl(\widetilde\lambda_{i\dot\alpha}\widetilde\partial_{i\dot\beta}+\widetilde\lambda_{i\dot\beta}\widetilde\partial_{i\dot\alpha}\bigr),\\[2pt]
    \mathcal D=N+\frac12\bigl(\widehat n+\widehat{\widetilde n}\bigr).
\end{gathered}
\label{eq:app-conformal-oscillator-generators}
\end{equation}
On a homogeneous polynomial of degrees $(n,\widetilde n)$ in $(\lambda,\widetilde\lambda)$, $\mathcal D$ has eigenvalue $\Delta=N+(n+\widetilde n)/2$, as in \Eq{eq:dimension-helicity-ssyt-counts}.
The auxiliary $\U(N)$ generators and the central $\U(1)$ generator are
\begin{equation}
    E^i{}_j
    =\lambda_i^\alpha\partial_{j\alpha}
    -\widetilde\lambda_j^{\dot\alpha}\widetilde\partial_{i\dot\alpha}, \qquad
    \widehat H=\sum_i E^i{}_i=\widehat n-\widehat{\widetilde n},
    \label{eq:app-row-un-generators}
\end{equation}
where by \Eq{eq:dimension-helicity-ssyt-counts}, $\widehat H$ has eigenvalue $2h$.

In our conventions, the conformal quadratic Casimir reads~\cite{Barbarin:1996bv,Simmons-Duffin:2016gjk}
\begin{equation}
    C_{\SU(2,2)}^2
    =-\operatorname{tr}(M^2)
    -\operatorname{tr}(\widetilde M^2)
    +\mathcal D(\mathcal D-4)
    +2P^{\alpha\dot\alpha}K_{\alpha\dot\alpha}.
    \label{eq:app-conformal-casimir}
\end{equation}
Explicitly written in terms of Lorentz indices, the traces are $\operatorname{tr}(M^2)=M_\alpha{}^\beta M_\beta{}^\alpha$, and analogously in the dotted sector.
With this normalization, the eigenvalue on a conformal primary of dimension $\Delta$ and Lorentz spins $(j_1,j_2)$ is $\Delta(\Delta-4)+2j_1(j_1+1)+2j_2(j_2+1)$. 

For $\U(N)$, we separate the traceless $\SU(N)$ generators from the central $\U(1)$ generator and use the following normalization of the quadratic Casimirs~\cite{Haber,Lukas}
\begin{equation}
    C_{\SU(N)}^2
    =\frac{1}{2N}\left(\sum_{i,j}E^i{}_jE^j{}_i-\frac{\widehat H^2}{N}\right),
    \qquad
    C_{\U(1)}^2=\widehat H^2.
    \label{eq:app-row-casimir-normalizations}
\end{equation}
With this normalization, $C_{\SU(N)}^2$ has eigenvalue $(N^2-1)/(2N^2)$ in the fundamental representation.

Next, we relate these quadratic Casimirs. To do so, we define the following normal-ordered operators, which are second-order differential operators
\begin{equation}
    S_\lambda
    =\sum_{i,j}\lambda_i^\alpha\lambda_j^\beta\partial_{j\alpha}\partial_{i\beta},
    \quad
    S_{\widetilde\lambda}
    =\sum_{i,j}\widetilde\lambda_i^{\dot\alpha}\widetilde\lambda_j^{\dot\beta}\widetilde\partial_{j\dot\alpha}\widetilde\partial_{i\dot\beta}, \quad
    R
    =\sum_{i,j}\lambda_i^\alpha\widetilde\lambda_i^{\dot\alpha}\partial_{j\alpha}\widetilde\partial_{j\dot\alpha}.
\label{eq:app-casimir-normal-ordered-terms}
\end{equation}
Commuting each derivative past the spinor immediately to its right gives for the sum in \Eq{eq:app-row-casimir-normalizations} the expression
\begin{equation}
    \sum_{i,j}E^i{}_jE^j{}_i
    =N\bigl(\widehat n+\widehat{\widetilde n}\bigr)+S_\lambda+S_{\widetilde\lambda}-2R,
    \label{eq:app-casimir-normal-ordering-sum}
\end{equation}
while for the relevant terms in \Eq{eq:app-conformal-casimir}, we find
\begin{equation}
\begin{aligned}
    -\operatorname{tr}(M^2)
    &=S_\lambda+2\widehat n-\frac12\widehat n^2,
    \qquad
    -\operatorname{tr}(\widetilde M^2)
    =S_{\widetilde\lambda}+2\widehat{\widetilde n}-\frac12\widehat{\widetilde n}^{\,2},\\
    \mathcal D(\mathcal D-4)
    &=N(N-4)+(N-2)\bigl(\widehat n+\widehat{\widetilde n}\bigr)+\frac14\bigl(\widehat n+\widehat{\widetilde n}\bigr)^2,\\
    P^{\alpha\dot\alpha}K_{\alpha\dot\alpha}&=-R.
\end{aligned}
\label{eq:app-casimir-normal-ordering}
\end{equation}
Inserting these expressions into \Eq{eq:app-conformal-casimir} yields
\begin{equation}
\begin{aligned}
    C_{\SU(2,2)}^2
    &=N(N-4)+N\bigl(\widehat n+\widehat{\widetilde n}\bigr)+S_\lambda+S_{\widetilde\lambda}-2R-\frac14\bigl(\widehat n-\widehat{\widetilde n}\bigr)^2\\
    &=N(N-4)+\sum_{i,j}E^i{}_jE^j{}_i-\frac14\widehat H^2.
\end{aligned}
\label{eq:app-conformal-row-trace}
\end{equation}
From \Eq{eq:app-row-casimir-normalizations} we see that this gives the identity
\begin{equation}
    \boxed{
    C_{\SU(2,2)}^2-N(N-4)
    =2N C_{\SU(N)}^2-\frac{N-4}{4N}C_{\U(1)}^2
    }.
    \label{eq:app-casimir-identity}
\end{equation}
Within the oscillator representation, the reductive dual pair structure underlying conformal-helicity duality establishes a one-to-one correspondence between the allowed $\U(N)$ representations and the conformal $\SU(2,2)$ representations together with their total helicity~\cite{Henning:2019mcv,Basile:2020gqi}. 
Equation~\eqref{eq:app-casimir-identity} gives the corresponding relation between the quadratic Casimirs.

For $N=4$, the constant and $\U(1)$ terms in \Eq{eq:app-casimir-identity} vanish, giving
\begin{equation}
    C_{\SU(2,2)}^2=8C_{\SU(4)}^2.
    \label{eq:app-four-particle-casimir-identity}
\end{equation}
For four particles, the $\SU(4)$ representation alone therefore determines the conformal quadratic Casimir, without requiring the total helicity as an additional label. For $N=2$, the Casimir relation provides a direct connection to the Pauli--Lubanski operator, which we derive next.

\subsection{Two-particle channels and the Pauli--Lubanski Casimir}

For $N=2$, \Eq{eq:app-casimir-identity} becomes
\begin{equation}
    C_{\SU(2,2)}^2+4-\frac14 C_{\U(1)}^2
    =4C_{\SU(2)}^2.
    \label{eq:app-two-particle-conformal-casimir}
\end{equation}
To relate this result to $W^2$, we take the total two-particle momentum $P$ to be timelike, with $s=P^2>0$. For a particle set $\mathcal I$, let $M_{\mathcal I}$ and $\widetilde M_{\mathcal I}$ denote the generators of \Eq{eq:app-conformal-oscillator-generators} with the sums restricted to $i\in\mathcal I$. The explicit expression for the Pauli--Lubanski vector squared of \Ref{Jiang:2020rwz} in the normalization of the generators used in this appendix reads
\begin{equation}
    W_{\mathcal I}^2
    =\frac{s}{2}\left[\operatorname{tr}(M_{\mathcal I}^2)+\operatorname{tr}(\widetilde M_{\mathcal I}^2)\right]
    -P^{\alpha\dot\alpha}P^{\beta\dot\beta}M_{\mathcal I,\alpha\beta}\widetilde M_{\mathcal I,\dot\alpha\dot\beta}.
    \label{eq:app-pauli-lubanski-spinor-form}
\end{equation}

This expression holds for any particle set $\mathcal I$, with $P=\sum_{i\in\mathcal I}p_i$.

For $\mathcal I=\{1,2\}$, substituting the two-particle momentum into the mixed term gives
\begin{equation}
    P^{\alpha\dot\alpha}P^{\beta\dot\beta}M_{\mathcal I,\alpha\beta}\widetilde M_{\mathcal I,\dot\alpha\dot\beta}
    =\frac{s}{2}\left(\widehat n\,\widehat{\widetilde n}-2R\right).
    \label{eq:app-two-particle-mixed-contraction}
\end{equation}
Combining this expression with \Eq{eq:app-casimir-normal-ordering} in \Eq{eq:app-pauli-lubanski-spinor-form}, and using \Eqs{eq:app-casimir-normal-ordering-sum}{eq:app-row-casimir-normalizations} for $N=2$, gives
\begin{equation}
    \boxed{
    W_{\mathcal I}^2=-2sC_{\SU(2)}^2
    =-\frac{s}{2}\left(C_{\SU(2,2)}^2+4-\frac14 C_{\U(1)}^2\right)
    }
    \qquad(N=2).
    \label{eq:app-two-particle-casimir-identity}
\end{equation}
On a two-particle state of total spin $J$, $W_{\mathcal I}^2$ has eigenvalue $-sJ(J+1)$.

For two particles, the matrix $Z$ in \Eq{eq:bp-Lambda-Zx} is a $2\times2$ matrix, so the fixed-$P$ condition $Z^\dagger Z=\mathbf1_2$ makes it unitary. The allowed spinor configurations therefore form $V_2(\mathbb C^2)=\U(2)$.
The corresponding left $\SU(2)$ and right $\SU(2)_P$ representations have the same spin $J$.
Thus, each $\U(2)$ representation fixes the physical total spin, as discussed in \App{appsec:joint_basis}.

\newpage

\bibliographystyle{JHEP}
\addcontentsline{toc}{section}{Bibliography}
\bibliography{bibliography}

\end{document}